\documentclass[pdflatex,sn-mathphys]{sn-jnl} 

\jyear{2022}%

\theoremstyle{thmstyleone}%
\theoremstyle{thmstyletwo}%

\theoremstyle{thmstylethree}%

\usepackage{siunitx} 
\usepackage{amsmath,amstext,color,tabu}
\usepackage{gensymb}
\usepackage{natbib}
\usepackage{hyperref}
\usepackage{longtable}
\usepackage{graphicx}
\usepackage{tablefootnote}

\usepackage{lineno}

\begin{document}

\makeatletter
\let\jnl@style=\rm
\def\ref@jnl#1{{\jnl@style#1}}

\def\aj{\ref@jnl{AJ}}                   
\def\actaa{\ref@jnl{Acta Astron.}}      
\def\araa{\ref@jnl{ARA\&A}}             
\def\apj{\ref@jnl{ApJ}}                 
\def\apjl{\ref@jnl{ApJ}}                
\def\apjs{\ref@jnl{ApJS}}               
\def\ao{\ref@jnl{Appl.~Opt.}}           
\def\apss{\ref@jnl{Ap\&SS}}             
\def\aap{\ref@jnl{A\&A}}                
\def\aapr{\ref@jnl{A\&A~Rev.}}          
\def\aaps{\ref@jnl{A\&AS}}              
\def\azh{\ref@jnl{AZh}}                 
\def\baas{\ref@jnl{BAAS}}               
\def\bac{\ref@jnl{Bull. astr. Inst. Czechosl.}}
\def\caa{\ref@jnl{Chinese Astron. Astrophys.}}
\def\cjaa{\ref@jnl{Chinese J. Astron. Astrophys.}}
\def\icarus{\ref@jnl{Icarus}}           
\def\jcap{\ref@jnl{J. Cosmology Astropart. Phys.}}
\def\jrasc{\ref@jnl{JRASC}}             
\def\memras{\ref@jnl{MmRAS}}            
\def\mnras{\ref@jnl{MNRAS}}             
\def\na{\ref@jnl{New A}}                
\def\nar{\ref@jnl{New A Rev.}}          
\def\pra{\ref@jnl{Phys.~Rev.~A}}        
\def\prb{\ref@jnl{Phys.~Rev.~B}}        
\def\prc{\ref@jnl{Phys.~Rev.~C}}        
\def\prd{\ref@jnl{Phys.~Rev.~D}}        
\def\pre{\ref@jnl{Phys.~Rev.~E}}        
\def\prl{\ref@jnl{Phys.~Rev.~Lett.}}    
\def\pasa{\ref@jnl{PASA}}               
\def\pasp{\ref@jnl{PASP}}               
\def\pasj{\ref@jnl{PASJ}}               
\def\rmxaa{\ref@jnl{Rev. Mexicana Astron. Astrofis.}}%
\def\qjras{\ref@jnl{QJRAS}}             
\def\skytel{\ref@jnl{S\&T}}             
\def\solphys{\ref@jnl{Sol.~Phys.}}      
\def\sovast{\ref@jnl{Soviet~Ast.}}      
\def\ssr{\ref@jnl{Space~Sci.~Rev.}}     
\def\zap{\ref@jnl{ZAp}}                 
\def\nat{\ref@jnl{Nature}}              
\def\iaucirc{\ref@jnl{IAU~Circ.}}       
\def\aplett{\ref@jnl{Astrophys.~Lett.}} 
\def\apspr{\ref@jnl{Astrophys.~Space~Phys.~Res.}}
\def\bain{\ref@jnl{Bull.~Astron.~Inst.~Netherlands}} 
\def\fcp{\ref@jnl{Fund.~Cosmic~Phys.}}  
\def\gca{\ref@jnl{Geochim.~Cosmochim.~Acta}}   
\def\grl{\ref@jnl{Geophys.~Res.~Lett.}} 
\def\jcp{\ref@jnl{J.~Chem.~Phys.}}      
\def\jgr{\ref@jnl{J.~Geophys.~Res.}}    
\def\jqsrt{\ref@jnl{J.~Quant.~Spec.~Radiat.~Transf.}}
\def\memsai{\ref@jnl{Mem.~Soc.~Astron.~Italiana}}
\def\nphysa{\ref@jnl{Nucl.~Phys.~A}}   
\def\physrep{\ref@jnl{Phys.~Rep.}}   
\def\physscr{\ref@jnl{Phys.~Scr}}   
\def\planss{\ref@jnl{Planet.~Space~Sci.}}   
\def\procspie{\ref@jnl{Proc.~SPIE}}   
\makeatother
\let\astap=\aap
\let\apjlett=\apjl
\let\apjsupp=\apjs
\let\applopt=\ao



\title[GRB\,191019A]{A long-duration gamma-ray burst of dynamical origin from the nucleus of an ancient galaxy}


\author[1,2]{\fnm{Andrew J.} \sur{Levan}}\email{a.levan@astro.ru.nl}
\author[1,3,4,5]{\fnm{Daniele B.} \sur{Malesani}}
\author[6]{\fnm{Benjamin P.} \sur{Gompertz}}
\author[7]{\fnm{Anya E.} \sur{Nugent}}
\author[6]{\fnm{Matt} \sur{Nicholl}}
\author[6]{\fnm{Samantha R.} \sur{Oates}}
\author[8]{\fnm{Daniel A.} \sur{Perley}}
\author[7]{\fnm{Jillian} \sur{Rastinejad}}
\author[9,10]{\fnm{Brian D.} \sur{Metzger}}
\author[11]{\fnm{Steve} \sur{Schulze}}
\author[2]{\fnm{Elizabeth R.} \sur{Stanway}}
\author[1,12]{\fnm{Anne} \sur{Inkenhaag}}
\author[13,14]{\fnm{Tayyaba} \sur{Zafar}}
\author[15]{\fnm{J. Feliciano}
\sur{Ag\"u\'i Fern\'andez}}
\author[1]{\fnm{Ashley A.} \sur{Chrimes}}
\author[16]{\fnm{Kornpob} \sur{Bhirombhakdi}}
\author[17]{\fnm{Antonio} \sur{de Ugarte Postigo}}
\author[7]{\fnm{Wen-fai} \sur{Fong}}
\author[16]{\fnm{Andrew S.} \sur{Fruchter}}
\author[7]{\fnm{Giacomo} \sur{Fragione}}
\author[3,4]{\fnm{Johan P. U.} \sur{Fynbo}}
\author[1]{\fnm{Nicola} \sur{Gaspari}}
\author[3,4]{\fnm{Kasper E.} \sur{Heintz}}
\author[18]{\fnm{Jens} \sur{Hjorth}}
\author[19]{\fnm{Pall} \sur{Jakobsson}}
\author[1,12]{\fnm{Peter G.} \sur{Jonker}}
\author[8,20]{\fnm{Gavin P.} \sur{Lamb}}
\author[21,22]{\fnm{Ilya} \sur{Mandel}}
\author[23]{\fnm{Soheb} \sur{Mandhai}}
\author[1,24]{\fnm{Maria E.} \sur{Ravasio}}
\author[25]{\fnm{Jesper} \sur{Sollerman}}
\author[20]{\fnm{Nial R.} \sur{Tanvir}}

\affil[1]{\orgdiv{Department of Astrophysics/IMAPP}, \orgname{Radboud University}, \orgaddress{\street{6525 AJ Nijmegen}, \country{The Netherlands}}}

\affil[2]{\orgdiv{Department of Physics}, \orgname{University of Warwick}, \orgaddress{\street{Coventry, CV4 7AL}, \country{UK}}}

\affil[3]{\orgdiv{Cosmic Dawn Center (DAWN)}, \orgaddress{\country{Denmark}}}

\affil[4]{\orgdiv{Niels Bohr Institute}, \orgname{University of Copenhagen}, \orgaddress{\street{Jagtvej 128}, \postcode{2200}, \city{Copenhagen N}, \country{Denmark}}}

\affil[5]{\orgdiv{DTU Space}, \orgname{Technical University of Denmark}, \orgaddress{\street{Elektrovej 327}, \postcode{2800}, \city{Kongens Lyngby}, \country{Denmark}}}

\affil[6]{\orgdiv{Institute for Gravitational Wave Astronomy and School of Physics and Astronomy}, \orgname{University of Birmingham}, \orgaddress{\street{B15 2TT}, \country{UK}}}

\affil[7]{\orgdiv{Center for Interdisciplinary Exploration and Research in Astrophysics and Department of Physics and Astronomy}, \orgname{Northwestern University}, \orgaddress{\street{2145 Sheridan Road}, \city{Evanston}, \postcode{60208-3112}, \state{IL}, \country{USA}}}


\affil[8]{\orgdiv{Astrophysics Research Institute}, \orgname{Liverpool John Moores University},  \orgaddress{\street{Liverpool Science Park, 146 Brownlow Hill}, \city{Liverpool}, \country{UK}, \postcode{L3 5RF}}}

\affil[9]{\orgdiv{Center for Computational Astrophysics}, \orgname{Flatiron Institute}, \orgaddress{\street{162 W. 5th Avenue}, \city{New York}, \postcode{10011}, \state{NY}, \country{USA}}}

\affil[10]{\orgdiv{Department of Physics and Columbia Astrophysics Laboratory}, \orgname{Columbia University}, \orgaddress{\city{New York}, \postcode{10027}, \state{NY}, \country{USA}}}

\affil[11]{\orgdiv{Department of Physics, The Oskar Klein Center}, \orgname{Stockholm University}, \orgaddress{\street{AlbaNova} \city{Stockholm},   \country{Sweden}}}

\affil[12]{\orgdiv{SRON}, \orgname{Netherlands Institute for Space Research}, \orgaddress{\street{Niels Bohrweg 4}, \postcode{NL-2333 CA}, \city{Leiden}, \country{the Netherlands}}}

\affil[13]{\orgname{Australian Astronomical Optics, Macquarie University,  \orgaddress{\street{105 Delhi Road, \city{North Ryde}, NSW \postcode{2113}, \country{Australia}}}}}

\affil[14]{\orgname{Astronomy, Astrophysics and Astrophotonics Research Centre, Macquarie University, \city{Sydney}, NSW \postcode{2109}, \country{Australia}}}

\affil[15]{\orgname{Instituto de Astrof\'isica de Andaluc\'ia (IAA-CSIC)}, \orgaddress{\street{Glorieta de la Astronom\'ia s/n}, \postcode{18008}, \city{Granada}, \country{Spain}}}

\affil[16]{\orgname{Space Telescope Science Institute}, \orgaddress{\street{3700 San Martin Drive} \city{Baltimore}, \postcode{21218}, \state{MD}, \country{USA}}}

\affil[17]{\orgdiv{Université Côte D'Azur}, \orgname{Observatoire de la Côte D'Azur, CNRS, Artemis}, \orgaddress{\city{Nice}, \postcode{F-063004}, \country{France}}}


\affil[18]{\orgdiv{DARK}, \orgname{Niels Bohr Institute, University of Copenhagen}, \orgaddress{\street{Jagtvej 128}, \postcode{2200} \city{Copenhagen}, \country{Denmark}}}

\affil[19]{\orgdiv{Centre for Astrophysics and Cosmology, Science Institute}, \orgname{University of Iceland}, \orgaddress{\street{Dunhagi 5}, \city{Reykjav\'ik}, \postcode{107}, \country{Iceland}}}

\affil[20]{\orgdiv{School of Physics and Astronomy}, \orgname{University of Leicester}, \orgaddress{\street{LE1 7RH, University Road}, \city{Leicester}, \country{UK}}}

\affil[21]{\orgdiv{Monash Centre for Astrophysics, School of Physics and Astronomy }, \orgname{Monash University}, \orgaddress{\street{3800, Clayton},  \state{VIC}, \country{Australia}}}

\affil[22]{\orgdiv{OzGrav}, \orgname{ARC Centre of Excellence for Gravitational Wave Discovery}, \orgaddress{\street{Hawthorn},   \country{Australia}}}

\affil[23]{School of Physics and Astronomy, University of Leicester, University Road, Leicester LE1 7RH, UK; Jodrell Bank Centre for Astrophysics}

\affil[24]{\orgdiv{INAF}, \orgname{Astronomical Observatory of Brera}, \orgaddress{\street{via E. Bianchi 46}, \postcode{23807}, \city{Merate (LC)}, \country{Italy}}}

\affil[25]{\orgdiv{Department of Astronomy, The Oskar Klein Center}, \orgname{Stockholm University}, \orgaddress{\street{AlbaNova},  \city{Stockholm},   \country{Sweden}}}

\abstract{
The majority of long duration ($>2$ s) gamma-ray bursts (GRBs) are believed to arise from the collapse of massive stars \cite{Hjorth+03}, with a small proportion created from the merger of compact objects \cite{rastinejad22,troja22,yang22}. Most of these systems are likely formed via standard stellar evolution pathways.
However, it has long been thought that a fraction of GRBs 
may instead be an outcome of
dynamical interactions in dense environments \cite{grindlay06,fragione19,mckernan20}, channels which could also contribute significantly to 
the samples of compact object mergers detected as gravitational wave sources \cite{oleary16}.
Here we report the case of GRB\,191019A, a long GRB ($T_{90} = 64.4 \pm 4.5$ s) which we pinpoint close ($\lessapprox 100$ pc projected) to the nucleus of
an ancient ($>1$~Gyr old) host galaxy at $z=0.248$. The lack of evidence for star formation and deep limits on any supernova emission make a massive star origin difficult to reconcile with observations, while
the timescales of the emission rule out a direct interaction with the supermassive black hole in the nucleus of the galaxy, 
We suggest that the most likely route for progenitor formation is via dynamical interactions in the dense nucleus of the host, 
consistent with the centres of such galaxies
exhibiting interaction rates up to two orders of magnitude larger than typical field galaxies \cite{stone16}. 
The burst properties could naturally be explained via compact object mergers involving white dwarfs (WD), neutron stars (NS) or black holes (BH). These may form dynamically in dense stellar clusters, or originate in a gaseous disc around the supermassive black hole \cite{perna21,lazzati22}. Future electromagnetic and gravitational-wave observations in tandem thus offer a route to probe the dynamical fraction and the details of dynamical interactions in galactic nuclei and other high density stellar systems. 

}

\maketitle

The evolution of most stars in the Universe is dominated by their stellar or binary evolution. However, for a small fraction in dense environments additional many-body interactions create new channels to the formation of exotic stellar systems, such as the progenitors of gamma-ray bursts. These bursts arise in at least two varieties. The first is formed from the collapse of massive stars and typically with duration $>$ 2s \cite{Galama+98}. The second arises from the mergers of compact objects, and typically have duration $<$2s \cite{Abbott+17a}, although recent evidence demonstrates some can be much longer \cite{rastinejad22,troja22,yang22}.

\begin{figure*}
\centerline{
\includegraphics[width=1.0\textwidth]{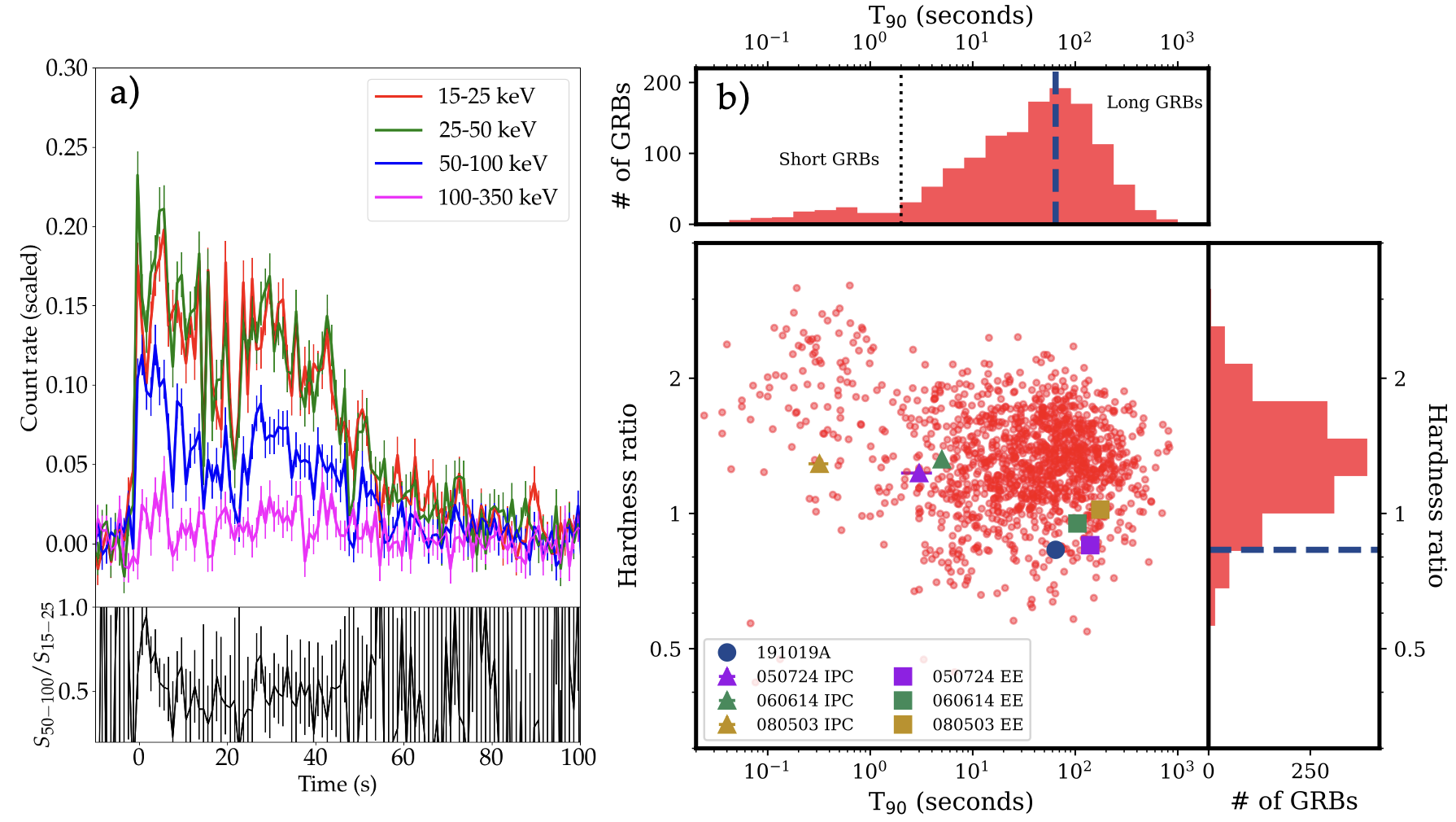}}
\caption{a) The $\gamma$-ray light curve of GRB\,191019A as observed by the {\em Swift}-BAT. The burst consists of a single emission episode, with additional 
intrinsic variability. The burst begins with a short spike, but it is not especially hard, nor separated from the bulk of the emission. The lower panel shows the hardness ratio between the 50--100 and 15--25 keV bands, demonstrating some degree of spectral softening, with the initial peak being the hardest emission episode. b) The location of GRB\,191019A on the hardness--duration plane. The background red points represent bursts from the {\em Swift}-BAT catalog \cite{Lien+16}, while GRB\,191019A is indicated with the dark blue circle. Also marked are the locations of bursts identified as short+EE based on the duration of their initial complex and extended emission (EE) separately. The properties of GRB\,191019A are comparable with the properties of the EE-component in other bursts.  }
\label{fig:bat_lc}
\end{figure*}

GRB\,191019A was detected by the {\em Neil Gehrels Swift Observatory} (hereafter {\em Swift}) at 15:12:33 UT on 19 October 2019 \cite{simpson19}. 
The burst is characterised by a fast rise and slower decay with additional variability superimposed (Figure~\ref{fig:bat_lc}). The duration is measured to be $T_{90} = 64.4 \pm 4.5$ s \cite{krimm19}, hence classified as a long GRB.
The burst is relatively soft with a power law photon 
index of $\Gamma = 2.25 \pm 0.05$. Its fluence is $S = (1.00 \pm 0.03) \times 10^{-7}$ erg cm$^{-2}$ (15--150 keV) \cite{krimm19}. 

\begin{figure*}
\centerline{
\includegraphics[width=1.0\textwidth]{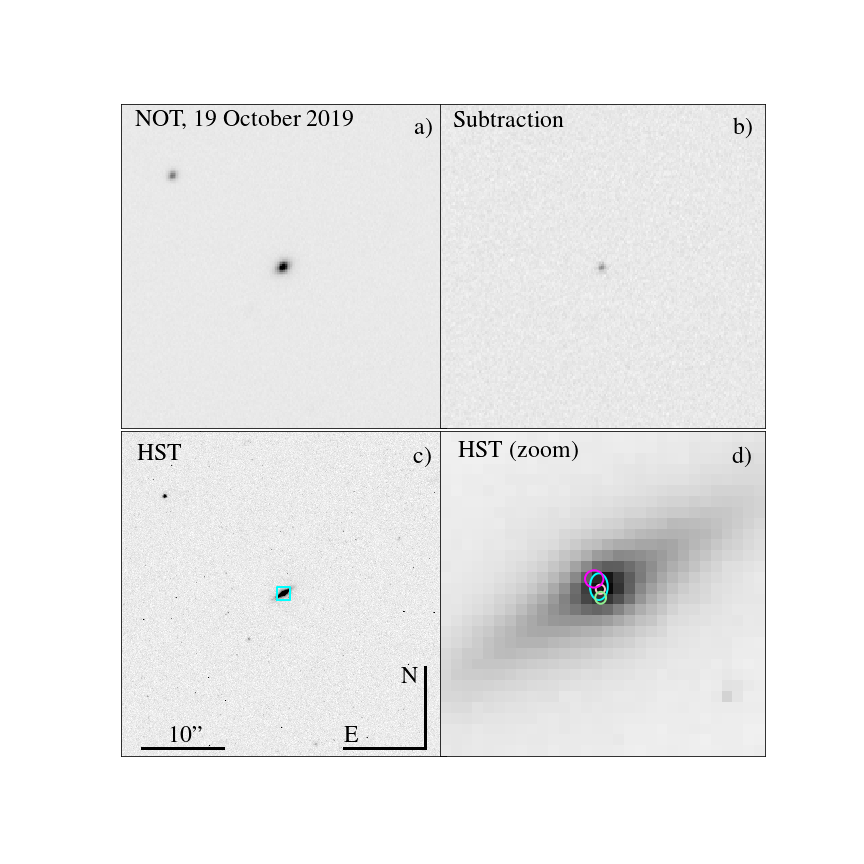}}
\caption{Optical images of the afterglow of GRB\,191019A and its host galaxy. a) The $i$-band afterglow discovery image from the NOT. b) The result of a PSF-matched image subtraction with an image taken on 29 October. A residual  is clearly visible at the centre of the galaxy. c) The field as observed by {\em HST} in April 2020, matched to the NOT images. d) A zoomed in region around the host galaxy of GRB 191019A as seen with {\em HST} (as indicated with the cyan box in panel c). The ellipses indicate the $2\sigma$ uncertainty regions for the optical afterglow on the host as inferred from the NOT $g$ (cyan), $r$ (green), $i$ (yellow) and $z$ (magenta). The location of the afterglow is consistent with the nucleus of the host galaxy with a projected offset, based on the $i$-band measurement, of $r_\text{proj} = 78 \pm 109$ pc.  }
\label{fig:finder}
\end{figure*}

\begin{figure*}
\centerline{
\includegraphics[width=1.0\textwidth]{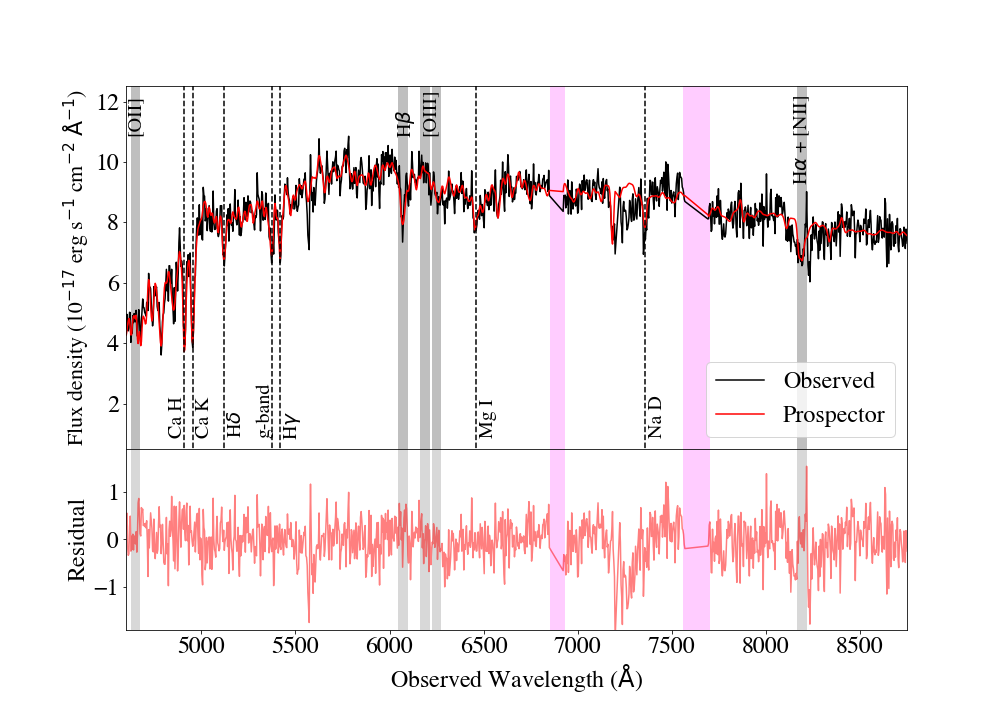}}
\caption{ The optical spectrum of the host galaxy of GRB\,191019A
as observed with the NOT. The spectrum shows no
emission lines associated with star formation (the expected locations of strong emission lines are marked with grey bands, and telluric absorption in pink). 
There is weak evidence for emission from
[N\,{\sc ii}] (6584~\AA). The locations of prominent absorption features from which the redshift is determined are marked with
dashed lines. Also shown are the results of a \texttt{Prospector} fit to the stellar spectrum (e.g. omitting any emission lines). Any lines would appear in the residuals. }
\label{fig:spec}
\end{figure*}

Space-craft constraints prevented a prompt slew from {\em Swift}, and observations with the X-ray Telescope (XRT) and the Ultraviolet and Optical Telescope (UVOT) began 52 minutes after the burst. These revealed an X-ray and UV afterglow \cite{evans19}. We obtained optical observations of the field with the Nordic Optical Telescope (NOT) beginning 4.52 hours after the burst \cite{perley19}.
Comparison with later epochs reveals a faint afterglow positionally consistent with the nucleus of the host galaxy visible in each of the $g,r,i$ and $z$-bands (Figure~\ref{fig:finder}). Spectroscopy  obtained with the NOT on 19 October 2019, and confirmed with the Gemini-South telescope on 1 December 2019, found a redshift of $z=0.248$ based on several absorption lines, including Ca H\&K and the hydrogen Balmer series (Figure~\ref{fig:spec}). The standard star-forming emission lines are notably absent from these spectra, suggesting an old galaxy.

\begin{figure*}
\centerline{
\includegraphics[width=1.2\textwidth]{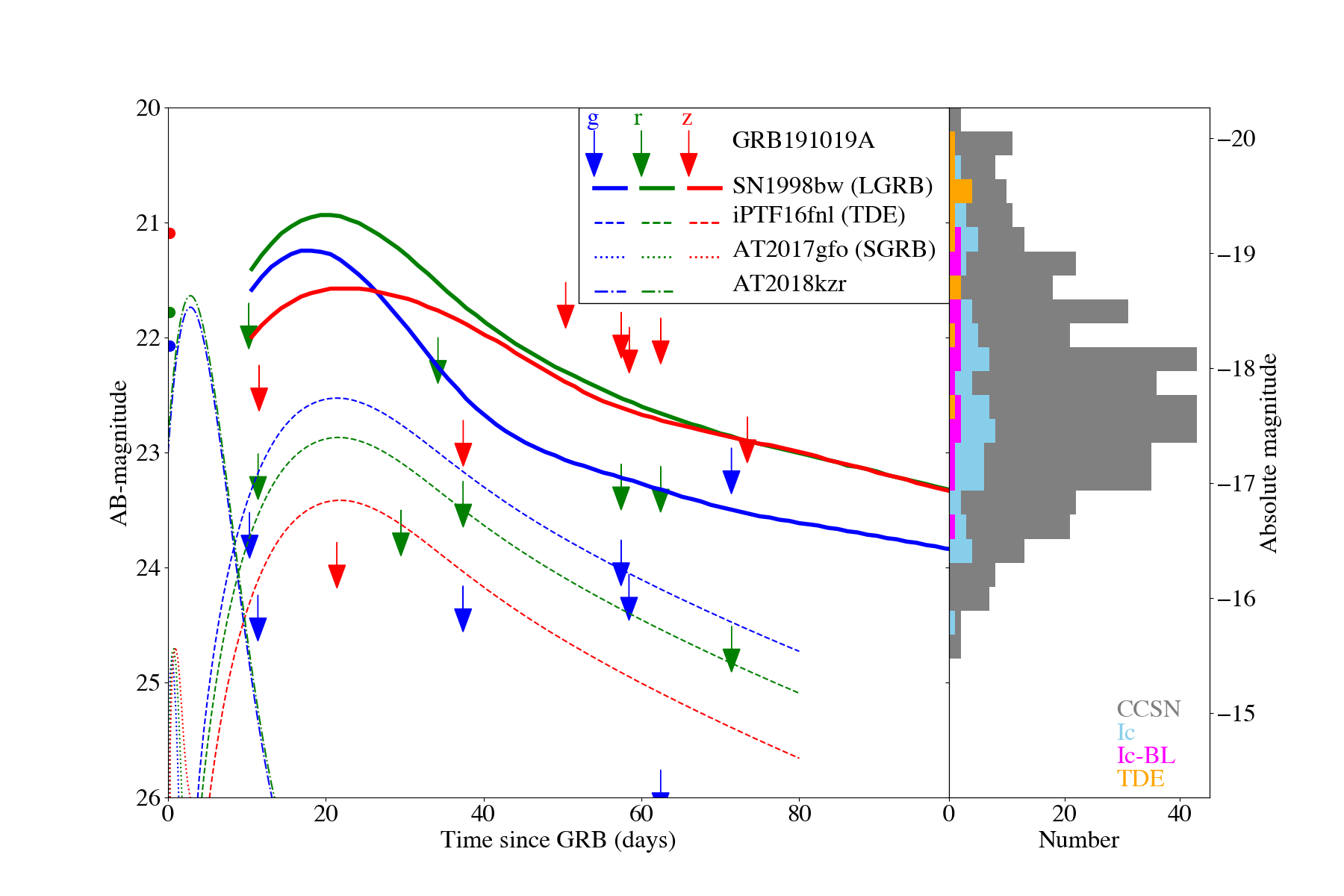}}
\caption{Comparison between the upper-limits obtained from our targeted observations of GRB 191019A and the expectations of the lightcurve from supernova or tidal disruption events. The upper limits represent the depth of our NOT, Gemini and {\em HST} 
observations, while the solid lines correspond to the expectations of SN~1998bw at $z=0.248$, based on the light curves of \cite{clocchiatti11}. 
The right hand panel shows histograms
of the peak absolute magnitude distributions of supernovae and tidal disruption events found by ZTF \cite{perley20,hammerstein22}.  Also shown are the faintest and fastest evolving tidal disruption event iPTF16fnl \cite{blagorodnova17,nicholl22}, AT2018kzr, suggested to form via a BH-WD mergers \cite{mcbrien19} and AT2017gfo, associated with GW170817 \cite{villar+17}. Our optical observations reach a depth where we would have expected to observe the vast majority of supernovae or tidal disruption events. However, we do not have sensitivity to detect kilonovae like AT2017gfo.}
\label{fig:sn}
\end{figure*}

Following these observations, we obtained 
deep imaging 
in the $g$, $r$ and $z$-bands from the NOT and the Gemini-South telescope from 2 to 73 days after the burst, and optical imaging with the {\em Hubble Space Telescope} at 30 and 184 days. None of these images
reveal any evidence for transient emission to limits of
typically $g>24$, $r>23.5$, $z>22$ (see Figure~\ref{fig:sn}).

The non-detection of optical light between 2 and 70 days places stringent limits on any associated supernova  to levels $\sim 20$ times fainter than SN~1998bw  (Figure~\ref{fig:sn}; see also Methods). In fact, the deepest $r$-band/F606W limits reach absolute magnitudes of  $M \sim -16$. This is comparable to the faint end of the core-collapse supernova distribution and fainter than any known stripped-envelope events found in the
large sample from the Zwicky Transient Factory \cite{perley20}. It is also fainter than optically selected tidal disruption events \cite{nicholl22,hammerstein22}. The limiting luminosity is comparable to the peak luminosity of kilonovae. However, our observations probe much longer timescales than those of kilonovae, 
such that we could not rule out events equivalent to AT2017gfo \cite{villar+17}.
The lack of a SN detection cannot readily be ascribed to dust extinction since the spectral energy distribution of the afterglow constrains this to be small (see Methods).

Combining {\em HST} UV observations with our spectroscopy and archival imaging, we fit the available photometric and spectroscopic data with the stellar population inference code \texttt{Prospector} (Figure~\ref{fig:spec} and see Methods). The results favour an old stellar population for the host, with the majority of stellar mass forming $>1$ Gyr ago, and
little ongoing star formation ($\approx 0.05~M_\odot$~yr$^{-1}$). 
The stellar mass itself is found to be $\approx 3 \times 10^{10}~M_\odot$.

The location of GRB 191019A in the galaxy nucleus could indicate an origin associated with the supermassive black hole which resides there, with scaling relations implying a black hole with a mass of a $\text{few} \times 10^7$ $M_\odot$ \cite{kormendy13}. However, the timescales for the emission are too short for either variability in an active galactic nucleus (AGN) or a tidal disruption event (TDE) (see Methods). Instead, the burst most likely arises from a stellar progenitor. The lack of a supernova and the location in an old population rule out a massive star. Instead, it appears that GRB 191019A belongs to the population of apparently long GRBs formed from compact object mergers \cite{rastinejad22,troja22,yang22}. Its apparent energy release and afterglow luminosity are consistent with this group of GRBs (Methods). 

However, the nuclear location of the GRB on its host galaxy differs from compact object merger expectations. Systems formed via standard stellar evolution channels involve two supernovae; at each supernova, the combination of natal kicks and those induced from mass loss combined to give the binary a substantial (50--500 km s$^{-1}$) 
systemic velocity. Indeed, no short GRB with sub-arcsecond localisation is consistent with the nucleus of its host galaxy \cite{fong22}. 

We suggest instead that the binary which created GRB 191019A formed via dynamical interactions in the dense nucleus of its host galaxy. Dynamical channels for compact object formation may be due to many-body interactions in dense stellar systems such as globular clusters \cite{grindlay06,ye20} or nuclear star clusters in galaxies \cite{antonini16,fragione19}. Alternatively, they may also form at a significantly enhanced rate in the gaseous discs that surround supermassive black holes \cite{mckernan20,tagawa20}.

The host galaxy of GRB 191019A appears similar to those that preferentially host tidal disruption events, with a very compact core and Balmer absorption lines. The Lick indices for H$\delta$ in absorption and H$\alpha$ in emission are $1.54^{+1.44}_{-0.74}$ and $2.51^{+1.81}_{-2.51}$, consistent with those of the TDE population which make up only $\sim 2$\% of SDSS galaxies, but 75\% of the TDE hosts \cite{French17}. The TDE rate effectively measures the
stellar interaction rate close to the black hole. Scattering events are responsible for placing stars on paths which cross closer than the tidal radius for the star around the SMBH. 
The preference for TDEs in galaxies of certain types is related directly to their dense stellar environments, and interaction rates \cite{stone16}. At face value, then, the host galaxy of GRB\,191019A may have a dynamical interaction rate one to two orders of magnitude larger than typical galaxies.

If GRB\,191019A results from a dynamically formed compact object merger, it may arise from several possible merger products, including NS-NS, NS-BH, NS-WD and BH-WD. The nature of the merger product and its location (e.g. stellar cluster versus gas disc) should have a direct impact on
the observed properties of the burst, particularly concerning duration, spectral hardness and energetics. 

In the case of NS-NS or NS-BH systems, one may wonder why no apparent short ($< 2$~s) spike is observed in the prompt lightcurve, as in short GRBs with extended emission. The detection of the kilonova in GRB 211211A demonstrates that such a short spike is not necessarily required, although GRB 211211A appears to show other similarities to extended emission (EE) bursts \cite{gompertz22}. However, GRB 191019A may arise from a similar population where the contrast between ``spike" and ``extended" emission is smaller \cite{perley09}, or that the extended emission is beamed with a larger opening angle than the initial spike and is unseen in this case
\cite{Bucciantini12,lu22}. Alternatively, mergers involving white dwarfs have longer timescales naturally \cite{king07}, 
and such an event is also possible here. Indeed, interactions in dense clusters tend to leave the more massive components in binaries, so BH-NS or BH-WD mergers may be favoured \cite{ye20}. White dwarf-containing systems should yield rapid, relatively faint transients, with one event, AT2018kzr \cite{mcbrien19,Gillanders20}, suggested to arise from the merger of a white dwarf with a black hole. Our observations are not sufficiently sensitive to constrain the presence of such a signal in 
GRB 191019A. 

Alternatively, the nuclear location also allows compact object mergers within a disc around the supermassive black hole, although there is no
strong evidence for AGN activity in the host (see Methods). In these discs, the compact object binaries are frequently formed by ``gas capture" mergers, which can substantially enhance the rate, despite the relatively small number of stars within the disc \cite{tagawa20}.
In this scenario, the long duration may well be expected, even for an intrinsically short engine. The higher densities within the disc cause the external shock to form and slow much closer to the progenitor than in bursts with a normal interstellar medium density. This extra baryon loading may effectively choke the jet \cite{perna21} for very high densities. However, this interaction's effect smears out the prompt emission over an extended period. A very recent and explicit prediction of compact object mergers within discs is that intrinsically short-hard GRBs should become longer and softer \cite{lazzati22}, with a notable hard-soft evolution. This is exactly what is seen in GRB\,191019A. 

It is relevant to consider whether similar events exist within the GRB population but have been hitherto unrecognised. The vast majority of long GRB hosts are star-forming galaxies and, where searches are possible, usually show the signatures of broad-lined type Ic supernovae. There is a small population of long bursts without supernova
signatures \cite{Fynbo+06,GalYam+06}. Some of these have already been classified as short GRBs with extended emission \citep{Gehrels+06}, however there are additional bursts which bear further scrutiny. GRB 111005A \cite{Michalowski+18} was localised only via its radio afterglow but has deep limits on associated supernova emission. It lies in a local galaxy at only 55 Mpc and is also 
close to the nucleus. It could well have arisen from a compact object merger as suggested by \cite{le22} and its location raises the prospect of dynamical formation. 
GRB 050219A does not have a sub-arcsecond localisation, but is likely associated with a post-starburst galaxy whose properties are similar to the host of GRB\,191019A \cite{rossi14}. Finally, there are
several long GRBs whose locations are consistent with their host nucleus \cite{fruchter06}, although most of these are in star-forming hosts and likely arise from massive star collapse. 
Overall, the observational evidence suggests that, at most, a few per cent of the observed (long and short) GRB population forms via dynamical channels and that most of the observed systems arise via stellar (binary) evolution.

Identifying a likely dynamically produced GRB offers some of the first evidence for forming stellar-mass compact objects via dynamical channels in galactic nuclei. The mergers of such systems
have received significant attention as a possible explanation for a fraction of the observed gravitational-wave 
population, particularly with regard to more massive black holes which can be formed via successive mergers \cite{fragione22}.
The
gamma-ray bright population of mergers may be dwarfed by those that do not emit such high-energy flashes. In 
particular, very high densities within gaseous discs can result in the choking of any GRB-like emission
\cite{perna21}, and BH-BH mergers are generally expected to be EM-dark. 
GRBs in dense galactic nuclei therefore offer a unique new route to probe exotic compact object formation channels.

\clearpage

\section*{Methods}
\label{sec:methods}

\subsection*{\textit{Swift} Observations}

\subsubsection*{BAT}

BAT data were downloaded from the UK \emph{Swift} Science Data Centre (UKSSDC; \cite{Evans07,Evans09}). Reduction was performed using the dedicated pipeline \texttt{batgrbproduct} v2.48 from the High Energy Astrophysics Software package (HEAsoft v6.28; \cite{HEAsoft})\footnote{\url{http://heasarc.gsfc.nasa.gov/ftools}}. We extract count-rate light curves in four energy bands: 15--25~keV, 25--50~keV, 50--100~keV and 100--150~keV, using the \texttt{batbinevt} routine with 64~ms time bins. Spectral lag in the $T_{90}$ interval is calculated with the Python routine \texttt{signal.correlate} from the \texttt{scipy} package \cite{SciPy}. The time lag is taken to be the value corresponding to the peak of the correlation coefficient, and the confidence interval as $2 / \sqrt{n-d}$, where $n$ is the size of the data array and $d$ is the measured lag \cite{Tonry79}.

To obtain the hardness ratios presented in Figure~\ref{fig:bat_lc}, BAT spectra in the energy range 15--150~keV were extracted with \texttt{batbinevt}. Spectra were produced for the duration of the initial pulse complex (IPC; see Figure~\ref{fig:bat_lc}), and from the end of the IPC to $T_{90}$ (marked `EE' in Figure~\ref{fig:bat_lc}), following the definitions of these epochs in \cite{Barthelmy05,Gehrels+06,perley09} for GRBs 080503, 060614 and 050724, respectively. Spectra were then fit in \texttt{xspec} v12.11.1 with an absorbed power-law model of the form \texttt{cflux*tbabs*ztbabs*pow} \cite{Wilms00}, where \texttt{cflux} is used to measure the time-averaged flux in the 25--50\,keV and 50--100\,keV bands in each spectrum. Absorption in the Milky Way is fixed to the values derived in \cite{Willingale13}, while flux, photon index and redshifted absorption are free parameters.

\subsubsection*{XRT}

XRT data for light curves and spectral parameters are taken directly from the UKSSDC \cite{Evans07,Evans09}.

\subsubsection*{UVOT}
The {\it Swift}/UVOT began settled observations of the field of GRB\,191019A
3294 s after the \textit{Swift}/BAT trigger. The source counts were extracted initially
using a source region of $5''$ radius. When the count rate dropped to below 0.5
counts per second, we used a source region of $3''$ radius. In order to be consistent
with the UVOT calibration, these count rates were then corrected to $5''$ using the curve of growth contained
in the calibration files. Background counts were extracted using 3 circular regions of radius $15''$ located in source-free regions.
The count rates were obtained from the image lists using the {\it Swift} tools \texttt{uvotevtlc} and \texttt{uvotsource}, respectively.
At late times the light curves are contaminated by the underlying host galaxy. In order to estimate the level
of contamination, for each filter we combined the late time exposures (beyond $10^7$~s) until the end of observations. We
extracted the count rate in the late combined exposures using the same $3''$ and $5''$ radii apertures, aperture correcting where appropriate. These were subtracted from the source count rates derived with the
same size aperture to obtain the afterglow count rates. The afterglow count rates were converted to magnitudes
using the UVOT photometric zero points (Poole et al. 2008; Breeveld et al. 2011). To improve the
signal-to-noise ratio, the count rates in each filter were binned using $\Delta t/t = 0.2$.

\subsection*{Nordic Optical Telescope}
We obtained multiple epochs of observation of GRB\,191019A with the Nordic Optical Telescope (NOT) and ALFOSC imaging spectrograph. Our first night observations were obtained in the $griz$ bands, beginning 0.19 days after the burst. Images were reduced using standard procedures. To search for transient emission we undertook PSF matched 
image subtraction \cite{alard98}. This revealed a clear transient source in the first epoch in all four bands. Further observations were obtained at 2.4, 3.2, 10.2, 34 and 245 days. However, these observations did not reveal any transient emission. A full log of imaging observations is shown in Table~\ref{tab:afterglow}.

In addition to imaging observations we also obtained a spectrum of GRB 191019A on 19 October 2019, approximately 6 hours after the GRB. The spectrum was processed through \texttt{IRAF} for flat-fielding, wavelength and flux calibration. 

\subsection*{Gemini South}
We obtained a series of observations of the location of GRB\,191019A from the Gemini-South Observatory using GMOS. Imaging observations were obtained in the $g$, $r$ and $z$-bands at 8 epochs between 
11 and 70 days after the burst, with the primary aim of detecting and characterising any associated supernova. Data were bias subtracted, flat-field corrected and combined via the Gemini {\tt IRAF} package.  
To determine any transient contribution we use two different approaches. The first is the standard approach of image subtractions which we attempted via the \texttt{HOTPANTS} code. These images reveal no evidence for transient emission. However, because of the compact nature of the host galaxy core which is unresolved in ground based resolution, not all epochs yielded clean subtractions. Therefore, to determine limits across all epochs we utilise the simpler approach of direct photometry in a large (3 arcsecond) apertures. There is no evidence for any variation in the galaxy with the RMS between the different epochs corresponding to 1.3\% in $g$, 1.0\% in $r$ and 1.5\% in $z$. This suggests that there is no variation
in the source across the 11--70 day period of observations. To obtain limits for individual epochs we set the host galaxy value as the mean of all epochs and subtract this from each individual epoch to obtain measured fluxes at the time of each observation. These values are tabulated in Table~\ref{tab:afterglow} and are plotted as $3\sigma$ upper limits in Figure~\ref{fig:sn}. Photometric calibration is performed against Pan-STARRS. 

\subsection*{\textit{Hubble Space Telescope} Observations}

We observed GRB\,191019A with the {\em Hubble Space Telescope} (\textit{HST}) at two epochs on 19 November 2019 and 24 April 2020. At each epoch we obtained imaging observations in the F606W (exposure times of 180 and 680 s, respectively) filter and grism
spectroscopy with G800L. We reduced the imaging with the {\tt astrodrizzle} software, and subtracted the first epoch from the second. Such an analysis is complicated because
in the first epoch the first image was short (180 s) and intended to act as a direct image for the grism spectroscopy. Subsequently multiple cosmic rays are present that cannot be removed by the addition of multiple images. This complicates direct photometry of the galaxy. However, subtraction of the two epochs of imaging reveals no evidence for any transient emission at the burst location. Inserting artificial stars suggests these would be readily visible should they be brighter than $\text{F606W} > 23.5$ AB. 

In addition to these observations we also obtained UV observations in F225W and F275W with exposure times of 2200 s. The data were reduced via {\tt astrodrizzle} and aligned to our NOT and Gemini observations. The host galaxy is well detected in both filters, and appears extended. The resulting 
photometry is shown in Table~\ref{tab:host}.

\subsection*{Astrometry}
To determine the location of GRB\,191019A on its host galaxy we performed astrometry between the images taken with the NOT on 19 October 2019 and that with \textit{HST} on 24 April 2020. We chose 20 compact
sources in common to each image and derived a map between the two sets of pixel co-ordinates via the {\tt IRAF} task \texttt{geomap} in each of the $g$,$r$,$i$ and $z$-bands. The resulting uncertainties
arise from the astrometric fit and the uncertainty in the centroid of the afterglow in the NOT subtracted images.
We estimate the centroid error be 0.3 ACS pixels (appropriate for a S/N=30 detection of the source with a seeing of $\sim 1.0$ arcseconds). This is typically (but not always) smaller than the error from the astrometric fit. The resulting positions are shown in Figure~\ref{fig:finder}. In the $i-$band (which has the tightest astrometric fit) we find that the offset from the centroid of the host galaxy is 
$\delta_{x(i)} = 0.30 \pm 0.41$ and
$\delta_{y(i)} = 0.27 \pm 0.41$ pixels. In
$g$, $r$ and $z$ the corresponding values are 
$\delta_{x(g)} = 0.44 \pm 0.82$,
$\delta_{y(g)} = 0.03 \pm 1.21$, 
$\delta_{x(r)} = 0.43 \pm 0.50$,
$\delta_{y(r)} = 1.48 \pm 0.54$,
$\delta_{x(z)} = 0.85 \pm 0.91$.
$\delta_{y(z)} = -0.68 \pm 0.87$. We therefore conclude that the source is consistent with the nucleus of the host galaxy at a projected offset (based on the $i-$band astrometry) of $r=0.020 \pm 0.029$ arcseconds or $78 \pm 109$ pc at
$z=0.248$.

\subsection*{Chance alignment} 
It is relevant to consider the probability of chance alignment of a given position with a galaxy. Such chance alignments are inevitable in large samples of transient sources, such as the {\em Swift}-BAT 
catalog. However, the location of GRB\,191019A, so close to the nucleus of a relatively bright ($r \sim 19$) galaxy, leads to an extremely small chance probability. Formally, following \cite{Bloom+02}, the probability of lying within $0.04''$ of such a host galaxy is $\sim 10^{-6}$.  Therefore, even considering the $\sim 1000$ long GRBs observed by {\em Swift}, the likelihood of a chance alignment of GRB\,191019A with the nucleus of this galaxy
is very small. We therefore consider the association to be robust.

\subsection*{Afterglow properties}

\subsubsection*{Light curve}
The optical afterglow is detected only at a single epoch at $\sim 0.2$ days. All observations beyond this point are upper limits.

The X-ray light curve parameters, obtained from the UKSSDC, show that
the X-ray afterglow can be adequately modelled by a 
single power-law with index $\alpha_1=1.27^{+0.17}_{-0.15}$.
Alternatively, a broken power-law with $\alpha_1=-0.14^{+0.54}_{-0.16}$, $\alpha_2 =1.6^{+0.5}_{-0.4}$ and a break time of $t_b=(5.9^{+4.2}_{-1.8}) \times 10^3$ s also provides a good fit, although not statistically required (chance improvement probability of 4.5\%, or $\sim 2\sigma$). 

\begin{figure*}
\centerline{
\includegraphics[width=.49\textwidth]{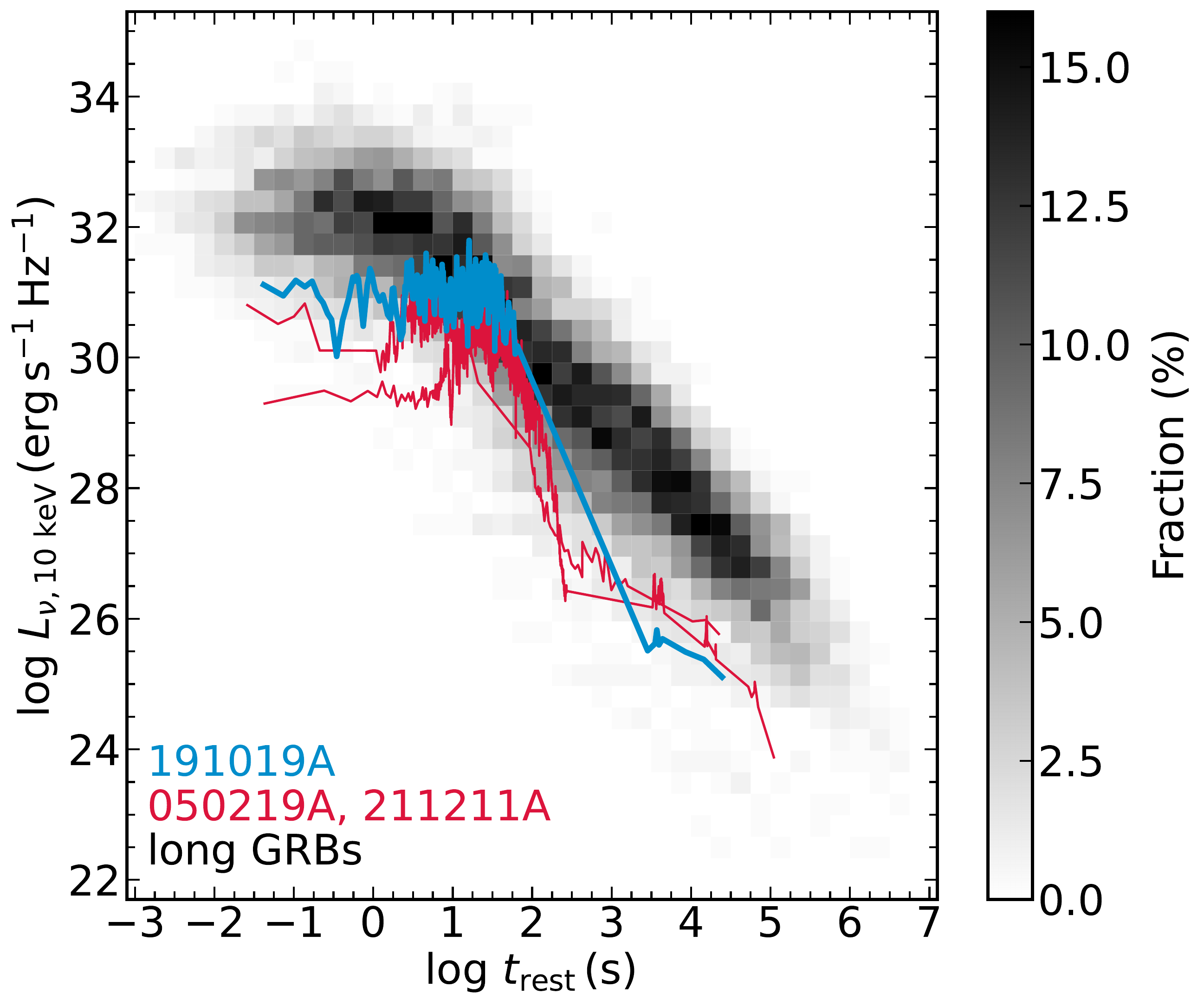}
\hspace{0.1cm}
\includegraphics[width=.49\textwidth]{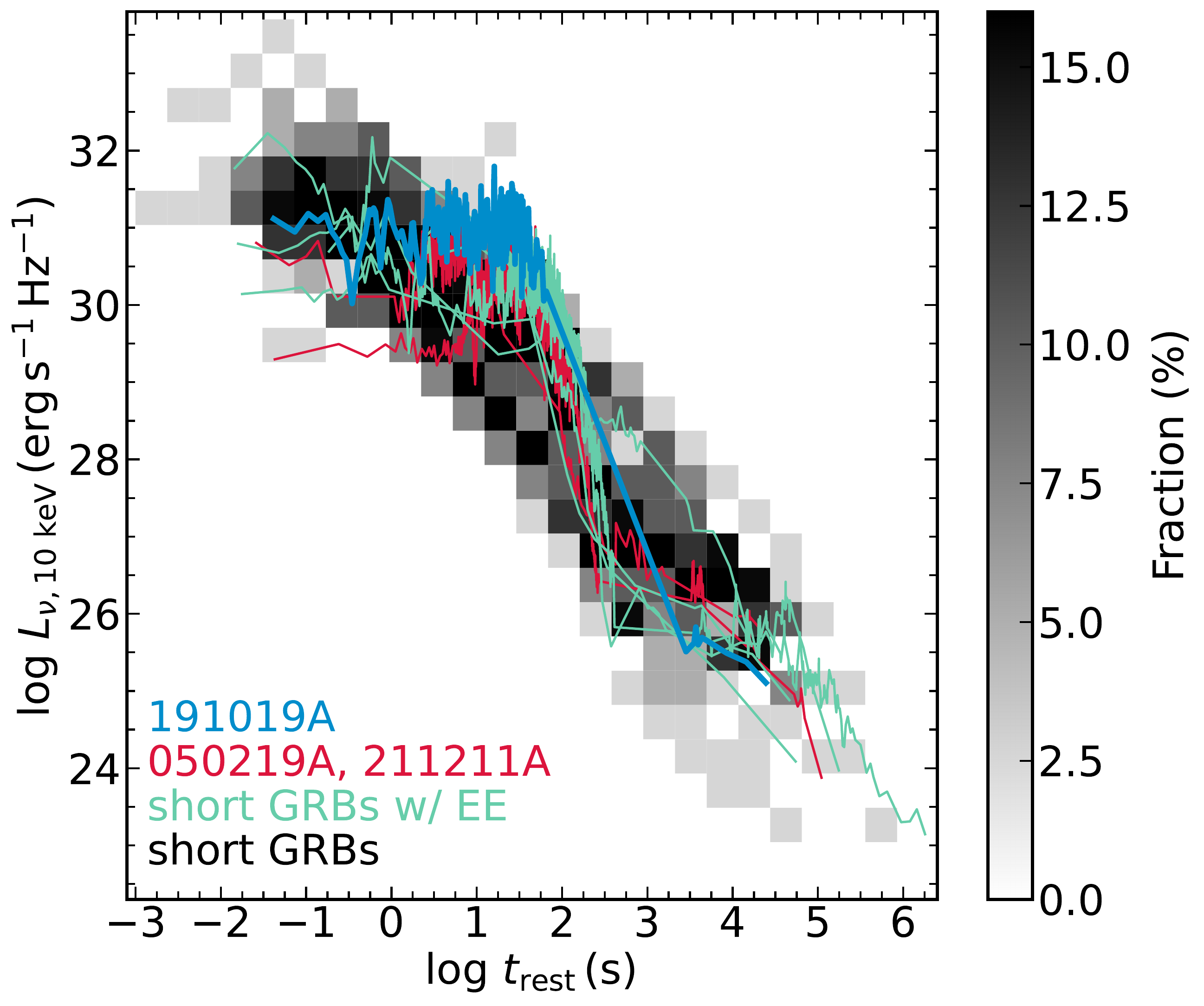}}
\caption{The X-ray afterglow of GRB\,191019A in context with other GRBs. The left hand panel shows a
comparison to the long duration GRBs, and the right hand panel to the short bursts. The greyscale background indicates the fraction of bursts at a given luminosity. GRB\,191019A lies at the fainter end of the prompt emission for long GRBs (i.e. within the first hundred seconds), and has an X-ray afterglow which is significantly underluminous for long bursts. However, it has a luminosity entirely consistent with short GRBs.}
\label{fig:xrt_lum}
\end{figure*}

To place the X-ray (and early $\gamma$-ray data)
in context with the overall {\em Swift} population,
we retrieve from the \textit{Swift} Burst Analyser \cite{Evans+09} the $\gamma$-ray and X-ray light curves of all \textit{Swift} GRBs detected up until 9 October 2022. We select all GRBs with at least 2 detections by BAT and XRT each and a measured redshift with an accuracy of $\leq0.1$ in redshift space. To divide the final input sample of 395 GRBs into long and short GRBs, we follow \cite{fong22}. In total, our sample consists of 356 long and 39 short GRBs. We processed their light curve data and moved them to their rest-frames following \cite{Schulze+14}. Figure~\ref{fig:xrt_lum} shows the parameter space occupied by the long (left) and short (right) GRBs as a density plot and the BAT+XRT light curve of 191019A in blue. In both plots, we also display the light curves of GRB 050219A and 211211A (in red) and, in the right hand panel, also highlight the short GRBs with extended emission \cite{Gompertz+20}.

The X-ray light curve of GRB\,191019A is relatively poorly sampled, but its evolution in luminosity space is consistent with the population of short-GRBs with extended emission (see Figure~\ref{fig:xrt_lum}), while
being far less consistent with the long-GRB population.  This offers further support of the
interpretation of GRB\,191019A as belonging to
the population of GRBs created via compact object mergers.

\subsubsection*{Spectral energy distribution and extinction}

A straightforward way to explain the non-detection of any supernova emission would be to invoke dust extinction. To explain the non-detection of the supernova in our observations would require
$A_V > 3$ mag. However, the afterglow in this case would also 
be subject to extinction and would be red. The detection in the
UVW2 ultraviolet filter offers a strong indication that the extinction is low.

\begin{figure*}
\centerline{
\includegraphics[width=.8\textwidth]{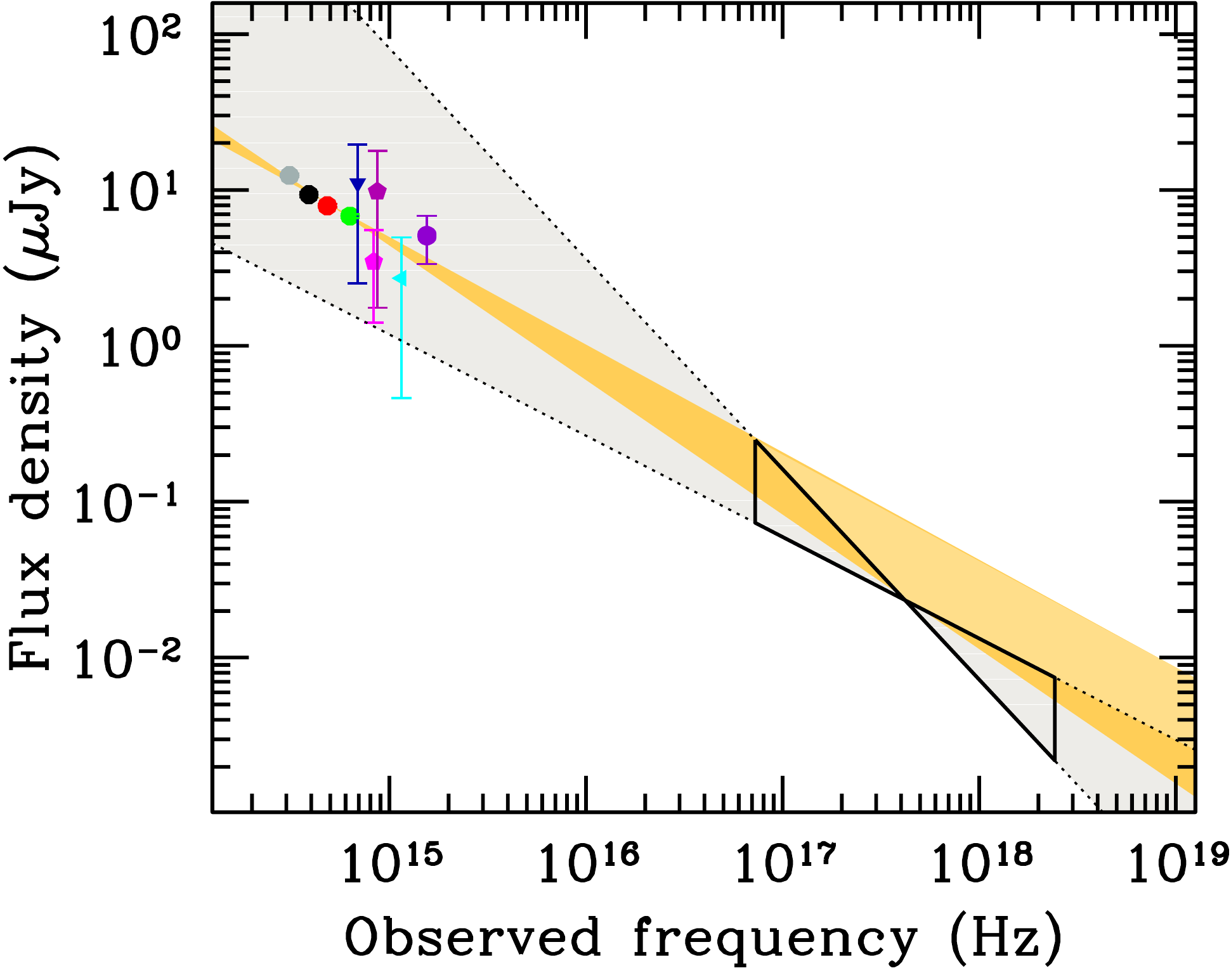}}
\caption{The X-ray to optical spectral energy distribution of the GRB\,191019 afterglow, 0.21 days after the detection (average time of the first-night NOT observations). The optical data ($z$ to UVW2) are corrected for the (small) Galactic extinction corresponding to $A_V = 0.10$~mag. The solid butterfly represents the observed X-ray spectrum and uncertainty (0.3--10 keV). Optical/UV and X-ray points have been extrapolated to the same time assuming they follow the observed X-ray decay. In practice, this is a small change for the $griz$ NOT data, which are driving the optical fit. The yellow/orange shaded area represents the best-fit to the optical/UV data, and its extrapolation to the X-ray range, while the grey region shows the same extrapolation of the X-ray to the optical/UV regime. The X-ray and UV/optical spectral indices are consistent with each other, indicating no requirement for any additional extinction.}
\label{fig:sed}
\end{figure*}

Moreover, corrected for the (small) Galactic extinction, the optical and UV data do not show a reddened spectrum (Fig.~\ref{fig:sed}), and the spectral energy distribution is well described with a power-law with spectral index $\beta_{\rm opt} = 0.78 \pm 0.08$, which is consistent both with the X-ray value $\beta_{\rm X}=1.0^{+0.4}_{-0.3}$ and especially with the optical-to-X-ray slope $\beta_{\rm OX} = 0.82$ \cite{DAi19}. This indicates that the optical flux is not suppressed, indicating negligible extinction, and that a single power law can describe the entire data set. Supporting this finding is the X-ray derived hydrogen-equivalent column density $N_{\rm H}$ = 1.2$^{+1.6}_{-1.2} \times 10^{21}$ cm$^{-2}$, consistent with zero. All of these diagnostics argue against any role of extinction.

\subsection*{Host galaxy properties} 
\begin{figure*}
\centerline{
\includegraphics[width=.8\textwidth]{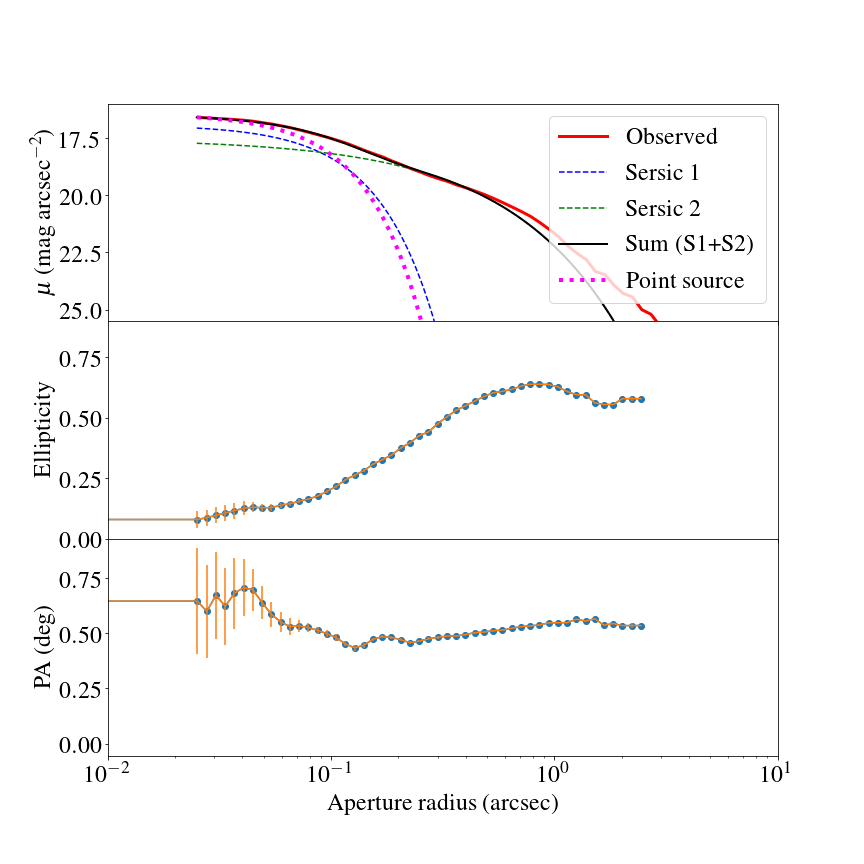}}
\caption{The surface brightness profile of the GRB\,191019A host as determined via elliptical isophote fitting in {\tt ellipse}. The host has a very compact, almost point-like core, although its surface brightness profile is not well fit with either a point-source plus a Sersic profile, nor the sum of two Sersic profiles, especially beyond $1''$. }
\label{fig:sb}
\end{figure*}

\begin{figure*}
\centerline{
\includegraphics[width=.8\textwidth]{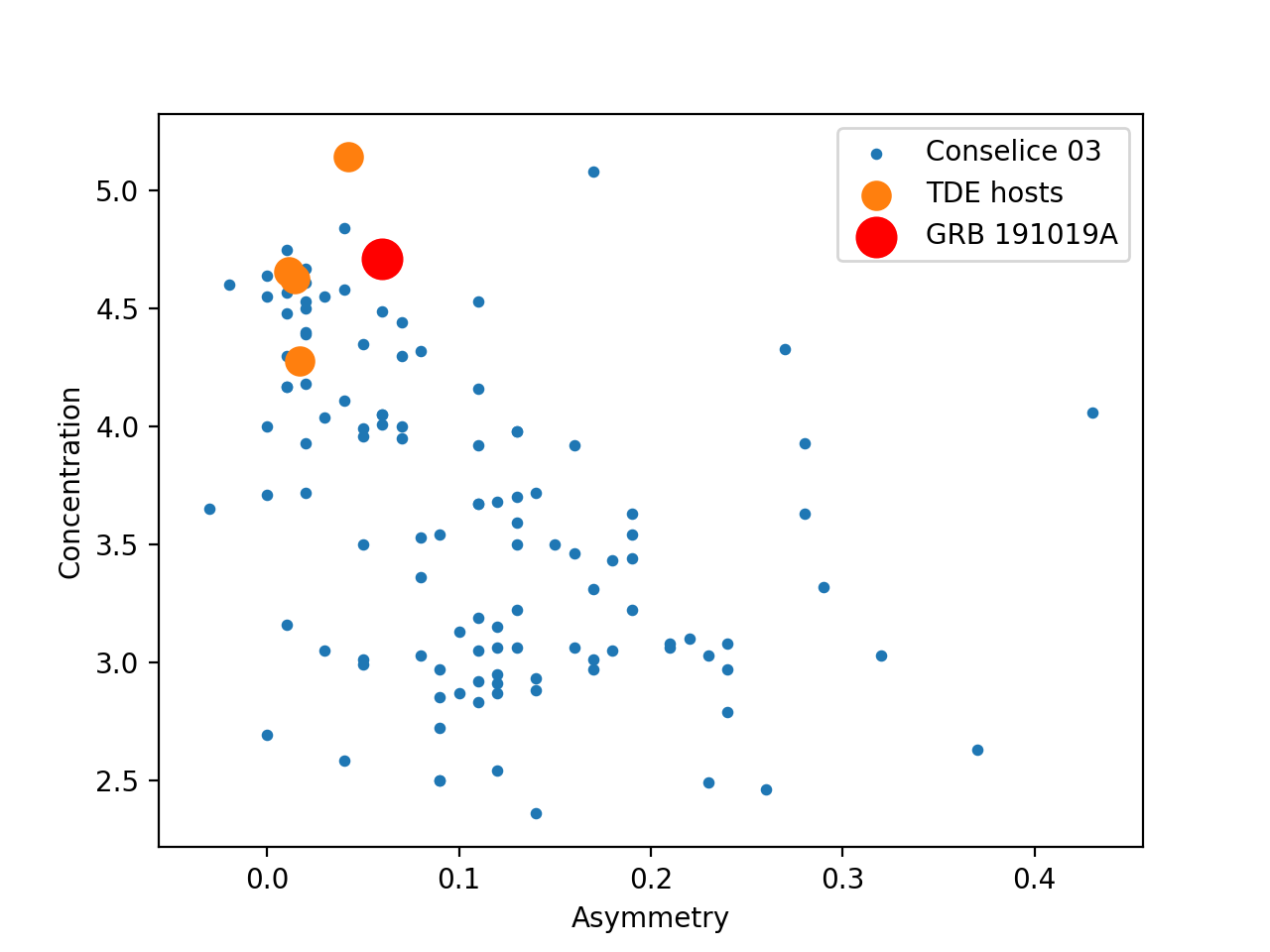}}
\caption{Concentration and asymmetry measurements for the host of GRB\,191019A compared with those of a sample of normal galaxies from \cite{conselice03}, and the hosts of tidal disruption events from \cite{french20}. Morphologically, the GRB\,191019A host appears very similar to the TDE systems, consistent with an origin in dense environments where stellar interactions are common.}
\label{fig:cas}
\end{figure*}

The host galaxy is morphologically smooth and highly centrally concentrated (Figure~\ref{fig:sb}). We determine the surface brightness profile via
fitting elliptical isophotes to the late time {\em HST} observations. The peak surface brightness is $\sim 16.5$ mag arcsec$^{-2}$, almost a magnitude brighter than, for example, the central 
surface brightness of the very luminous host of the short GRB 050509B (at $z=0.22$, a similar redshift). The surface brightness profile is not well modelled by a single Sersic profile, but constitutes a near point-like source with lower surface brightness emission. Its 20, 50 and 80\% light radii are 0.09, 0.27 and 0.75 arcsec. Notably, its concentration index $r_{20} / r_{80}$ is extreme compared to 
most samples of galaxies \cite{conselice03}, but
comparable to those of TDE hosts (see Figure~\ref{fig:cas}).  It is relevant to consider if some of this light could arise from an AGN. However, we cannot confirm this in the absence of any AGN-like emission lines
in the optical spectrum of the source. The presence of a weak [N\,{\sc ii}] line is apparent in both the NOT and Gemini spectra, and the absence of oxygen or hydrogen emission lines may favour a more AGN-like set of line ratios, but such an interpretation is not conclusive. A late time observation with the {\em Swift} X-ray Telescope suggests an upper limit of $F_{\rm X} < 3 \times 10^{-14}$ erg s$^{-1}$ cm$^{-2}$, corresponding to 
a luminosity of $L_{\rm X} < 6 \times 10^{42}$ erg s$^{-1}$.  This rules out X-ray luminous AGN, but not fainter, low luminosity examples. Finally, the colours in the WISE catalog of $\text{W1}-\text{W2} = 0.25 \pm 0.12$ lie far from the expected colours of AGN in these bands ($\text{W1}-\text{W2} > 0.8$). 

\begin{figure*}
\includegraphics[width=0.5\textwidth]{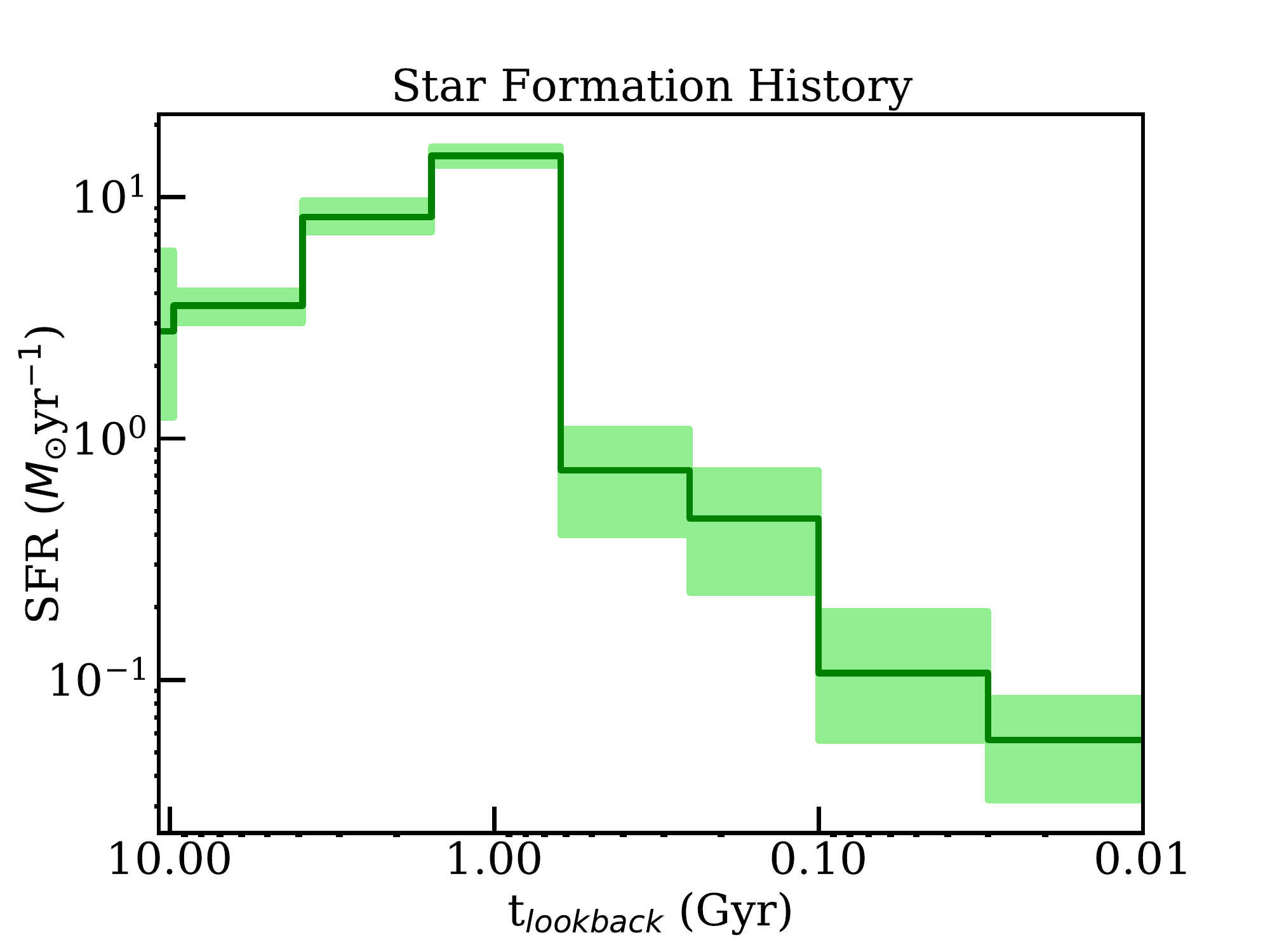}
\includegraphics[width=0.5\textwidth]{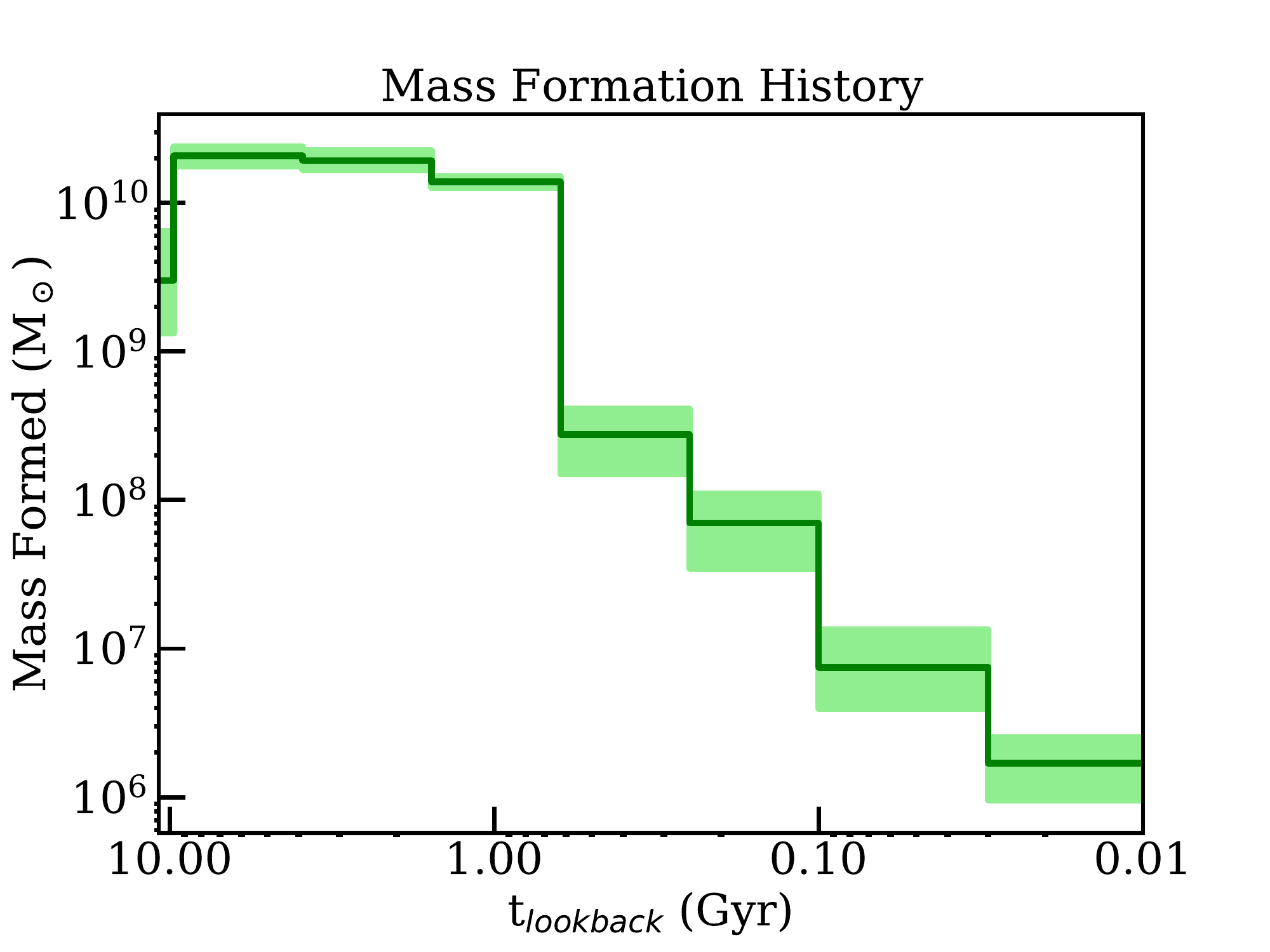}
\caption{The star (left) and mass (right) formation histories of the host of GRB\,191019A, determined through the \texttt{Prospector} fitting. The dark green lines indicate the median SFR and mass formed in each bin, and the light green regions represent the $1\sigma$ uncertainty. We find that the majority of stars and mass in the galaxy formed at $t_\textrm{lookback} > 1$~Gyr, and that the host has transitioned into a quiescent galaxy, with a low present-day SFR.}
\label{fig:sfh}
\end{figure*}

We fit the optical NOT/ALFOSC spectrum and broader-band photometry of the host galaxy with \texttt{Prospector} \citep{Leja+17, jlc+2021}, a stellar population modeling inference code, to determine its stellar population properties, such as stellar population age, mass formation history, and star formation history (SFH). \texttt{Prospector} samples each property parameter space with a nested sampling fitting routine, \texttt{dynesty} \citep{Dynesty}, and produces model spectral energy distributions with \texttt{FSPS} and \texttt{Python-fsps} \citep{FSPS_2009, FSPS_2010}. We apply a Milky Way extinction law \citep{MilkyWay}, Chabrier IMF \citep{Chabrier2003}, and a non-parametric SFH to the fit. We choose a non-parametric SFH model as we can then more accurately determine when the majority of stars formed in the galaxy's history, and thus when the progenitor likely formed. However, we note that most stellar population modeling to date uses a parametric SFH that tends to result in lower stellar masses and stellar population ages. We use a non-parametric SFH with seven age bins; the first two are between 0 and 30 Myr and 30 and 100 Myr, and the final five are log-spaced from 100 Myr to the age of the universe at GRB\,191019A's redshift ($z=0.248$, $t_\textrm{univ} \sim 10.78$ Gyr). We further apply a mass-metallicity relation \citep{gcb+05} to sample realistic masses and stellar metallicities, and a dust 2:1 ratio between the old and young stellar populations \citep{pkb+14, Leja2019, cab+20}. We fit the model spectral continuum with a 10$^\mathrm{th}$ order Chebyshev polynomial and include a nebular emission model with gas-phase metallicity and a gas-ionization parameter in the fit to measure spectral line strengths. Since the host may also contain an AGN we also add two AGN components, that dictate the mid-IR optical depth and the fraction of AGN luminosity in the galaxy.

We find that the host of GRB\,191019A has a stellar population age of $4.34^{+0.88}_{-0.47}$~Gyr (median and $1\sigma$), stellar mass with $\log(M/M_\odot) = 10.57^{+0.02}_{-0.01}$, and current-day SFR of $0.06^{+0.08}_{-0.03}~M_\odot$~yr$^{-1}$, thus is currently a quiescent galaxy, given the sSFR and redshift. From a limit of the H$\alpha$ flux, we determine an H$\alpha$ $\text{SFR} < 0.12^{+0.07}_{-0.06}~M_\odot$~yr$^{-1}$.  We report the SFH and mass formation history of the host in terms of the lookback time ($t_{\textrm{lookback}}$), and show the subsequent histories in Figure \ref{fig:sfh}. We find that the majority of stellar mass and stars formed at $t_{\textrm{lookback}} \gtrsim 1$~Gyr, with a steep decline in mass and star formation to present-day, $\sim 99$\% of the stellar mass was assembled $>1$ Gyr before the merger (Figure \ref{fig:sfh}, right). Thus, the progenitor of GRB\,191019A has a higher a priori probability of forming $> 1$~Gyr ago, making it unlikely to originate from a young stellar progenitor.

As an independent check of the absence of 
emission lines in the host galaxy of GRB 191019A we also fit the NOT spectrum with penalised pixel fitting \texttt{pPXF} \cite{Cappellari2022}, where we fit only the stellar component and no emission lines following
\cite{inkenhaag22}. As with our {\em Prospector} fitting the resulting residuals provide no evidence for emission features. 

\subsubsection*{Comparison with short and long GRB host galaxies}
\begin{figure*}
\centerline{
\includegraphics[width=.8\textwidth]{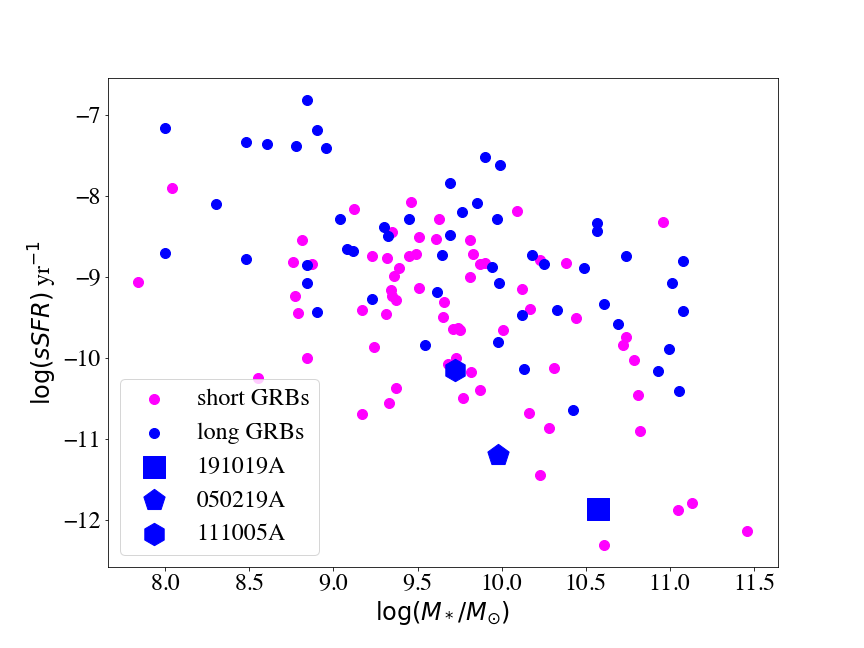}}
\caption{A comparison of the host properties (stellar mass and specific star formation rate) of GRB\,191019A with those of long GRBs (from \cite{Perley2013}) and short duration GRBs (from \cite{nugent22}). The GRB\,191019A host has a very low star formation rate, and lies in a region devoid of other long GRB hosts. However, this region is populated by short-GRB host galaxies. The same is true for the host of GRB 050219A \cite{rossi14}. Another plausibly similar event, GRB 110005A \cite{Michalowski+18} is in a region populated by both long and short GRB hosts.  }
\label{fig:host_prop}
\end{figure*}

We can compare the properties of the host of GRB\,191019A with those of 
other long and short duration GRBs. A bulk comparison is often done utilizing the
stellar mass and star formation rate of these galaxies. This is plotted for a sample of long \cite{Perley2013} and short \cite{nugent22} host galaxies in Figure~\ref{fig:host_prop}. The long GRBs overwhelmingly favour actively star forming hosts, with high specific star formation rates. In contrast the short GRBs span a
wide range of star formation rates including a modest fraction apparently in 
quiescent systems. 

There are two long GRB host galaxies which stand out from the apparent trend. One is the host of GRB\,191019A. The other is the host of GRB 050219A \cite{rossi14}. This burst is only localised via its X-ray afterglow, but has a comparable redshift to GRB\,191019A and similar energetics ($E_{\rm iso} \sim 10^{51}$ erg). 
With only an X-ray position it is not possible to accurately determine if the
burst is nuclear, and indeed the probability of chance alignment is larger due to the poor localisation ($P_\text{chance} \sim 0.8\%$). However, it also lies in a galaxy showing Balmer absorption lines but little evidence for star formation. Rossi et al. \cite{rossi14} also classify it as a post starburst system. The similarities with
GRB\,191019A are striking, and we consider it a possible example of a similar event.

\subsection*{GRB\,191019A as a merger-GRB}

\subsubsection*{Collapsars}
At first sight, GRB\,191019A appears a relatively normal, if soft, long duration GRB. It consists of a fast rise with a slower decay with some variability super-imposed on this decay. 
It is not a short GRB, nor is it obviously a member of the population of short GRBs with extended emission, where an initial spike is followed by longer-lived, softer emission. However, the location in an old host galaxy and the lack of any visible supernova emission strongly disfavour the presence of any massive stars which could produce collapsar-like events. Although previously identified supernova-less GRBs have been suggested to arise from direct collapse of massive stars to black holes \cite{Fynbo+06,GalYam+06,DellaValle+06}, in those cases the bursts were associated with star forming host galaxies, and in one case with a highly active star forming region within them \cite{Fynbo+06}. In principle massive stars can be built by the merger of lower mass stars in dense environments \cite{pg04}. However, with a $>1$ Gyr old population, the turn off mass is $<2$ $M_\odot$. It would take multiple mergers to build a star sufficiently massive to directly collapse to a black hole.

\subsubsection*{Tidal disruption events}
The location in the galaxy nucleus could be indicative of an origin associated with the supermassive black hole which resides there. In particular, a population of relativistic tidal disruption events
have been identified by the {\em Swift}-BAT \cite{bloom11,burrows11,levan11,cenko12,brown15}. However, these events are typically of very long duration and were visible to the {\em Swift}-BAT
for several days at a luminosity of $>10^{47}$ erg s$^{-1}$. Alternatively, more 
typical tidal disruption events do not generate detectable $\gamma$-ray emission but are found with long-lived  ($>10^7$~s), lower luminosity X-ray emission ($\sim 10^{42}$--$10^{44}$ erg s$^{-1}$ \cite{auchettl17,saxton20}), as well as long-lived optical/UV thermal signatures significantly brighter than our limits for GRB\,191019A (Figure~\ref{fig:sn}) \cite{perley20}. 

The timescale for tidal disruption events is generally thought to be related to the return time for the most bound ejecta of the disrupted star. It can be much shorter
in the case of white dwarfs disrupted by intermediate mass black holes \cite{maguire20}. Such a black hole mass is inconsistent with that inferred for the black hole mass in the GRB 191019A host via scaling relations (few $\times 10^7$ $M_\odot$). Furthermore, even these systems have generally been discussed in the context of long GRBs with extremely long durations \citep{levan14,macleod14}, and simulations suggest they should give rise to detectable X-ray and optical emission for tens of days following the disruption \cite{macleod16}. There is also a suggested population of micro-TDEs, in which a main sequence star is disrupted by a stellar mass black hole \cite{Perets16}. However, these events are also suggested as an explanation for the longest duration (so-called ultra-long) GRBs \cite{levan14} and do not naturally match the timescales here.

\subsubsection*{Compact object mergers}
The recent identification of a 
kilonova in GRB 211211A demonstrated that much longer lived bursts can arise from
compact object mergers \cite{rastinejad22,yang22,troja22}. For GRB 191019A, the older host and lack of supernova would be consistent with this expectation. However, it is relevant to consider if its properties are consistent with previous examples of long-lived merger-GRBs. 

The $\gamma$-ray light curve of GRB\,191019A does start with a short lived (0.5~s), somewhat harder pulse, but this is not clearly distinct from the overall prompt emission. However, in the population of short GRBs with extended emission there is a range of contrast ratios between the short spike and the extended emission ranging from 0.2 to $>50$ \cite{perley09}. There is also a substantial variation in the time between the short spike and the appearance of extended emission from $\sim 0$ to $\sim 20$ s \cite{NorrisBonnell06,gompertz22}. The combination of a low contrast and short delay could readily mean that some bursts are not easily identifiable as short+EE based
on their light curves. While at durations of $\sim 1$ minute the majority of GRBs are likely to arise from collapsars, a small fraction from mergers is entirely plausible. Indeed, the softer than usual spectrum for GRB 191019A places it in a region of hardness duration space which is comparable to extended emission see in other GRBs (see Figure \ref{fig:bat_lc}).

We also measure the spectral lag between the harder and softer emission. In collapsar GRBs there is a noticable spectral lag in which the softer emission is delayed (lags) the harder emission. Such a measurement is not seen in merger-GRBs \cite{Gehrels+06}. For GRB 191019A, formally the soft emission leads the harder emission with a measured spectral lag of $\tau = -96 \pm 63$ ms (between the 50--100 and 15--25 keV bands, with 16~ms binning). This lag measurement is atypical for long GRBs, and suggests a merger-origin is plausible. 

It may also be the case that the burst arises from a merger, but there is no short spike in this case (hidden or otherwise). This may be the case if, for example, there are different beaming angles for the short spike and extended emission. Indeed, it has been suggested that a population of orphan extended emission bursts should also be present \cite{Bucciantini12,lu22}. Furthermore, there are claims that a population of fast X-ray transients seen in narrow field X-ray observations (e.g with \textit{Chandra}) arise
from compact object mergers \cite{xue19,qv22}. None of these systems show initial short spikes (albeit in a much softer energy band than the {\em Swift}-BAT). 

Finally, mergers within the discs surrounding supermassive black holes create jets with very different dynamics to those in a tenuous interstellar medium. In particular, the high densities result in the formation of an external shock very close to the progenitor, and the dissipation of energy within this shock causes bursts which are 
intrinsically short and hard to be smeared out and softened \cite{lazzati22}. Although not 
conclusive, the similarity of 
GRB 191019A to these models is remarkable.

\subsubsection*{Dynamical formation channels}

There are various ways in which dynamical interactions create
compact object binaries. Firstly, the interactions can create new compact
object binaries via 2+1 interactions which tend to leave the more massive components within a binary. Once formed these binaries can be further hardened 
(have their separations reduced) by additional interactions and eventually merger within the Hubble time \cite{heggie03}. The SMBH can
also act as a perturbation, in particular via the creation of Kozai-Lidov cycles,
which increase the eccentricity and decrease the periapsis separation of the binary, thus increasing GW energy losses and ultimately enhancing the merger rate \cite{hoang18}. Fragione et al. \cite{fragione19} find the rates of compact object binary formation in galactic nuclei to be of order 1 Gpc$^{-3}$ yr$^{-1}$, although these are highly sensitive to the details of the dynamics within the galactic nuclei. If indeed galaxies similar to the hosts of GRB 191019A have much higher interaction rates than MW-like galaxies, this could be enhanced significantly. 

A challenge for dynamical production via many-body interactions is that these 
interactions tend to leave the most massive objects within the binaries \cite{ye20}.
Hence, the expected rates of NS-NS mergers within globular clusters are much
smaller than the BH-BH rate \cite{ye20}.  
This is somewhat in conflict with the 1-3 out of $\sim 20$ BNS systems in the Milky Way which are found within globular clusters (the range reflects the uncertainty in the nature of the compact companions in some cases). It may be that in these systems the BNS is
a transitory state, and further interactions may yet take place before the GW-induced merger. 
However, the rate of dynamical formation is higher than the rate of mergers in clusters. Many NS-NS and BH-NS systems which are dynamically formed in globular clusters are ejected, and ultimately merge far from the cluster. The escape velocities of globular clusters are small ($\sim 50$ km s$^{-1}$), in contrast, the escape velocities from the centre of the GRB\,191019A host are much larger. The central regions cannot be directly resolved, but assuming the half-light radius is also half the stellar mass the escape velocity from $\sim 1$ kpc would be $>300$ km s$^{-1}$, so that binaries ejected via dynamical formation would also likely merge close to the nucleus, especially since these ejected systems likely have lower velocities than field binaries \cite{ye20}. Of course, should systems formed in the field also be formed while the 
galactic potential is so centrally concentrated they are also less likely to escape, although would be placed on eccentric orbits, spending most of their time at larger projected radii. 

In addition to these star-star (or star-SMBH) interactions there are also 
interactions with the gaseous disk which may exist around the SMBH.  The rates of such systems are subject to significant uncertainty and debate; upper limits
as high as 400 Gpc$^{-1}$ yr$^{-1}$ \cite{mckernan20} have
been suggested. Indeed, associations with unusual AGN outbursts have been claimed for some gravitational wave (BBH) sources
at moderate significance \cite{graham20,graham22}.There remain significant model and observational uncertainties associated with these results \cite{tagawa20}, however, if they are correct, then a rather large fraction of BBH mergers would be taking place within the discs of AGN. 

\subsubsection*{Dynamical versus field rates}
If we assume that GRB\,191019A were created by a compact object
binary it is also relevant to consider the probability that it
arises via field binary evolution, or via dynamical interactions.
Published rates \cite{mandel22} for field interactions span a range
from 0.3--8900 Gpc$^{-3}$ yr$^{-1}$ and nuclear interactions from
0.004--1.5 Gpc$^{-3}$ yr$^{-1}$ or $<400$ Gpc$^{-3}$ yr$^{-1}$
considering gaseous discs. This large range of rates makes it
difficult to make any robust prediction as to the fraction of mergers which form via each channel. However, median values would imply
a ratio of several hundred field binary mergers per dynamically formed nuclear system.

This is consistent with the fact that no well localised  (sub-arcsecond)
short GRBs are apparently consistent with the nuclei of their hosts.
Indeed, population synthesis predictions suggest that very few
mergers should be nuclear \cite{Belczynski06,fong22}. Wiggins et
al. \cite{Wiggins18} find that less than 1\% of mergers originate
at $<100$ pc of the nucleus of their host galaxies.  Therefore, within the uncertainty of the models, the fraction of mergers formed via binary evolution is similar to the
fraction of mergers that form dynamically in the nuclear region of host galaxies.

However, the properties of the galaxy in the case of GRB 191019A
offer strong suggestions that the merger is dynamical, and not of
a field binary.  Firstly, in ancient galaxies we primarily observe
the population of field compact object mergers with long delay
times. These will travel further from their birth sites, and so
have larger offsets than for binaries with short merger times.
Indeed, in the short-GRB population there are apparently larger
offsets to early type hosts \cite{fong22,nugent22}. Therefore, the
presence of GRB 191019A in the nucleus of an ancient host already suggests a more
likely dynamical origin.

Secondly, not all galaxies have identical interaction rates in
their cores. Those interaction rates are a sensitive function of
the core density, of the properties of its nuclear star cluster (e.g. mass, the presence of a disc), and of the characteristics of its black hole population.  Galaxies with similar properties
to the host of GRB 191019A (in terms of Lick indices) are
over-represented as the hosts of tidal disruption events by up to
two orders of magnitude (representing 2.3\% of the observed SDSS
population but 75\% of TDE hosts \cite{French17}). The TDE rate is influenced by the
scattering of stars into the loss cone of the supermassive black
hole. It is not straightforward to scale from the TDE rate to the
dynamical formation rate of compact objects, however, it is
reasonable to assume that the enhanced TDE rate also leads to 
increased rates of other dynamical interactions.

Hence, the probability of a field binary arising in the nuclear
regions of an ancient galaxy in which the interaction rates are
very high, is substantially lower than arising in {\em any} galactic
nucleus. The dynamical probability would also increase for systems including black holes, relative to those with neutron stars (BH-NS, BH-WD).  We therefore conclude that in the case of GRB 191019A a
dynamical origin is more likely than a field system.

\section*{Declarations}

\bmhead{Acknowledgments}

A.J. Levan, D.B. Malesani and N.R. Tanvir are supported by the European Research Council (ERC) under the European Union’s Horizon 2020 research and innovation programme (grant agreement No.~725246). D.B.M. also acknowledges research grant 19054 from VILLUM FONDEN.
M. Nicholl and B. Gompertz are supported by the European Research Council (ERC) under the European Union’s Horizon 2020 research and innovation programme (grant agreement No.~948381). M. Nicholl acknowledges a Turing Fellowship.
The Fong Group at Northwestern acknowledges support by the National Science Foundation under grant Nos. AST-1814782, AST-1909358 and CAREER grant No. AST-2047919. W.F. gratefully acknowledges support by the David and Lucile Packard Foundation. 
K.E. Heintz acknowledges support from the Carlsberg Foundation Reintegration Fellowship Grant CF21-0103.
J. Hjorth was supported by a VILLUM FONDEN Investigator grant (project number 16599).
G. Lamb is supported by the UK Science Technology and Facilities Council grant, ST/S000453/1.
A. Inkenhaag acknowledges the research programme Athena with project number 184.034.002, which is financed by the Dutch Research Council (NWO).
K. Bhirombhakdi was supported by STScI HST General Observer Programs 16548 and 16051 through a grant from STScI under NASA contract NAS5-26555.
The Cosmic Dawn Center is funded by the Danish National Research Foundation under grant No{.} 140.
G. Fragione acknowledges support from NASA Grant 80NSSC21K1722 and NSF Grant AST-2108624 at Northwestern University. I.~Mandel acknowledges support from the Australian Research Council through the Centre of Excellence for Gravitational  Wave  Discovery (OzGrav), project number CE17010004, and through Future Fellowship FT190100574.

Partly based on observations made with the Nordic Optical Telescope, under programs 58-502 and 61-503, owned in collaboration by the University of Turku and Aarhus University, and operated jointly by Aarhus University, the University of Turku and the University of Oslo, representing Denmark, Finland and Norway, the University of Iceland and Stockholm University at the Observatorio del Roque de los Muchachos, La Palma, Spain, of the Instituto de Astrof\'isica de Canarias. And on observations obtained at the international Gemini Observatory (program IDs GS-2019B-DD-106 and GS-2019B-FT-209), a program of NOIRLab, which is managed by the Association of Universities for Research in Astronomy (AURA) under a cooperative agreement with the National Science Foundation on behalf of the Gemini Observatory partnership: the National Science Foundation (United States), National Research Council (Canada), Agencia Nacional de Investigaci\'{o}n y Desarrollo (Chile), Ministerio de Ciencia, Tecnolog\'{i}a e Innovaci\'{o}n (Argentina), Minist\'{e}rio da Ci\^{e}ncia, Tecnologia, Inova\c{c}\~{o}es e Comunica\c{c}\~{o}es (Brazil), and Korea Astronomy and Space Science Institute (Republic of Korea). Processed using the Gemini \texttt{IRAF} package and \texttt{DRAGONS} (Data Reduction for Astronomy from Gemini Observatory North and South). And on observations made with the NASA/ESA \textit{Hubble Space Telescope} obtained from the Space Telescope Science Institute, which is operated by the Association of Universities for Research in Astronomy, Inc., under NASA contract NAS 5–26555.  These observations are associated with program \#16051 and 16458. This work made use of data supplied by the UK \textit{Swift} Science Data Centre at the University of Leicester.

The Pan-STARRS1 Surveys (PS1) and the PS1 public science archive have been made possible through contributions by the Institute for Astronomy, the University of Hawaii, the Pan-STARRS Project Office, the Max-Planck Society and its participating institutes, the Max Planck Institute for Astronomy, Heidelberg and the Max Planck Institute for Extraterrestrial Physics, Garching, The Johns Hopkins University, Durham University, the University of Edinburgh, the Queen's University Belfast, the Harvard-Smithsonian Center for Astrophysics, the Las Cumbres Observatory Global Telescope Network Incorporated, the National Central University of Taiwan, the Space Telescope Science Institute, the National Aeronautics and Space Administration under Grant No. NNX08AR22G issued through the Planetary Science Division of the NASA Science Mission Directorate, the National Science Foundation Grant No. AST-1238877, the University of Maryland, Eotvos Lorand University (ELTE), the Los Alamos National Laboratory, and the Gordon and Betty Moore Foundation. Partly based on observations obtained as part of the VISTA Hemisphere Survey, ESO Progam, 179.A-2010 (PI: McMahon). This publication makes use of data products from the Wide-field Infrared Survey Explorer, which is a joint project of the University of California, Los Angeles, and the Jet Propulsion Laboratory/California Institute of Technology, funded by the National Aeronautics and Space Administration. 

\bmhead{Data Availability}

The majority of data generated or analysed during this study are included in this published article (and its supplementary information files). Gamma-ray and X-ray data from \emph{Swift} may be downloaded from the UK \emph{Swift} Science Data Centre at \url{https://www.swift.ac.uk/}. {\em HST}, Gemini and NOT data can be downloaded from
the relevant archives at  \url{https://archive.stsci.edu}, \url{https://archive.gemini.edu}, \url{https://www.not.iac.es/observing/forms/fitsarchive/}.

\bmhead{Code Availability}

The {\tt Prospector} stellar population modeling code is available at \url{https://github.com/bd-j/prospector}. The \texttt{IRAF} and \texttt{python} scripts necessary for \textit{HST} data reduction can be obtained via \texttt{astroconda}, and \texttt{IRAF} (including the relevant Gemini \texttt{IRAF} packages) from \url{http://www.gemini.edu/observing/phase-iii/understanding-and-processing-data/data-processing-software/gemini-iraf-general}. 

\bmhead{Conflict of interest}

We declare no conflicts of interests.

\bmhead{Authors' Contributions}

AJL obtained, reduced and analysed observations and wrote the text. DBM obtained and reduced the NOT observations, performed subtractions and photometry and contributed to analysis and interpretation. BPG undertook the \textit{Swift} BAT and XRT analysis and contributed to analysis and interpretation. AEN performed analysis of the host galaxy with \texttt{Prospector}. MN contributed to the light curve and TDE sections and spectral analysis. SRO analysed the UVOT data. DAP contributed to the NOT observations and interpretation. JR worked with the Gemini observations, photometry and subtractions. BDM contributed to the theoretical discussion and interpretation. SS provided the comparison between the X-ray/gamma-ray light curves of GRB\,191019A and other bursts. ERS worked on population modelling of the host galaxy. AI performed the fit of the spectrum with \texttt{pPXF}. AAC investigated the host population and contributed to interpretation. KB and AF worked on the {\em HST} observations and provided comments. AdUP worked on the NOT observations and commented on the text. WF worked on the interpretation. GF provided theoretical interpretation. JPUF led the first NOT observations and provided comments. NG worked on the offset implications. KEH worked on the NOT data and interpretation. JH worked on  interpretation and text. PGJ worked on the interpretation, in particular with regard to TDE possibilities. GL worked on the interpretation, IM provided theoretical input. JS and PJ worked on the NOT data and provided comments. NRT was involved in the NOT, Gemini and \text{HST} observations.

\bmhead{Supplementary Information}

Supplementary Information is available for this paper.


\section*{Supplementary Methods}

\section*{Supplementary Discussion}

\section*{Supplementary Tables}

\begin{table*}[htp]
\footnotesize
\caption{Optical observations of the counterpart of GRB\,191019A. The magnitude given is the 
integrated magnitude of host + afterglow, while the afterglow column provides
the afterglow flux. An (s) indicates this flux
is measured in a subtracted image while
other magnitudes are based on the subtraction of the mean host galaxy flux in a large
aperture. Magnitudes have not been corrected for foreground extinction
of $E(B-V) = 0.04$~mag}. 
\footnotesize
\begin{center}
\begin{tabular}{llllllllll}
\hline
Date & MJD & $\Delta T$ & Telescope &  Band & Exptime & Magnitude  & Afterglow   \\
     &     & (days)     &           &       & (s)  & (AB) & ($\mu$Jy)\\
\hline

2019-10-19 & 58775.82559  & 0.188   & NOT & $i$ & 600 & 18.585 $\pm$ 0.022   & 9.12$\pm$0.36 (s)\\
2019-10-19 & 58775.83911 & 0.205 & NOT & $g$ & 900 & 19.806 $\pm$ 0.014  & $5.40 \pm 0.16$ (s) \\
2019-10-19 & 58775.83911 & 0.217 & NOT & $r$ & 900 & 19.028 $\pm$ 0.018 & $7.05 \pm 0.21$ (s) \\
2019-10-19 & 58775.86312 & 0.229 & NOT & $z$ & 1000 & 19.212 $\pm$ 0.014 & $13.30 \pm 0.40$ (s) \\
2019-10-22 & 58778.87524 & 3.242 & NOT & $i$ & 1500 & 18.646 $\pm$ 0.028 & 1.30 $\pm$ 3.66 \\
2019-10-29 & 58785.85191 & 10.218 & NOT & $i$ & 900 & 18.678 $\pm$ 0.029 & -2.36 $\pm$ 3.69 \\
2019-10-29 & 58785.86350 & 10.230 & NOT & $g$ & 900 & 19.998 $\pm$ 0.014 & 0.017 $\pm$ 0.509 \\
2019-10-29 & 58785.87499 & 10.241 & NOT & $r$ & 900 & 19.093 $\pm$ 0.021 & 1.57 $\pm$ 1.93 \\
2019-10-29 & 58785.88748 & 10.254 & NOT & $z$ & 1000 & 19.391 $\pm$ 0.017 & --- \\
2019-10-31 & 58787.03036 &  11.397 & Gemini-S & $g$ &  900 & 19.940 $\pm$ 0.006  &  -0.14 $\pm$ 0.29\\
2019-10-31 & 58787.04344 &  11.410 & Gemini-S & $r$ & 900 & 19.039  $\pm$ 0.004 &  0.82 $\pm$ 0.48 \\
2019-10-31 & 58787.18344  &  11.550 & Gemini-S & $z$ & 900 &18.348 $\pm$  0.008  & 1.04 $\pm$ 1.18 \\
2019-11-10 & 58797.04364 & 21.410 & Gemini-S & $z$ & 720 & 18.383 $\pm$  0.010  & -4.23 $\pm$ 1.78\\
2019-11-22 & 58809.82539 & 34.192 & NOT & $i$ & 3000 & 18.648 $\pm$ 0.034 & 1.06 $\pm$  4.39 \\
2019-11-22 & 58809.84448 & 34.211 & NOT & $r$ & 3000 & 19.110 $\pm$ 0.021 & 0.27 $\pm$ 1.91 \\
2019-11-26 & 58813.02799  & 37.394 &  Gemini-S & $g$ & 900 & 19.938 $\pm$ 0.006 & -0.08 $\pm$ 0.29\\
2019-11-26 & 58813.03911 & 37.405 & Gemini-S & $r$ & 900 & 19.060 $\pm$ 0.004   &  -0.85 $\pm$ 0.46 \\
2019-11-26 & 58813.05029 & 37.417 & Gemini-S & $z$ & 600 & 18.368 $\pm$  0.009 & -1.99 $\pm$ 1.65\\
2019-12-09 & 58826.02618 & 50.393 & Gemini-S & $z$ & 780 & 18.336 $\pm$  0.011 & 2.88 $\pm$ 2.00\\
2019-12-16 & 58833.04820& 57.415 &  Gemini-S & $g$ & 750  & 19.931 $\pm$ 0.007 & 0.17 $\pm$ 0.32 \\
2019-12-16 & 58833.05932   & 57.426 & Gemini-S &  $r$ & 750 & 19.047 $\pm$ 0.005 & 0.18 $\pm$ 0.54 \\
2019-12-16 & 58833.07047 &  57.437 & Gemini-S & $z$ & 750 & 18.342 $\pm$  0.009 & 1.95 $\pm$ 1.68 \\
2019-12-17 & 58834.05827 & 58.425 &  Gemini-S & $g$ & 750  & 19.936 $\pm$ 0.006 & -0.01 $\pm$ 0.29\\
2019-12-17 & 58834.07159&  58.438 & Gemini-S & $r$ & 750 & 19.041  $\pm$ 0.004  & 0.66 $\pm$ 0.47 \\
2019-12-17 & 58834.08246 & 58.449 & Gemini-S & $z$ & 750 & 18.353 $\pm$  0.011  & 0.27 $\pm$ 1.97\\
2019-12-21 & 58838.05466 & 62.421 & Gemini-S & $g$ & 750  &  19.955 $\pm$ 0.006 & -0.67 $\pm$ 0.29\\
2019-12-21 & 58838.06358 & 62.430 & Gemini-S & $r$ & 750  & 19.044 $\pm$ 0.005 & 0.42 $\pm$ 0.54 \\
2019-12-21 & 58838.07468 & 62.441 & Gemini-S & $z$ & 750 & 18.347 $\pm$  0.010 & 1.18 $\pm$ 1.83 \\
2019-12-30 & 58847.03322  & 71.400 & Gemini-S & $g$ & 800  & 19.915 $\pm$ 0.013 & 0.74 $\pm$ 0.53 \\
2019-12-30 & 58847.04607  & 71.412 & Gemini-S & $r$ & 800 &19.065 $\pm$ 0.006  & -1.25 $\pm$ 0.61\\
2020-01-01 & 58849.04549 & 73.412 & Gemini-S &  $z$ & 750 & 18.362 $\pm$  0.007 & -1.10 $\pm$ 1.38  \\
2020-06-21 & 59021.15875 & 245.525 & NOT & $r$ & 1800 & 19.138 $\pm$ 0.010 & -1.83 $\pm$ 1.13 \\
2020-06-21 & 59021.18441 & 245.551 & NOT & $g$ & 2400 & 19.999 $\pm$ 0.010 & -0.017 $\pm$ 0.364\\

\hline

\end{tabular}
\end{center}
\label{tab:afterglow}
\end{table*}%

\begin{table*}[htp]
\footnotesize
\caption{Log of {\em Swift}-UVOT observations of GRB\,191019A. The given magnitudes are a combination of
afterglow + host galaxy. The afterglow flux densities are the remaining afterglow flux after the subtraction of the final epoch values for the White, $V$, $B$ and $U$ bands. For the UV filters (UVW1, UVM2, UVW2) we have {\em HST} observations which demonstrate that the host galaxy is much fainter that these detections. Therefore, 
rather than subtract the (low signal to noise) late UVOT observations, we subtract instead a source of
$1.7 \pm 0.5$ $\mu$Jy (with the error larger than those measured in the \textit{HST} observations
to reflect uncertainties in the precise band). The magnitudes are indicated in the table with an (a).
Magnitudes have not been corrected for foreground extinction
of $E(B-V) = 0.04$ mag.}
\footnotesize
\begin{center}
\begin{tabular}{lllllllll}
\hline
 MJD & $\Delta T$ & Telescope &  Band & $t_{1/2}$ (s) & Magnitude  & Afterglow   \\
     & (days)     &           &       & (s)  & (AB) & ($\mu$Jy)\\

\hline
 58775.673  & 0.0389 & UVOT & White & 75 & $20.85^{+0.39}_{-0.29}$ & $9.26 \pm 4.29$ \\
 58775.690  & 0.0562 & UVOT & White & 100 & $20.48^{+0.16}_{-0.14}$ & $14.59 \pm 3.73$ \\
58775.962  & 0.3279 & UVOT & White & 343 & $21.02^{+0.13}_{-0.12}$ & $4.01  \pm 1.50$ \\
 58781.722  & 6.0880 & UVOT & White & 40666 & $21.24^{+0.04}_{-0.04}$ & $1.52 \pm 0.53$\\
58961.436  & 185.8020 & UVOT & White & 37620 & $21.43^{+0.14}_{-0.13}$ & --- \\
58775.675  & 0.0410 & UVOT & $V$ & 100 & $18.90^{+0.29}_{-0.23}$ & $44.92 \pm 26.64$ \\
58775.695 & 0.0609 & UVOT & $V$ & 100 & $19.11^{+0.31}_{-0.24}$ & $27.4 \pm 24.17$ \\
58775.811 & 0.1769 & UVOT & $V$ & 80 & $19.93^{+0.82}_{-0.46}$ & $-15.58 \pm 24.38$ \\
58955.963  & 179.3290 & UVOT & $V$ & 589 & $19.55^{+0.30}_{-0.24}$ & --- \\
58775.688  & 0.0538 & UVOT & $B$ & 100 & $20.52^{+0.50}_{-0.34}$ & $10.09 \pm 10.71$ \\
58775.829  & 0.1952 & UVOT & $B$ & 333 & $20.52^{+0.25}_{-0.21}$ & $ 9.74 \pm 6.31$ \\
58775.952  & 0.3186 & UVOT & $B$ & 453 & $20.54^{ +0.22}_{-0.18}$ & $9.52 \pm 5.56$ \\
58776.781  & 1.1474 & UVOT & $B$ & 9405 & $20.68^{+0.16}_{-0.14}$ & $7.21 \pm 3.54$ \\
58959.945  & 184.3120 & UVOT & $B$ & 444 & $20.95^{+0.66}_{-0.41}$ & --- \\
58775.685  & 0.0514 & UVOT & $U$ & 100 & $21.10^{+0.41}_{-0.30}$ & $9.89 \pm 5.17$ \\
58775.743  & 0.1096 & UVOT & $U$ & 73 & $21.22^{+0.57}_{-0.37}$ & $8.47 \pm 5.72$ \\
58775.942  & 0.3081 & UVOT & $U$ & 453 & $21.81^{+0.39}_{-0.29}$ & $3.58 \pm 2.75$ \\
58776.781  & 1.1475 & UVOT & $U$ & 9670 & $22.28^{+0.31}_{-0.24}$ & $1.27 \pm  1.50$ \\
58961.638  & 186.0050 & UVOT & $U$ & 26238 & $22.42^{+1.42}_{-0.60}$ & --- \\
58775.683  & 0.0491 & UVOT & UVW1 & 100 & $21.282^{+0.41}_{-0.30}$ & $9.31 \pm 3.46$a \\
58775.699  & 0.0654 & UVOT & UVW1 & 74 & $21.27^{+0.48}_{-0.33}$ & $9.44 \pm 3.98$a \\
58775.890  & 0.2564 & UVOT & UVW1 & 789 & $22.40^{+0.50}_{-0.34}$ & $2.22 \pm 1.54$a \\
58776.006 & 0.3724 & UVOT & UVW1 & 222 & $21.83^{+0.46}_{-0.32}$ & $4.94  \pm 2.33$a \\
58962.195  & 186.5620 & UVOT & UVW1 & 500 & $23.21^{+1.74}_{-0.64}$ & --- \\
58775.677  & 0.0435 & UVOT & UVM2 & 100 & $22.36^{+1.12}_{-0.54}$ &  $2.43 \pm 2.69$a \\
58775.697  & 0.0633 & UVOT & UVM2 & 100 & $21.68^{+0.57}_{-0.37}$ & $6.04 \pm 3.21$a\\
58775.876  & 0.2419 & UVOT & UVM2 & 450 & $23.52^{+1.37}_{-0.59}$ & $-0.28 \pm 1.13$a \\
58968.894  & 193.2610 & UVOT & UVM2 & 400 & $23.11^{+0.94}_{-0.50}$ & --- \\ 
58775.692  & 0.0585 & UVOT & UVW2 & 100 & $22.06^{+0.59}_{-0.38}$ & $3.73 \pm 2.33$a \\
58775.762  & 0.1280 & UVOT & UVW2 & 404 & $22.02^{+0.26}_{-0.21}$ & $3.94 \pm  1.3$a \\
58959.813  & 184.1790 & UVOT & UVW2 & 444 & $23.46^{+1.40}_{-0.59}$ & --- \\
\hline
\end{tabular}
\end{center}
\label{tab:UVOT}
\end{table*}%


\begin{table*}[htp]
\caption{Multiband photometry of the host galaxy of GRB\,191019A.}
\begin{center}
\begin{tabular}{llll}
\hline
Filter & $\lambda$ (nm) &  Magnitude (AB) & Origin \\
\hline
F225W &235.8 & 23.32 $\pm$  0.07  & This work\\
F275W &270.3 & 23.43 $\pm$ 0.04  & This work \\
$g$ & 481.0 & 20.17 $\pm$ 0.02 & Pan-STARRS \cite{Chambers+16} \\
$r$ &  617.0 & 19.18 $\pm$ 0.01 & Pan-STARRS \cite{Chambers+16} \\
$i$ &  752.0 & 18.76 $\pm$ 0.01 & Pan-STARRS \cite{Chambers+16} \\
$z$ &  866.0  & 18.58 $\pm$ 0.01 & Pan-STARRS \cite{Chambers+16} \\
$y$ &  962.0  & 18.53 $\pm$ 0.025 & Pan-STARRS \cite{Chambers+16} \\
$Y$ & 1020.0 & 18.21 $\pm$ 0.057 & VHS \cite{McMahon+2013}  \\ 
$J$ & 1252 & 18.096 $\pm$ 0.079 & VHS \cite{McMahon+2013}  \\ 
$K_s$ & 2147 & 17.547 $\pm$ 0.09 & VHS \cite{McMahon+2013}  \\ 
WISE W1 & 3353 & 18.338 $\pm$ 0.047  & WISE \cite{WISE_Wright+10} \\ 
WISE W2 & 4603 & 18.729 $\pm$ 0.112 & WISE \cite{WISE_Wright+10} \\ 
\hline
\end{tabular}
\end{center}
\label{tab:host}
\end{table*}%



\clearpage





\begin{thebibliography}{103}
\ifx \bisbn   \undefined \def \bisbn  #1{ISBN #1}\fi
\ifx \binits  \undefined \def \binits#1{#1}\fi
\ifx \bauthor  \undefined \def \bauthor#1{#1}\fi
\ifx \batitle  \undefined \def \batitle#1{#1}\fi
\ifx \bjtitle  \undefined \def \bjtitle#1{#1}\fi
\ifx \bvolume  \undefined \def \bvolume#1{\textbf{#1}}\fi
\ifx \byear  \undefined \def \byear#1{#1}\fi
\ifx \bissue  \undefined \def \bissue#1{#1}\fi
\ifx \bfpage  \undefined \def \bfpage#1{#1}\fi
\ifx \blpage  \undefined \def \blpage #1{#1}\fi
\ifx \burl  \undefined \def \burl#1{\textsf{#1}}\fi
\ifx \doiurl  \undefined \def \doiurl#1{\url{https://doi.org/#1}}\fi
\ifx \betal  \undefined \def \betal{\textit{et al.}}\fi
\ifx \binstitute  \undefined \def \binstitute#1{#1}\fi
\ifx \binstitutionaled  \undefined \def \binstitutionaled#1{#1}\fi
\ifx \bctitle  \undefined \def \bctitle#1{#1}\fi
\ifx \beditor  \undefined \def \beditor#1{#1}\fi
\ifx \bpublisher  \undefined \def \bpublisher#1{#1}\fi
\ifx \bbtitle  \undefined \def \bbtitle#1{#1}\fi
\ifx \bedition  \undefined \def \bedition#1{#1}\fi
\ifx \bseriesno  \undefined \def \bseriesno#1{#1}\fi
\ifx \blocation  \undefined \def \blocation#1{#1}\fi
\ifx \bsertitle  \undefined \def \bsertitle#1{#1}\fi
\ifx \bsnm \undefined \def \bsnm#1{#1}\fi
\ifx \bsuffix \undefined \def \bsuffix#1{#1}\fi
\ifx \bparticle \undefined \def \bparticle#1{#1}\fi
\ifx \barticle \undefined \def \barticle#1{#1}\fi
\bibcommenthead
\ifx \bconfdate \undefined \def \bconfdate #1{#1}\fi
\ifx \botherref \undefined \def \botherref #1{#1}\fi
\ifx \url \undefined \def \url#1{\textsf{#1}}\fi
\ifx \bchapter \undefined \def \bchapter#1{#1}\fi
\ifx \bbook \undefined \def \bbook#1{#1}\fi
\ifx \bcomment \undefined \def \bcomment#1{#1}\fi
\ifx \oauthor \undefined \def \oauthor#1{#1}\fi
\ifx \citeauthoryear \undefined \def \citeauthoryear#1{#1}\fi
\ifx \endbibitem  \undefined \def \endbibitem {}\fi
\ifx \bconflocation  \undefined \def \bconflocation#1{#1}\fi
\ifx \arxivurl  \undefined \def \arxivurl#1{\textsf{#1}}\fi
\csname PreBibitemsHook\endcsname

\bibitem{Hjorth+03}
\begin{barticle}
\bauthor{\bsnm{{Hjorth}}, \binits{J.}},
\bauthor{\bsnm{{Sollerman}}, \binits{J.}},
\bauthor{\bsnm{{M{\o}ller}}, \binits{P.}},
\bauthor{\bsnm{{Fynbo}}, \binits{J.P.U.}},
\bauthor{\bsnm{{Woosley}}, \binits{S.E.}},
\bauthor{\bsnm{{Kouveliotou}}, \binits{C.}},
\bauthor{\bsnm{{Tanvir}}, \binits{N.R.}},
\bauthor{\bsnm{{Greiner}}, \binits{J.}},
\bauthor{\bsnm{{Andersen}}, \binits{M.I.}},
\bauthor{\bsnm{{Castro-Tirado}}, \binits{A.J.}},
\bauthor{\bsnm{{Castro Cer{\'o}n}}, \binits{J.M.}},
\bauthor{\bsnm{{Fruchter}}, \binits{A.S.}},
\bauthor{\bsnm{{Gorosabel}}, \binits{J.}},
\bauthor{\bsnm{{Jakobsson}}, \binits{P.}},
\bauthor{\bsnm{{Kaper}}, \binits{L.}},
\bauthor{\bsnm{{Klose}}, \binits{S.}},
\bauthor{\bsnm{{Masetti}}, \binits{N.}},
\bauthor{\bsnm{{Pedersen}}, \binits{H.}},
\bauthor{\bsnm{{Pedersen}}, \binits{K.}},
\bauthor{\bsnm{{Pian}}, \binits{E.}},
\bauthor{\bsnm{{Palazzi}}, \binits{E.}},
\bauthor{\bsnm{{Rhoads}}, \binits{J.E.}},
\bauthor{\bsnm{{Rol}}, \binits{E.}},
\bauthor{\bsnm{{van den Heuvel}}, \binits{E.P.J.}},
\bauthor{\bsnm{{Vreeswijk}}, \binits{P.M.}},
\bauthor{\bsnm{{Watson}}, \binits{D.}},
\bauthor{\bsnm{{Wijers}}, \binits{R.A.M.J.}}:
\batitle{{A very energetic supernova associated with the
  {\ensuremath{\gamma}}-ray burst of 29 March 2003}}.
\bjtitle{\nat}
\bvolume{423}(\bissue{6942}),
\bfpage{847}--\blpage{850}
(\byear{2003})
{\href{https://arxiv.org/abs/astro-ph/0306347}{{arXiv:astro-ph/0306347}}}
{[astro-ph]}.
\doiurl{10.1038/nature01750}
\end{barticle}
\endbibitem

\bibitem{rastinejad22}
\begin{botherref}
\oauthor{\bsnm{{Rastinejad}}, \binits{J.C.}},
\oauthor{\bsnm{{Gompertz}}, \binits{B.P.}},
\oauthor{\bsnm{{Levan}}, \binits{A.J.}},
\oauthor{\bsnm{{Fong}}, \binits{W.}},
\oauthor{\bsnm{{Nicholl}}, \binits{M.}},
\oauthor{\bsnm{{Lamb}}, \binits{G.P.}},
\oauthor{\bsnm{{Malesani}}, \binits{D.B.}},
\oauthor{\bsnm{{Nugent}}, \binits{A.E.}},
\oauthor{\bsnm{{Oates}}, \binits{S.R.}},
\oauthor{\bsnm{{Tanvir}}, \binits{N.R.}},
\oauthor{\bsnm{{de Ugarte Postigo}}, \binits{A.}},
\oauthor{\bsnm{{Kilpatrick}}, \binits{C.D.}},
\oauthor{\bsnm{{Moore}}, \binits{C.J.}},
\oauthor{\bsnm{{Metzger}}, \binits{B.D.}},
\oauthor{\bsnm{{Ravasio}}, \binits{M.E.}},
\oauthor{\bsnm{{Rossi}}, \binits{A.}},
\oauthor{\bsnm{{Schroeder}}, \binits{G.}},
\oauthor{\bsnm{{Jencson}}, \binits{J.}},
\oauthor{\bsnm{{Sand}}, \binits{D.J.}},
\oauthor{\bsnm{{Smith}}, \binits{N.}},
\oauthor{\bsnm{{Ag{\"u}{\'\i} Fern{\'a}ndez}}, \binits{J.F.}},
\oauthor{\bsnm{{Berger}}, \binits{E.}},
\oauthor{\bsnm{{Blanchard}}, \binits{P.K.}},
\oauthor{\bsnm{{Chornock}}, \binits{R.}},
\oauthor{\bsnm{{Cobb}}, \binits{B.E.}},
\oauthor{\bsnm{{De Pasquale}}, \binits{M.}},
\oauthor{\bsnm{{Fynbo}}, \binits{J.P.U.}},
\oauthor{\bsnm{{Izzo}}, \binits{L.}},
\oauthor{\bsnm{{Kann}}, \binits{D.A.}},
\oauthor{\bsnm{{Laskar}}, \binits{T.}},
\oauthor{\bsnm{{Marini}}, \binits{E.}},
\oauthor{\bsnm{{Paterson}}, \binits{K.}},
\oauthor{\bsnm{{Rouco Escorial}}, \binits{A.}},
\oauthor{\bsnm{{Sears}}, \binits{H.M.}},
\oauthor{\bsnm{{Th{\"o}ne}}, \binits{C.C.}}:
{A Kilonova Following a Long-Duration Gamma-Ray Burst at 350 Mpc}.
arXiv e-prints,
2204--10864
(2022)
{\href{https://arxiv.org/abs/2204.10864}{{arXiv:2204.10864}}}
{[astro-ph.HE]}
\end{botherref}
\endbibitem

\bibitem{troja22}
\begin{botherref}
\oauthor{\bsnm{{Troja}}, \binits{E.}},
\oauthor{\bsnm{{Fryer}}, \binits{C.L.}},
\oauthor{\bsnm{{O'Connor}}, \binits{B.}},
\oauthor{\bsnm{{Ryan}}, \binits{G.}},
\oauthor{\bsnm{{Dichiara}}, \binits{S.}},
\oauthor{\bsnm{{Kumar}}, \binits{A.}},
\oauthor{\bsnm{{Ito}}, \binits{N.}},
\oauthor{\bsnm{{Gupta}}, \binits{R.}},
\oauthor{\bsnm{{Wollaeger}}, \binits{R.}},
\oauthor{\bsnm{{Norris}}, \binits{J.P.}},
\oauthor{\bsnm{{Kawai}}, \binits{N.}},
\oauthor{\bsnm{{Butler}}, \binits{N.}},
\oauthor{\bsnm{{Aryan}}, \binits{A.}},
\oauthor{\bsnm{{Misra}}, \binits{K.}},
\oauthor{\bsnm{{Hosokawa}}, \binits{R.}},
\oauthor{\bsnm{{Murata}}, \binits{K.L.}},
\oauthor{\bsnm{{Niwano}}, \binits{M.}},
\oauthor{\bsnm{{Pandey}}, \binits{S.B.}},
\oauthor{\bsnm{{Kutyrev}}, \binits{A.}},
\oauthor{\bsnm{{van Eerten}}, \binits{H.J.}},
\oauthor{\bsnm{{Chase}}, \binits{E.A.}},
\oauthor{\bsnm{{Hu}}, \binits{Y.-D.}},
\oauthor{\bsnm{{Caballero-Garcia}}, \binits{M.D.}},
\oauthor{\bsnm{{Castro-Tirado}}, \binits{A.J.}}:
{A long gamma-ray burst from a merger of compact objects}.
arXiv e-prints,
2209--03363
(2022)
{\href{https://arxiv.org/abs/2209.03363}{{arXiv:2209.03363}}}
{[astro-ph.HE]}
\end{botherref}
\endbibitem

\bibitem{yang22}
\begin{botherref}
\oauthor{\bsnm{{Yang}}, \binits{J.}},
\oauthor{\bsnm{{Ai}}, \binits{S.}},
\oauthor{\bsnm{{Zhang}}, \binits{B.-B.}},
\oauthor{\bsnm{{Zhang}}, \binits{B.}},
\oauthor{\bsnm{{Liu}}, \binits{Z.-K.}},
\oauthor{\bsnm{{Wang}}, \binits{X.I.}},
\oauthor{\bsnm{{Yang}}, \binits{Y.-H.}},
\oauthor{\bsnm{{Yin}}, \binits{Y.-H.}},
\oauthor{\bsnm{{Li}}, \binits{Y.}},
\oauthor{\bsnm{{L{\"u}}}, \binits{H.-J.}}:
{A long-duration gamma-ray burst with a peculiar origin}.
arXiv e-prints,
2204--12771
(2022)
{\href{https://arxiv.org/abs/2204.12771}{{arXiv:2204.12771}}}
{[astro-ph.HE]}
\end{botherref}
\endbibitem

\bibitem{grindlay06}
\begin{barticle}
\bauthor{\bsnm{{Grindlay}}, \binits{J.}},
\bauthor{\bsnm{{Portegies Zwart}}, \binits{S.}},
\bauthor{\bsnm{{McMillan}}, \binits{S.}}:
\batitle{{Short gamma-ray bursts from binary neutron star mergers in globular
  clusters}}.
\bjtitle{Nature Physics}
\bvolume{2},
\bfpage{116}--\blpage{119}
(\byear{2006})
{\href{https://arxiv.org/abs/astro-ph/0512654}{{astro-ph/0512654}}}.
\doiurl{10.1038/nphys214}
\end{barticle}
\endbibitem

\bibitem{fragione19}
\begin{barticle}
\bauthor{\bsnm{{Fragione}}, \binits{G.}},
\bauthor{\bsnm{{Grishin}}, \binits{E.}},
\bauthor{\bsnm{{Leigh}}, \binits{N.W.C.}},
\bauthor{\bsnm{{Perets}}, \binits{H.B.}},
\bauthor{\bsnm{{Perna}}, \binits{R.}}:
\batitle{{Black hole and neutron star mergers in galactic nuclei}}.
\bjtitle{\mnras}
\bvolume{488}(\bissue{1}),
\bfpage{47}--\blpage{63}
(\byear{2019})
{\href{https://arxiv.org/abs/1811.10627}{{arXiv:1811.10627}}}
{[astro-ph.GA]}.
\doiurl{10.1093/mnras/stz1651}
\end{barticle}
\endbibitem

\bibitem{mckernan20}
\begin{barticle}
\bauthor{\bsnm{{McKernan}}, \binits{B.}},
\bauthor{\bsnm{{Ford}}, \binits{K.E.S.}},
\bauthor{\bsnm{{O'Shaughnessy}}, \binits{R.}}:
\batitle{{Black hole, neutron star, and white dwarf merger rates in AGN
  discs}}.
\bjtitle{\mnras}
\bvolume{498}(\bissue{3}),
\bfpage{4088}--\blpage{4094}
(\byear{2020})
{\href{https://arxiv.org/abs/2002.00046}{{arXiv:2002.00046}}}
{[astro-ph.HE]}.
\doiurl{10.1093/mnras/staa2681}
\end{barticle}
\endbibitem

\bibitem{oleary16}
\begin{barticle}
\bauthor{\bsnm{{O'Leary}}, \binits{R.M.}},
\bauthor{\bsnm{{Meiron}}, \binits{Y.}},
\bauthor{\bsnm{{Kocsis}}, \binits{B.}}:
\batitle{{Dynamical Formation Signatures of Black Hole Binaries in the First
  Detected Mergers by LIGO}}.
\bjtitle{\apjl}
\bvolume{824}(\bissue{1}),
\bfpage{12}
(\byear{2016})
{\href{https://arxiv.org/abs/1602.02809}{{arXiv:1602.02809}}}
{[astro-ph.HE]}.
\doiurl{10.3847/2041-8205/824/1/L12}
\end{barticle}
\endbibitem

\bibitem{stone16}
\begin{barticle}
\bauthor{\bsnm{{Stone}}, \binits{N.C.}},
\bauthor{\bsnm{{Metzger}}, \binits{B.D.}}:
\batitle{{Rates of stellar tidal disruption as probes of the supermassive black
  hole mass function}}.
\bjtitle{\mnras}
\bvolume{455}(\bissue{1}),
\bfpage{859}--\blpage{883}
(\byear{2016})
{\href{https://arxiv.org/abs/1410.7772}{{arXiv:1410.7772}}}
{[astro-ph.HE]}.
\doiurl{10.1093/mnras/stv2281}
\end{barticle}
\endbibitem

\bibitem{perna21}
\begin{barticle}
\bauthor{\bsnm{{Perna}}, \binits{R.}},
\bauthor{\bsnm{{Lazzati}}, \binits{D.}},
\bauthor{\bsnm{{Cantiello}}, \binits{M.}}:
\batitle{{Electromagnetic Signatures of Relativistic Explosions in the Disks of
  Active Galactic Nuclei}}.
\bjtitle{\apjl}
\bvolume{906}(\bissue{2}),
\bfpage{7}
(\byear{2021})
{\href{https://arxiv.org/abs/2011.08873}{{arXiv:2011.08873}}}
{[astro-ph.HE]}.
\doiurl{10.3847/2041-8213/abd319}
\end{barticle}
\endbibitem

\bibitem{lazzati22}
\begin{botherref}
\oauthor{\bsnm{{Lazzati}}, \binits{D.}},
\oauthor{\bsnm{{Soares}}, \binits{G.}},
\oauthor{\bsnm{{Perna}}, \binits{R.}}:
{Prompt Emission of Gamma-Ray Bursts in the High-density Environment of Active
  Galactic Nuclei Accretion Disks}.
arXiv e-prints,
2209--14308
(2022)
{\href{https://arxiv.org/abs/2209.14308}{{arXiv:2209.14308}}}
{[astro-ph.HE]}
\end{botherref}
\endbibitem

\bibitem{Galama+98}
\begin{barticle}
\bauthor{\bsnm{{Galama}}, \binits{T.J.}},
\bauthor{\bsnm{{Vreeswijk}}, \binits{P.M.}},
\bauthor{\bsnm{{van Paradijs}}, \binits{J.}},
\bauthor{\bsnm{{Kouveliotou}}, \binits{C.}},
\bauthor{\bsnm{{Augusteijn}}, \binits{T.}},
\bauthor{\bsnm{{B{\"o}hnhardt}}, \binits{H.}},
\bauthor{\bsnm{{Brewer}}, \binits{J.P.}},
\bauthor{\bsnm{{Doublier}}, \binits{V.}},
\bauthor{\bsnm{{Gonzalez}}, \binits{J.-F.}},
\bauthor{\bsnm{{Leibundgut}}, \binits{B.}},
\bauthor{\bsnm{{Lidman}}, \binits{C.}},
\bauthor{\bsnm{{Hainaut}}, \binits{O.R.}},
\bauthor{\bsnm{{Patat}}, \binits{F.}},
\bauthor{\bsnm{{Heise}}, \binits{J.}},
\bauthor{\bsnm{{in't Zand}}, \binits{J.}},
\bauthor{\bsnm{{Hurley}}, \binits{K.}},
\bauthor{\bsnm{{Groot}}, \binits{P.J.}},
\bauthor{\bsnm{{Strom}}, \binits{R.G.}},
\bauthor{\bsnm{{Mazzali}}, \binits{P.A.}},
\bauthor{\bsnm{{Iwamoto}}, \binits{K.}},
\bauthor{\bsnm{{Nomoto}}, \binits{K.}},
\bauthor{\bsnm{{Umeda}}, \binits{H.}},
\bauthor{\bsnm{{Nakamura}}, \binits{T.}},
\bauthor{\bsnm{{Young}}, \binits{T.R.}},
\bauthor{\bsnm{{Suzuki}}, \binits{T.}},
\bauthor{\bsnm{{Shigeyama}}, \binits{T.}},
\bauthor{\bsnm{{Koshut}}, \binits{T.}},
\bauthor{\bsnm{{Kippen}}, \binits{M.}},
\bauthor{\bsnm{{Robinson}}, \binits{C.}},
\bauthor{\bsnm{{de Wildt}}, \binits{P.}},
\bauthor{\bsnm{{Wijers}}, \binits{R.A.M.J.}},
\bauthor{\bsnm{{Tanvir}}, \binits{N.}},
\bauthor{\bsnm{{Greiner}}, \binits{J.}},
\bauthor{\bsnm{{Pian}}, \binits{E.}},
\bauthor{\bsnm{{Palazzi}}, \binits{E.}},
\bauthor{\bsnm{{Frontera}}, \binits{F.}},
\bauthor{\bsnm{{Masetti}}, \binits{N.}},
\bauthor{\bsnm{{Nicastro}}, \binits{L.}},
\bauthor{\bsnm{{Feroci}}, \binits{M.}},
\bauthor{\bsnm{{Costa}}, \binits{E.}},
\bauthor{\bsnm{{Piro}}, \binits{L.}},
\bauthor{\bsnm{{Peterson}}, \binits{B.A.}},
\bauthor{\bsnm{{Tinney}}, \binits{C.}},
\bauthor{\bsnm{{Boyle}}, \binits{B.}},
\bauthor{\bsnm{{Cannon}}, \binits{R.}},
\bauthor{\bsnm{{Stathakis}}, \binits{R.}},
\bauthor{\bsnm{{Sadler}}, \binits{E.}},
\bauthor{\bsnm{{Begam}}, \binits{M.C.}},
\bauthor{\bsnm{{Ianna}}, \binits{P.}}:
\batitle{{An unusual supernova in the error box of the
  {\ensuremath{\gamma}}-ray burst of 25 April 1998}}.
\bjtitle{\nat}
\bvolume{395}(\bissue{6703}),
\bfpage{670}--\blpage{672}
(\byear{1998})
{\href{https://arxiv.org/abs/astro-ph/9806175}{{arXiv:astro-ph/9806175}}}
{[astro-ph]}.
\doiurl{10.1038/27150}
\end{barticle}
\endbibitem

\bibitem{Abbott+17a}
\begin{barticle}
\bauthor{\bsnm{{Abbott}}, \binits{B.P.}},
\bauthor{\bsnm{{Abbott}}, \binits{R.}},
\bauthor{\bsnm{{Abbott}}, \binits{T.D.}},
\bauthor{\bsnm{{Acernese}}, \binits{F.}},
\bauthor{\bsnm{{Ackley}}, \binits{K.}},
\bauthor{\bsnm{{Adams}}, \binits{C.}},
\bauthor{\bsnm{{Adams}}, \binits{T.}},
\bauthor{\bsnm{{Addesso}}, \binits{P.}},
\bauthor{\bsnm{{Adhikari}}, \binits{R.X.}},
\bauthor{\bsnm{{Adya}}, \binits{V.B.}},
\bauthor{\bsnm{{Affeldt}}, \binits{C.}},
\bauthor{\bsnm{{Afrough}}, \binits{M.}},
\bauthor{\bsnm{{Agarwal}}, \binits{B.}},
\bauthor{\bsnm{{Agathos}}, \binits{M.}},
\bauthor{\bsnm{{Agatsuma}}, \binits{K.}},
\bauthor{\bsnm{{Aggarwal}}, \binits{N.}},
\bauthor{\bsnm{{Aguiar}}, \binits{O.D.}},
\bauthor{\bsnm{{Aiello}}, \binits{L.}},
\bauthor{\bparticle{et} \bsnm{al.}}:
\batitle{{GW170817: Observation of Gravitational Waves from a Binary Neutron
  Star Inspiral}}.
\bjtitle{\prl}
\bvolume{119}(\bissue{16}),
\bfpage{161101}
(\byear{2017})
{\href{https://arxiv.org/abs/1710.05832}{{arXiv:1710.05832}}}
{[gr-qc]}.
\doiurl{10.1103/PhysRevLett.119.161101}
\end{barticle}
\endbibitem

\bibitem{simpson19}
\begin{barticle}
\bauthor{\bsnm{{Simpson}}, \binits{K.K.}},
\bauthor{\bsnm{{Barthelmy}}, \binits{S.D.}},
\bauthor{\bsnm{{Gropp}}, \binits{J.D.}},
\bauthor{\bsnm{{Lien}}, \binits{A.Y.}},
\bauthor{\bsnm{{Page}}, \binits{K.L.}},
\bauthor{\bsnm{{Palmer}}, \binits{D.M.}},
\bauthor{\bsnm{{Tohuvavohu}}, \binits{A.}},
\bauthor{\bsnm{{Neil Gehrels Swift Observatory Team}}}:
\batitle{{GRB 191019A: Swift detection of a burst}}.
\bjtitle{GRB Coordinates Network}
\bvolume{26031},
\bfpage{1}
(\byear{2019})
\end{barticle}
\endbibitem

\bibitem{krimm19}
\begin{barticle}
\bauthor{\bsnm{{Krimm}}, \binits{H.A.}},
\bauthor{\bsnm{{Barthelmy}}, \binits{S.D.}},
\bauthor{\bsnm{{Cummings}}, \binits{J.R.}},
\bauthor{\bsnm{{Laha}}, \binits{S.}},
\bauthor{\bsnm{{Lien}}, \binits{A.Y.}},
\bauthor{\bsnm{{Markwardt}}, \binits{C.B.}},
\bauthor{\bsnm{{Palmer}}, \binits{D.M.}},
\bauthor{\bsnm{{Sakamoto}}, \binits{T.}},
\bauthor{\bsnm{{Simpson}}, \binits{K.K.}},
\bauthor{\bsnm{{Stamatikos}}, \binits{M.}},
\bauthor{\bsnm{{Ukwatta}}, \binits{T.N.}}:
\batitle{{GRB 191019A: Swift-BAT refined analysis}}.
\bjtitle{GRB Coordinates Network}
\bvolume{26046},
\bfpage{1}
(\byear{2019})
\end{barticle}
\endbibitem

\bibitem{evans19}
\begin{barticle}
\bauthor{\bsnm{{Evans}}, \binits{P.A.}},
\bauthor{\bsnm{{Osborne}}, \binits{J.P.}},
\bauthor{\bsnm{{Burrows}}, \binits{D.N.}},
\bauthor{\bsnm{{Kennea}}, \binits{J.A.}},
\bauthor{\bsnm{{Campana}}, \binits{S.}},
\bauthor{\bsnm{{Cusumano}}, \binits{G.}},
\bauthor{\bsnm{{Swift-XRT Team}}}:
\batitle{{GRB 191019A: Swift-XRT observations}}.
\bjtitle{GRB Coordinates Network}
\bvolume{26034},
\bfpage{1}
(\byear{2019})
\end{barticle}
\endbibitem

\bibitem{perley19}
\begin{barticle}
\bauthor{\bsnm{{Perley}}, \binits{D.A.}},
\bauthor{\bsnm{{Malesani}}, \binits{D.B.}},
\bauthor{\bsnm{{Levan}}, \binits{A.J.}},
\bauthor{\bsnm{{Fynbo}}, \binits{J.P.U.}},
\bauthor{\bsnm{{Djupvik}}, \binits{A.A.}}, \betal:
\batitle{{GRB 191019A: NOT optical afterglow and host association
  confirmation}}.
\bjtitle{GRB Coordinates Network}
\bvolume{26062},
\bfpage{1}
(\byear{2019})
\end{barticle}
\endbibitem

\bibitem{perley20}
\begin{barticle}
\bauthor{\bsnm{{Perley}}, \binits{D.A.}},
\bauthor{\bsnm{{Fremling}}, \binits{C.}},
\bauthor{\bsnm{{Sollerman}}, \binits{J.}},
\bauthor{\bsnm{{Miller}}, \binits{A.A.}},
\bauthor{\bsnm{{Dahiwale}}, \binits{A.S.}},
\bauthor{\bsnm{{Sharma}}, \binits{Y.}},
\bauthor{\bsnm{{Bellm}}, \binits{E.C.}},
\bauthor{\bsnm{{Biswas}}, \binits{R.}},
\bauthor{\bsnm{{Brink}}, \binits{T.G.}},
\bauthor{\bsnm{{Bruch}}, \binits{R.J.}},
\bauthor{\bsnm{{De}}, \binits{K.}},
\bauthor{\bsnm{{Dekany}}, \binits{R.}},
\bauthor{\bsnm{{Drake}}, \binits{A.J.}},
\bauthor{\bsnm{{Duev}}, \binits{D.A.}},
\bauthor{\bsnm{{Filippenko}}, \binits{A.V.}},
\bauthor{\bsnm{{Gal-Yam}}, \binits{A.}},
\bauthor{\bsnm{{Goobar}}, \binits{A.}},
\bauthor{\bsnm{{Graham}}, \binits{M.J.}},
\bauthor{\bsnm{{Graham}}, \binits{M.L.}},
\bauthor{\bsnm{{Ho}}, \binits{A.Y.Q.}},
\bauthor{\bsnm{{Irani}}, \binits{I.}},
\bauthor{\bsnm{{Kasliwal}}, \binits{M.M.}},
\bauthor{\bsnm{{Kim}}, \binits{Y.-L.}},
\bauthor{\bsnm{{Kulkarni}}, \binits{S.R.}},
\bauthor{\bsnm{{Mahabal}}, \binits{A.}},
\bauthor{\bsnm{{Masci}}, \binits{F.J.}},
\bauthor{\bsnm{{Modak}}, \binits{S.}},
\bauthor{\bsnm{{Neill}}, \binits{J.D.}},
\bauthor{\bsnm{{Nordin}}, \binits{J.}},
\bauthor{\bsnm{{Riddle}}, \binits{R.L.}},
\bauthor{\bsnm{{Soumagnac}}, \binits{M.T.}},
\bauthor{\bsnm{{Strotjohann}}, \binits{N.L.}},
\bauthor{\bsnm{{Schulze}}, \binits{S.}},
\bauthor{\bsnm{{Taggart}}, \binits{K.}},
\bauthor{\bsnm{{Tzanidakis}}, \binits{A.}},
\bauthor{\bsnm{{Walters}}, \binits{R.S.}},
\bauthor{\bsnm{{Yan}}, \binits{L.}}:
\batitle{{The Zwicky Transient Facility Bright Transient Survey. II. A Public
  Statistical Sample for Exploring Supernova Demographics}}.
\bjtitle{\apj}
\bvolume{904}(\bissue{1}),
\bfpage{35}
(\byear{2020})
{\href{https://arxiv.org/abs/2009.01242}{{arXiv:2009.01242}}}
{[astro-ph.HE]}.
\doiurl{10.3847/1538-4357/abbd98}
\end{barticle}
\endbibitem

\bibitem{nicholl22}
\begin{botherref}
\oauthor{\bsnm{{Nicholl}}, \binits{M.}},
\oauthor{\bsnm{{Lanning}}, \binits{D.}},
\oauthor{\bsnm{{Ramsden}}, \binits{P.}},
\oauthor{\bsnm{{Mockler}}, \binits{B.}},
\oauthor{\bsnm{{Lawrence}}, \binits{A.}},
\oauthor{\bsnm{{Short}}, \binits{P.}},
\oauthor{\bsnm{{Ridley}}, \binits{E.J.}}:
{Systematic light curve modelling of TDEs: statistical differences between the
  spectroscopic classes}.
arXiv e-prints,
2201--02649
(2022)
{\href{https://arxiv.org/abs/2201.02649}{{arXiv:2201.02649}}}
{[astro-ph.HE]}
\end{botherref}
\endbibitem

\bibitem{hammerstein22}
\begin{botherref}
\oauthor{\bsnm{{Hammerstein}}, \binits{E.}},
\oauthor{\bsnm{{van Velzen}}, \binits{S.}},
\oauthor{\bsnm{{Gezari}}, \binits{S.}},
\oauthor{\bsnm{{Cenko}}, \binits{S.B.}},
\oauthor{\bsnm{{Yao}}, \binits{Y.}},
\oauthor{\bsnm{{Ward}}, \binits{C.}},
\oauthor{\bsnm{{Frederick}}, \binits{S.}},
\oauthor{\bsnm{{Villanueva}}, \binits{N.}},
\oauthor{\bsnm{{Somalwar}}, \binits{J.J.}},
\oauthor{\bsnm{{Graham}}, \binits{M.J.}},
\oauthor{\bsnm{{Kulkarni}}, \binits{S.R.}},
\oauthor{\bsnm{{Stern}}, \binits{D.}},
\oauthor{\bsnm{{Bellm}}, \binits{E.C.}},
\oauthor{\bsnm{{Dekany}}, \binits{R.}},
\oauthor{\bsnm{{Drake}}, \binits{A.J.}},
\oauthor{\bsnm{{Groom}}, \binits{S.L.}},
\oauthor{\bsnm{{Kasliwal}}, \binits{M.M.}},
\oauthor{\bsnm{{Kool}}, \binits{E.C.}},
\oauthor{\bsnm{{Masci}}, \binits{F.J.}},
\oauthor{\bsnm{{Medford}}, \binits{M.S.}},
\oauthor{\bsnm{{van Roestel}}, \binits{J.}}:
{The Final Season Reimagined: 30 Tidal Disruption Events from the ZTF-I
  Survey}.
arXiv e-prints,
2203--01461
(2022)
{\href{https://arxiv.org/abs/2203.01461}{{arXiv:2203.01461}}}
{[astro-ph.HE]}
\end{botherref}
\endbibitem

\bibitem{villar+17}
\begin{barticle}
\bauthor{\bsnm{{Villar}}, \binits{V.A.}},
\bauthor{\bsnm{{Guillochon}}, \binits{J.}},
\bauthor{\bsnm{{Berger}}, \binits{E.}},
\bauthor{\bsnm{{Metzger}}, \binits{B.D.}},
\bauthor{\bsnm{{Cowperthwaite}}, \binits{P.S.}},
\bauthor{\bsnm{{Nicholl}}, \binits{M.}},
\bauthor{\bsnm{{Alexand er}}, \binits{K.D.}},
\bauthor{\bsnm{{Blanchard}}, \binits{P.K.}},
\bauthor{\bsnm{{Chornock}}, \binits{R.}},
\bauthor{\bsnm{{Eftekhari}}, \binits{T.}},
\bauthor{\bsnm{{Fong}}, \binits{W.}},
\bauthor{\bsnm{{Margutti}}, \binits{R.}},
\bauthor{\bsnm{{Williams}}, \binits{P.K.G.}}:
\batitle{{The Combined Ultraviolet, Optical, and Near-infrared Light Curves of
  the Kilonova Associated with the Binary Neutron Star Merger GW170817: Unified
  Data Set, Analytic Models, and Physical Implications}}.
\bjtitle{ApJL}
\bvolume{851}(\bissue{1}),
\bfpage{21}
(\byear{2017})
{\href{https://arxiv.org/abs/1710.11576}{{arXiv:1710.11576}}}
{[astro-ph.HE]}.
\doiurl{10.3847/2041-8213/aa9c84}
\end{barticle}
\endbibitem

\bibitem{kormendy13}
\begin{barticle}
\bauthor{\bsnm{{Kormendy}}, \binits{J.}},
\bauthor{\bsnm{{Ho}}, \binits{L.C.}}:
\batitle{{Coevolution (Or Not) of Supermassive Black Holes and Host Galaxies}}.
\bjtitle{\araa}
\bvolume{51}(\bissue{1}),
\bfpage{511}--\blpage{653}
(\byear{2013})
{\href{https://arxiv.org/abs/1304.7762}{{arXiv:1304.7762}}}
{[astro-ph.CO]}.
\doiurl{10.1146/annurev-astro-082708-101811}
\end{barticle}
\endbibitem

\bibitem{fong22}
\begin{botherref}
\oauthor{\bsnm{{Fong}}, \binits{W.-f.}},
\oauthor{\bsnm{{Nugent}}, \binits{A.E.}},
\oauthor{\bsnm{{Dong}}, \binits{Y.}},
\oauthor{\bsnm{{Berger}}, \binits{E.}},
\oauthor{\bsnm{{Paterson}}, \binits{K.}},
\oauthor{\bsnm{{Chornock}}, \binits{R.}},
\oauthor{\bsnm{{Levan}}, \binits{A.}},
\oauthor{\bsnm{{Blanchard}}, \binits{P.}},
\oauthor{\bsnm{{Alexander}}, \binits{K.D.}},
\oauthor{\bsnm{{Andrews}}, \binits{J.}},
\oauthor{\bsnm{{Cobb}}, \binits{B.E.}},
\oauthor{\bsnm{{Cucchiara}}, \binits{A.}},
\oauthor{\bsnm{{Fox}}, \binits{D.}},
\oauthor{\bsnm{{Fryer}}, \binits{C.L.}},
\oauthor{\bsnm{{Gordon}}, \binits{A.C.}},
\oauthor{\bsnm{{Kilpatrick}}, \binits{C.D.}},
\oauthor{\bsnm{{Lunnan}}, \binits{R.}},
\oauthor{\bsnm{{Margutti}}, \binits{R.}},
\oauthor{\bsnm{{Miller}}, \binits{A.}},
\oauthor{\bsnm{{Milne}}, \binits{P.}},
\oauthor{\bsnm{{Nicholl}}, \binits{M.}},
\oauthor{\bsnm{{Perley}}, \binits{D.}},
\oauthor{\bsnm{{Rastinejad}}, \binits{J.}},
\oauthor{\bsnm{{Rouco Escorial}}, \binits{A.}},
\oauthor{\bsnm{{Schroeder}}, \binits{G.}},
\oauthor{\bsnm{{Smith}}, \binits{N.}},
\oauthor{\bsnm{{Tanvir}}, \binits{N.}},
\oauthor{\bsnm{{Terreran}}, \binits{G.}}:
{Short GRB Host Galaxies I: Photometric and Spectroscopic Catalogs, Host
  Associations, and Galactocentric Offsets}.
arXiv e-prints,
2206--01763
(2022)
{\href{https://arxiv.org/abs/2206.01763}{{arXiv:2206.01763}}}
{[astro-ph.GA]}
\end{botherref}
\endbibitem

\bibitem{ye20}
\begin{barticle}
\bauthor{\bsnm{{Ye}}, \binits{C.S.}},
\bauthor{\bsnm{{Fong}}, \binits{W.-f.}},
\bauthor{\bsnm{{Kremer}}, \binits{K.}},
\bauthor{\bsnm{{Rodriguez}}, \binits{C.L.}},
\bauthor{\bsnm{{Chatterjee}}, \binits{S.}},
\bauthor{\bsnm{{Fragione}}, \binits{G.}},
\bauthor{\bsnm{{Rasio}}, \binits{F.A.}}:
\batitle{{On the Rate of Neutron Star Binary Mergers from Globular Clusters}}.
\bjtitle{\apjl}
\bvolume{888}(\bissue{1}),
\bfpage{10}
(\byear{2020})
{\href{https://arxiv.org/abs/1910.10740}{{arXiv:1910.10740}}}
{[astro-ph.HE]}.
\doiurl{10.3847/2041-8213/ab5dc5}
\end{barticle}
\endbibitem

\bibitem{antonini16}
\begin{barticle}
\bauthor{\bsnm{{Antonini}}, \binits{F.}},
\bauthor{\bsnm{{Rasio}}, \binits{F.A.}}:
\batitle{{Merging Black Hole Binaries in Galactic Nuclei: Implications for
  Advanced-LIGO Detections}}.
\bjtitle{\apj}
\bvolume{831}(\bissue{2}),
\bfpage{187}
(\byear{2016})
{\href{https://arxiv.org/abs/1606.04889}{{arXiv:1606.04889}}}
{[astro-ph.HE]}.
\doiurl{10.3847/0004-637X/831/2/187}
\end{barticle}
\endbibitem

\bibitem{tagawa20}
\begin{barticle}
\bauthor{\bsnm{{Tagawa}}, \binits{H.}},
\bauthor{\bsnm{{Haiman}}, \binits{Z.}},
\bauthor{\bsnm{{Kocsis}}, \binits{B.}}:
\batitle{{Formation and Evolution of Compact-object Binaries in AGN Disks}}.
\bjtitle{\apj}
\bvolume{898}(\bissue{1}),
\bfpage{25}
(\byear{2020})
{\href{https://arxiv.org/abs/1912.08218}{{arXiv:1912.08218}}}
{[astro-ph.GA]}.
\doiurl{10.3847/1538-4357/ab9b8c}
\end{barticle}
\endbibitem

\bibitem{French17}
\begin{barticle}
\bauthor{\bsnm{{French}}, \binits{K.D.}},
\bauthor{\bsnm{{Arcavi}}, \binits{I.}},
\bauthor{\bsnm{{Zabludoff}}, \binits{A.}}:
\batitle{{The Post-starburst Evolution of Tidal Disruption Event Host
  Galaxies}}.
\bjtitle{\apj}
\bvolume{835}(\bissue{2}),
\bfpage{176}
(\byear{2017})
{\href{https://arxiv.org/abs/1609.04755}{{arXiv:1609.04755}}}
{[astro-ph.GA]}.
\doiurl{10.3847/1538-4357/835/2/176}
\end{barticle}
\endbibitem

\bibitem{gompertz22}
\begin{botherref}
\oauthor{\bsnm{{Gompertz}}, \binits{B.P.}},
\oauthor{\bsnm{{Ravasio}}, \binits{M.E.}},
\oauthor{\bsnm{{Nicholl}}, \binits{M.}},
\oauthor{\bsnm{{Levan}}, \binits{A.J.}},
\oauthor{\bsnm{{Metzger}}, \binits{B.D.}},
\oauthor{\bsnm{{Oates}}, \binits{S.R.}},
\oauthor{\bsnm{{Lamb}}, \binits{G.P.}},
\oauthor{\bsnm{{Fong}}, \binits{W.}},
\oauthor{\bsnm{{Malesani}}, \binits{D.B.}},
\oauthor{\bsnm{{Rastinejad}}, \binits{J.C.}},
\oauthor{\bsnm{{Tanvir}}, \binits{N.R.}},
\oauthor{\bsnm{{Evans}}, \binits{P.A.}},
\oauthor{\bsnm{{Jonker}}, \binits{P.G.}},
\oauthor{\bsnm{{Page}}, \binits{K.L.}},
\oauthor{\bsnm{{Pe'er}}, \binits{A.}}:
{A minute-long merger-driven gamma-ray burst from fast-cooling synchrotron
  emission}.
arXiv e-prints,
2205--05008
(2022)
{\href{https://arxiv.org/abs/2205.05008}{{arXiv:2205.05008}}}
{[astro-ph.HE]}
\end{botherref}
\endbibitem

\bibitem{perley09}
\begin{barticle}
\bauthor{\bsnm{{Perley}}, \binits{D.A.}},
\bauthor{\bsnm{{Metzger}}, \binits{B.D.}},
\bauthor{\bsnm{{Granot}}, \binits{J.}},
\bauthor{\bsnm{{Butler}}, \binits{N.R.}},
\bauthor{\bsnm{{Sakamoto}}, \binits{T.}},
\bauthor{\bsnm{{Ramirez-Ruiz}}, \binits{E.}},
\bauthor{\bsnm{{Levan}}, \binits{A.J.}},
\bauthor{\bsnm{{Bloom}}, \binits{J.S.}},
\bauthor{\bsnm{{Miller}}, \binits{A.A.}},
\bauthor{\bsnm{{Bunker}}, \binits{A.}},
\bauthor{\bsnm{{Chen}}, \binits{H.-W.}},
\bauthor{\bsnm{{Filippenko}}, \binits{A.V.}},
\bauthor{\bsnm{{Gehrels}}, \binits{N.}},
\bauthor{\bsnm{{Glazebrook}}, \binits{K.}},
\bauthor{\bsnm{{Hall}}, \binits{P.B.}},
\bauthor{\bsnm{{Hurley}}, \binits{K.C.}},
\bauthor{\bsnm{{Kocevski}}, \binits{D.}},
\bauthor{\bsnm{{Li}}, \binits{W.}},
\bauthor{\bsnm{{Lopez}}, \binits{S.}},
\bauthor{\bsnm{{Norris}}, \binits{J.}},
\bauthor{\bsnm{{Piro}}, \binits{A.L.}},
\bauthor{\bsnm{{Poznanski}}, \binits{D.}},
\bauthor{\bsnm{{Prochaska}}, \binits{J.X.}},
\bauthor{\bsnm{{Quataert}}, \binits{E.}},
\bauthor{\bsnm{{Tanvir}}, \binits{N.}}:
\batitle{{GRB 080503: Implications of a Naked Short Gamma-Ray Burst Dominated
  by Extended Emission}}.
\bjtitle{\apj}
\bvolume{696}(\bissue{2}),
\bfpage{1871}--\blpage{1885}
(\byear{2009})
{\href{https://arxiv.org/abs/0811.1044}{{arXiv:0811.1044}}}
{[astro-ph]}.
\doiurl{10.1088/0004-637X/696/2/1871}
\end{barticle}
\endbibitem

\bibitem{Bucciantini12}
\begin{barticle}
\bauthor{\bsnm{{Bucciantini}}, \binits{N.}},
\bauthor{\bsnm{{Metzger}}, \binits{B.D.}},
\bauthor{\bsnm{{Thompson}}, \binits{T.A.}},
\bauthor{\bsnm{{Quataert}}, \binits{E.}}:
\batitle{{Short gamma-ray bursts with extended emission from magnetar birth:
  jet formation and collimation}}.
\bjtitle{\mnras}
\bvolume{419}(\bissue{2}),
\bfpage{1537}--\blpage{1545}
(\byear{2012})
{\href{https://arxiv.org/abs/1106.4668}{{arXiv:1106.4668}}}
{[astro-ph.HE]}.
\doiurl{10.1111/j.1365-2966.2011.19810.x}
\end{barticle}
\endbibitem

\bibitem{lu22}
\begin{botherref}
\oauthor{\bsnm{{Lu}}, \binits{W.}},
\oauthor{\bsnm{{Quataert}}, \binits{E.}}:
{Late-time accretion in neutron star mergers: implications for short gamma-ray
  bursts and kilonovae}.
arXiv e-prints,
2208--04293
(2022)
{\href{https://arxiv.org/abs/2208.04293}{{arXiv:2208.04293}}}
{[astro-ph.HE]}
\end{botherref}
\endbibitem

\bibitem{king07}
\begin{barticle}
\bauthor{\bsnm{{King}}, \binits{A.}},
\bauthor{\bsnm{{Olsson}}, \binits{E.}},
\bauthor{\bsnm{{Davies}}, \binits{M.B.}}:
\batitle{{A new type of long gamma-ray burst}}.
\bjtitle{\mnras}
\bvolume{374}(\bissue{1}),
\bfpage{34}--\blpage{36}
(\byear{2007})
{\href{https://arxiv.org/abs/astro-ph/0610452}{{arXiv:astro-ph/0610452}}}
{[astro-ph]}.
\doiurl{10.1111/j.1745-3933.2006.00259.x}
\end{barticle}
\endbibitem

\bibitem{mcbrien19}
\begin{barticle}
\bauthor{\bsnm{{McBrien}}, \binits{O.R.}},
\bauthor{\bsnm{{Smartt}}, \binits{S.J.}},
\bauthor{\bsnm{{Chen}}, \binits{T.-W.}},
\bauthor{\bsnm{{Inserra}}, \binits{C.}},
\bauthor{\bsnm{{Gillanders}}, \binits{J.H.}},
\bauthor{\bsnm{{Sim}}, \binits{S.A.}},
\bauthor{\bsnm{{Jerkstrand}}, \binits{A.}},
\bauthor{\bsnm{{Rest}}, \binits{A.}},
\bauthor{\bsnm{{Valenti}}, \binits{S.}},
\bauthor{\bsnm{{Roy}}, \binits{R.}},
\bauthor{\bsnm{{Gromadzki}}, \binits{M.}},
\bauthor{\bsnm{{Taubenberger}}, \binits{S.}},
\bauthor{\bsnm{{Fl{\"o}rs}}, \binits{A.}},
\bauthor{\bsnm{{Huber}}, \binits{M.E.}},
\bauthor{\bsnm{{Chambers}}, \binits{K.C.}},
\bauthor{\bsnm{{Gal-Yam}}, \binits{A.}},
\bauthor{\bsnm{{Young}}, \binits{D.R.}},
\bauthor{\bsnm{{Nicholl}}, \binits{M.}},
\bauthor{\bsnm{{Kankare}}, \binits{E.}},
\bauthor{\bsnm{{Smith}}, \binits{K.W.}},
\bauthor{\bsnm{{Maguire}}, \binits{K.}},
\bauthor{\bsnm{{Mandel}}, \binits{I.}},
\bauthor{\bsnm{{Prentice}}, \binits{S.}},
\bauthor{\bsnm{{Rodr{\'\i}guez}}, \binits{{\'O}.}},
\bauthor{\bsnm{{Pineda Garcia}}, \binits{J.}},
\bauthor{\bsnm{{Guti{\'e}rrez}}, \binits{C.P.}},
\bauthor{\bsnm{{Galbany}}, \binits{L.}},
\bauthor{\bsnm{{Barbarino}}, \binits{C.}},
\bauthor{\bsnm{{Clark}}, \binits{P.S.J.}},
\bauthor{\bsnm{{Sollerman}}, \binits{J.}},
\bauthor{\bsnm{{Kulkarni}}, \binits{S.R.}},
\bauthor{\bsnm{{De}}, \binits{K.}},
\bauthor{\bsnm{{Buckley}}, \binits{D.A.H.}},
\bauthor{\bsnm{{Rau}}, \binits{A.}}:
\batitle{{SN2018kzr: A Rapidly Declining Transient from the Destruction of a
  White Dwarf}}.
\bjtitle{\apjl}
\bvolume{885}(\bissue{1}),
\bfpage{23}
(\byear{2019})
{\href{https://arxiv.org/abs/1909.04545}{{arXiv:1909.04545}}}
{[astro-ph.HE]}.
\doiurl{10.3847/2041-8213/ab4dae}
\end{barticle}
\endbibitem

\bibitem{Gillanders20}
\begin{barticle}
\bauthor{\bsnm{{Gillanders}}, \binits{J.H.}},
\bauthor{\bsnm{{Sim}}, \binits{S.A.}},
\bauthor{\bsnm{{Smartt}}, \binits{S.J.}}:
\batitle{{AT2018kzr: the merger of an oxygen-neon white dwarf and a neutron
  star or black hole}}.
\bjtitle{\mnras}
\bvolume{497}(\bissue{1}),
\bfpage{246}--\blpage{262}
(\byear{2020})
{\href{https://arxiv.org/abs/2007.12110}{{arXiv:2007.12110}}}
{[astro-ph.HE]}.
\doiurl{10.1093/mnras/staa1822}
\end{barticle}
\endbibitem

\bibitem{Fynbo+06}
\begin{barticle}
\bauthor{\bsnm{{Fynbo}}, \binits{J.P.U.}},
\bauthor{\bsnm{{Watson}}, \binits{D.}},
\bauthor{\bsnm{{Th{\"o}ne}}, \binits{C.C.}},
\bauthor{\bsnm{{Sollerman}}, \binits{J.}},
\bauthor{\bsnm{{Bloom}}, \binits{J.S.}},
\bauthor{\bsnm{{Davis}}, \binits{T.M.}},
\bauthor{\bsnm{{Hjorth}}, \binits{J.}},
\bauthor{\bsnm{{Jakobsson}}, \binits{P.}},
\bauthor{\bsnm{{J{\o}rgensen}}, \binits{U.G.}},
\bauthor{\bsnm{{Graham}}, \binits{J.F.}},
\bauthor{\bsnm{{Fruchter}}, \binits{A.S.}},
\bauthor{\bsnm{{Bersier}}, \binits{D.}},
\bauthor{\bsnm{{Kewley}}, \binits{L.}},
\bauthor{\bsnm{{Cassan}}, \binits{A.}},
\bauthor{\bsnm{{Castro Cer{\'o}n}}, \binits{J.M.}},
\bauthor{\bsnm{{Foley}}, \binits{S.}},
\bauthor{\bsnm{{Gorosabel}}, \binits{J.}},
\bauthor{\bsnm{{Hinse}}, \binits{T.C.}},
\bauthor{\bsnm{{Horne}}, \binits{K.D.}},
\bauthor{\bsnm{{Jensen}}, \binits{B.L.}},
\bauthor{\bsnm{{Klose}}, \binits{S.}},
\bauthor{\bsnm{{Kocevski}}, \binits{D.}},
\bauthor{\bsnm{{Marquette}}, \binits{J.-B.}},
\bauthor{\bsnm{{Perley}}, \binits{D.}},
\bauthor{\bsnm{{Ramirez-Ruiz}}, \binits{E.}},
\bauthor{\bsnm{{Stritzinger}}, \binits{M.D.}},
\bauthor{\bsnm{{Vreeswijk}}, \binits{P.M.}},
\bauthor{\bsnm{{Wijers}}, \binits{R.A.M.}},
\bauthor{\bsnm{{Woller}}, \binits{K.G.}},
\bauthor{\bsnm{{Xu}}, \binits{D.}},
\bauthor{\bsnm{{Zub}}, \binits{M.}}:
\batitle{{No supernovae associated with two long-duration
  {\ensuremath{\gamma}}-ray bursts}}.
\bjtitle{\nat}
\bvolume{444}(\bissue{7122}),
\bfpage{1047}--\blpage{1049}
(\byear{2006})
{\href{https://arxiv.org/abs/astro-ph/0608313}{{arXiv:astro-ph/0608313}}}
{[astro-ph]}.
\doiurl{10.1038/nature05375}
\end{barticle}
\endbibitem

\bibitem{GalYam+06}
\begin{barticle}
\bauthor{\bsnm{{Gal-Yam}}, \binits{A.}},
\bauthor{\bsnm{{Fox}}, \binits{D.B.}},
\bauthor{\bsnm{{Price}}, \binits{P.A.}},
\bauthor{\bsnm{{Ofek}}, \binits{E.O.}},
\bauthor{\bsnm{{Davis}}, \binits{M.R.}},
\bauthor{\bsnm{{Leonard}}, \binits{D.C.}},
\bauthor{\bsnm{{Soderberg}}, \binits{A.M.}},
\bauthor{\bsnm{{Schmidt}}, \binits{B.P.}},
\bauthor{\bsnm{{Lewis}}, \binits{K.M.}},
\bauthor{\bsnm{{Peterson}}, \binits{B.A.}},
\bauthor{\bsnm{{Kulkarni}}, \binits{S.R.}},
\bauthor{\bsnm{{Berger}}, \binits{E.}},
\bauthor{\bsnm{{Cenko}}, \binits{S.B.}},
\bauthor{\bsnm{{Sari}}, \binits{R.}},
\bauthor{\bsnm{{Sharon}}, \binits{K.}},
\bauthor{\bsnm{{Frail}}, \binits{D.}},
\bauthor{\bsnm{{Moon}}, \binits{D.-S.}},
\bauthor{\bsnm{{Brown}}, \binits{P.J.}},
\bauthor{\bsnm{{Cucchiara}}, \binits{A.}},
\bauthor{\bsnm{{Harrison}}, \binits{F.}},
\bauthor{\bsnm{{Piran}}, \binits{T.}},
\bauthor{\bsnm{{Persson}}, \binits{S.E.}},
\bauthor{\bsnm{{McCarthy}}, \binits{P.J.}},
\bauthor{\bsnm{{Penprase}}, \binits{B.E.}},
\bauthor{\bsnm{{Chevalier}}, \binits{R.A.}},
\bauthor{\bsnm{{MacFadyen}}, \binits{A.I.}}:
\batitle{{A novel explosive process is required for the
  {\ensuremath{\gamma}}-ray burst GRB 060614}}.
\bjtitle{\nat}
\bvolume{444}(\bissue{7122}),
\bfpage{1053}--\blpage{1055}
(\byear{2006})
{\href{https://arxiv.org/abs/astro-ph/0608257}{{arXiv:astro-ph/0608257}}}
{[astro-ph]}.
\doiurl{10.1038/nature05373}
\end{barticle}
\endbibitem

\bibitem{Gehrels+06}
\begin{barticle}
\bauthor{\bsnm{{Gehrels}}, \binits{N.}},
\bauthor{\bsnm{{Norris}}, \binits{J.P.}},
\bauthor{\bsnm{{Barthelmy}}, \binits{S.D.}},
\bauthor{\bsnm{{Granot}}, \binits{J.}},
\bauthor{\bsnm{{Kaneko}}, \binits{Y.}},
\bauthor{\bsnm{{Kouveliotou}}, \binits{C.}},
\bauthor{\bsnm{{Markwardt}}, \binits{C.B.}},
\bauthor{\bsnm{{M{\'e}sz{\'a}ros}}, \binits{P.}},
\bauthor{\bsnm{{Nakar}}, \binits{E.}},
\bauthor{\bsnm{{Nousek}}, \binits{J.A.}},
\bauthor{\bsnm{{O'Brien}}, \binits{P.T.}},
\bauthor{\bsnm{{Page}}, \binits{M.}},
\bauthor{\bsnm{{Palmer}}, \binits{D.M.}},
\bauthor{\bsnm{{Parsons}}, \binits{A.M.}},
\bauthor{\bsnm{{Roming}}, \binits{P.W.A.}},
\bauthor{\bsnm{{Sakamoto}}, \binits{T.}},
\bauthor{\bsnm{{Sarazin}}, \binits{C.L.}},
\bauthor{\bsnm{{Schady}}, \binits{P.}},
\bauthor{\bsnm{{Stamatikos}}, \binits{M.}},
\bauthor{\bsnm{{Woosley}}, \binits{S.E.}}:
\batitle{{A new {\ensuremath{\gamma}}-ray burst classification scheme from
  GRB060614}}.
\bjtitle{\nat}
\bvolume{444}(\bissue{7122}),
\bfpage{1044}--\blpage{1046}
(\byear{2006})
{\href{https://arxiv.org/abs/astro-ph/0610635}{{arXiv:astro-ph/0610635}}}
{[astro-ph]}.
\doiurl{10.1038/nature05376}
\end{barticle}
\endbibitem

\bibitem{Michalowski+18}
\begin{barticle}
\bauthor{\bsnm{{Micha{\l}owskI}}, \binits{M.J.}},
\bauthor{\bsnm{{Xu}}, \binits{D.}},
\bauthor{\bsnm{{Stevens}}, \binits{J.}},
\bauthor{\bsnm{{Levan}}, \binits{A.}},
\bauthor{\bsnm{{Yang}}, \binits{J.}},
\bauthor{\bsnm{{Paragi}}, \binits{Z.}},
\bauthor{\bsnm{{Kamble}}, \binits{A.}},
\bauthor{\bsnm{{Tsai}}, \binits{A.-L.}},
\bauthor{\bsnm{{Dannerbauer}}, \binits{H.}},
\bauthor{\bsnm{{van der Horst}}, \binits{A.J.}},
\bauthor{\bsnm{{Shao}}, \binits{L.}},
\bauthor{\bsnm{{Crosby}}, \binits{D.}},
\bauthor{\bsnm{{Gentile}}, \binits{G.}},
\bauthor{\bsnm{{Stanway}}, \binits{E.}},
\bauthor{\bsnm{{Wiersema}}, \binits{K.}},
\bauthor{\bsnm{{Fynbo}}, \binits{J.P.U.}},
\bauthor{\bsnm{{Tanvir}}, \binits{N.R.}},
\bauthor{\bsnm{{Kamphuis}}, \binits{P.}},
\bauthor{\bsnm{{Garrett}}, \binits{M.}},
\bauthor{\bsnm{{Bartczak}}, \binits{P.}}:
\batitle{{The second-closest gamma-ray burst: sub-luminous GRB 111005A with no
  supernova in a super-solar metallicity environment}}.
\bjtitle{\aap}
\bvolume{616},
\bfpage{169}
(\byear{2018})
{\href{https://arxiv.org/abs/1610.06928}{{arXiv:1610.06928}}}
{[astro-ph.HE]}.
\doiurl{10.1051/0004-6361/201629942}
\end{barticle}
\endbibitem

\bibitem{le22}
\begin{barticle}
\bauthor{\bsnm{{Le{\'s}niewska}}, \binits{A.}},
\bauthor{\bsnm{{Micha{\l}owski}}, \binits{M.J.}},
\bauthor{\bsnm{{Kamphuis}}, \binits{P.}},
\bauthor{\bsnm{{Dziadura}}, \binits{K.}},
\bauthor{\bsnm{{Baes}}, \binits{M.}},
\bauthor{\bsnm{{Cer{\'o}n}}, \binits{J.M.C.}},
\bauthor{\bsnm{{Gentile}}, \binits{G.}},
\bauthor{\bsnm{{Hjorth}}, \binits{J.}},
\bauthor{\bsnm{{Hunt}}, \binits{L.K.}},
\bauthor{\bsnm{{Jespersen}}, \binits{C.K.}},
\bauthor{\bsnm{{Koprowski}}, \binits{M.P.}},
\bauthor{\bsnm{{Floc'h}}, \binits{E.L.}},
\bauthor{\bsnm{{Miraghaei}}, \binits{H.}},
\bauthor{\bsnm{{Guelbenzu}}, \binits{A.N.}},
\bauthor{\bsnm{{Oszkiewicz}}, \binits{D.}},
\bauthor{\bsnm{{Palazzi}}, \binits{E.}},
\bauthor{\bsnm{{Poli{\'n}ska}}, \binits{M.}},
\bauthor{\bsnm{{Rasmussen}}, \binits{J.}},
\bauthor{\bsnm{{Schady}}, \binits{P.}},
\bauthor{\bsnm{{Watson}}, \binits{D.}}:
\batitle{{The Interstellar Medium in the Environment of the Supernova-less
  Long-duration GRB 111005A}}.
\bjtitle{\apjs}
\bvolume{259}(\bissue{2}),
\bfpage{67}
(\byear{2022})
{\href{https://arxiv.org/abs/2202.01188}{{arXiv:2202.01188}}}
{[astro-ph.GA]}.
\doiurl{10.3847/1538-4365/ac5022}
\end{barticle}
\endbibitem

\bibitem{rossi14}
\begin{barticle}
\bauthor{\bsnm{{Rossi}}, \binits{A.}},
\bauthor{\bsnm{{Piranomonte}}, \binits{S.}},
\bauthor{\bsnm{{Savaglio}}, \binits{S.}},
\bauthor{\bsnm{{Palazzi}}, \binits{E.}},
\bauthor{\bsnm{{Micha{\l}owski}}, \binits{M.J.}},
\bauthor{\bsnm{{Klose}}, \binits{S.}},
\bauthor{\bsnm{{Hunt}}, \binits{L.K.}},
\bauthor{\bsnm{{Amati}}, \binits{L.}},
\bauthor{\bsnm{{Elliott}}, \binits{J.}},
\bauthor{\bsnm{{Greiner}}, \binits{J.}},
\bauthor{\bsnm{{Guidorzi}}, \binits{C.}},
\bauthor{\bsnm{{Japelj}}, \binits{J.}},
\bauthor{\bsnm{{Kann}}, \binits{D.A.}},
\bauthor{\bsnm{{Lo Faro}}, \binits{B.}},
\bauthor{\bsnm{{Nicuesa Guelbenzu}}, \binits{A.}},
\bauthor{\bsnm{{Schulze}}, \binits{S.}},
\bauthor{\bsnm{{Vergani}}, \binits{S.D.}},
\bauthor{\bsnm{{Arnold}}, \binits{L.A.}},
\bauthor{\bsnm{{Covino}}, \binits{S.}},
\bauthor{\bsnm{{D'Elia}}, \binits{V.}},
\bauthor{\bsnm{{Ferrero}}, \binits{P.}},
\bauthor{\bsnm{{Filgas}}, \binits{R.}},
\bauthor{\bsnm{{Goldoni}}, \binits{P.}},
\bauthor{\bsnm{{K{\"u}pc{\"u} Yolda{\c{s}}}}, \binits{A.}},
\bauthor{\bsnm{{Le Borgne}}, \binits{D.}},
\bauthor{\bsnm{{Pian}}, \binits{E.}},
\bauthor{\bsnm{{Schady}}, \binits{P.}},
\bauthor{\bsnm{{Stratta}}, \binits{G.}}:
\batitle{{A quiescent galaxy at the position of the long GRB 050219A}}.
\bjtitle{\aap}
\bvolume{572},
\bfpage{47}
(\byear{2014})
{\href{https://arxiv.org/abs/1409.0017}{{arXiv:1409.0017}}}
{[astro-ph.HE]}.
\doiurl{10.1051/0004-6361/201423865}
\end{barticle}
\endbibitem

\bibitem{fruchter06}
\begin{barticle}
\bauthor{\bsnm{{Fruchter}}, \binits{A.S.}},
\bauthor{\bsnm{{Levan}}, \binits{A.J.}},
\bauthor{\bsnm{{Strolger}}, \binits{L.}},
\bauthor{\bsnm{{Vreeswijk}}, \binits{P.M.}},
\bauthor{\bsnm{{Thorsett}}, \binits{S.E.}},
\bauthor{\bsnm{{Bersier}}, \binits{D.}},
\bauthor{\bsnm{{Burud}}, \binits{I.}},
\bauthor{\bsnm{{Castro Cer{\'o}n}}, \binits{J.M.}},
\bauthor{\bsnm{{Castro-Tirado}}, \binits{A.J.}},
\bauthor{\bsnm{{Conselice}}, \binits{C.}},
\bauthor{\bsnm{{Dahlen}}, \binits{T.}},
\bauthor{\bsnm{{Ferguson}}, \binits{H.C.}},
\bauthor{\bsnm{{Fynbo}}, \binits{J.P.U.}},
\bauthor{\bsnm{{Garnavich}}, \binits{P.M.}},
\bauthor{\bsnm{{Gibbons}}, \binits{R.A.}},
\bauthor{\bsnm{{Gorosabel}}, \binits{J.}},
\bauthor{\bsnm{{Gull}}, \binits{T.R.}},
\bauthor{\bsnm{{Hjorth}}, \binits{J.}},
\bauthor{\bsnm{{Holland}}, \binits{S.T.}},
\bauthor{\bsnm{{Kouveliotou}}, \binits{C.}},
\bauthor{\bsnm{{Levay}}, \binits{Z.}},
\bauthor{\bsnm{{Livio}}, \binits{M.}},
\bauthor{\bsnm{{Metzger}}, \binits{M.R.}},
\bauthor{\bsnm{{Nugent}}, \binits{P.E.}},
\bauthor{\bsnm{{Petro}}, \binits{L.}},
\bauthor{\bsnm{{Pian}}, \binits{E.}},
\bauthor{\bsnm{{Rhoads}}, \binits{J.E.}},
\bauthor{\bsnm{{Riess}}, \binits{A.G.}},
\bauthor{\bsnm{{Sahu}}, \binits{K.C.}},
\bauthor{\bsnm{{Smette}}, \binits{A.}},
\bauthor{\bsnm{{Tanvir}}, \binits{N.R.}},
\bauthor{\bsnm{{Wijers}}, \binits{R.A.M.J.}},
\bauthor{\bsnm{{Woosley}}, \binits{S.E.}}:
\batitle{{Long {\ensuremath{\gamma}}-ray bursts and core-collapse supernovae
  have different environments}}.
\bjtitle{\nat}
\bvolume{441}(\bissue{7092}),
\bfpage{463}--\blpage{468}
(\byear{2006})
{\href{https://arxiv.org/abs/astro-ph/0603537}{{arXiv:astro-ph/0603537}}}
{[astro-ph]}.
\doiurl{10.1038/nature04787}
\end{barticle}
\endbibitem

\bibitem{fragione22}
\begin{barticle}
\bauthor{\bsnm{{Fragione}}, \binits{G.}},
\bauthor{\bsnm{{Kocsis}}, \binits{B.}},
\bauthor{\bsnm{{Rasio}}, \binits{F.A.}},
\bauthor{\bsnm{{Silk}}, \binits{J.}}:
\batitle{{Repeated Mergers, Mass-gap Black Holes, and Formation of
  Intermediate-mass Black Holes in Dense Massive Star Clusters}}.
\bjtitle{\apj}
\bvolume{927}(\bissue{2}),
\bfpage{231}
(\byear{2022})
{\href{https://arxiv.org/abs/2107.04639}{{arXiv:2107.04639}}}
{[astro-ph.GA]}.
\doiurl{10.3847/1538-4357/ac5026}
\end{barticle}
\endbibitem

\bibitem{Lien+16}
\begin{barticle}
\bauthor{\bsnm{{Lien}}, \binits{A.}},
\bauthor{\bsnm{{Sakamoto}}, \binits{T.}},
\bauthor{\bsnm{{Barthelmy}}, \binits{S.D.}},
\bauthor{\bsnm{{Baumgartner}}, \binits{W.H.}},
\bauthor{\bsnm{{Cannizzo}}, \binits{J.K.}},
\bauthor{\bsnm{{Chen}}, \binits{K.}},
\bauthor{\bsnm{{Collins}}, \binits{N.R.}},
\bauthor{\bsnm{{Cummings}}, \binits{J.R.}},
\bauthor{\bsnm{{Gehrels}}, \binits{N.}},
\bauthor{\bsnm{{Krimm}}, \binits{H.A.}},
\bauthor{\bsnm{{Markwardt}}, \binits{C.B.}},
\bauthor{\bsnm{{Palmer}}, \binits{D.M.}},
\bauthor{\bsnm{{Stamatikos}}, \binits{M.}},
\bauthor{\bsnm{{Troja}}, \binits{E.}},
\bauthor{\bsnm{{Ukwatta}}, \binits{T.N.}}:
\batitle{{The Third Swift Burst Alert Telescope Gamma-Ray Burst Catalog}}.
\bjtitle{\apj}
\bvolume{829}(\bissue{1}),
\bfpage{7}
(\byear{2016})
{\href{https://arxiv.org/abs/1606.01956}{{arXiv:1606.01956}}}
{[astro-ph.HE]}.
\doiurl{10.3847/0004-637X/829/1/7}
\end{barticle}
\endbibitem

\bibitem{clocchiatti11}
\begin{barticle}
\bauthor{\bsnm{{Clocchiatti}}, \binits{A.}},
\bauthor{\bsnm{{Suntzeff}}, \binits{N.B.}},
\bauthor{\bsnm{{Covarrubias}}, \binits{R.}},
\bauthor{\bsnm{{Candia}}, \binits{P.}}:
\batitle{{The Ultimate Light Curve of SN 1998bw/GRB 980425}}.
\bjtitle{\aj}
\bvolume{141}(\bissue{5}),
\bfpage{163}
(\byear{2011})
{\href{https://arxiv.org/abs/1106.1695}{{arXiv:1106.1695}}}
{[astro-ph.HE]}.
\doiurl{10.1088/0004-6256/141/5/163}
\end{barticle}
\endbibitem

\bibitem{blagorodnova17}
\begin{barticle}
\bauthor{\bsnm{{Blagorodnova}}, \binits{N.}},
\bauthor{\bsnm{{Gezari}}, \binits{S.}},
\bauthor{\bsnm{{Hung}}, \binits{T.}},
\bauthor{\bsnm{{Kulkarni}}, \binits{S.R.}},
\bauthor{\bsnm{{Cenko}}, \binits{S.B.}},
\bauthor{\bsnm{{Pasham}}, \binits{D.R.}},
\bauthor{\bsnm{{Yan}}, \binits{L.}},
\bauthor{\bsnm{{Arcavi}}, \binits{I.}},
\bauthor{\bsnm{{Ben-Ami}}, \binits{S.}},
\bauthor{\bsnm{{Bue}}, \binits{B.D.}},
\bauthor{\bsnm{{Cantwell}}, \binits{T.}},
\bauthor{\bsnm{{Cao}}, \binits{Y.}},
\bauthor{\bsnm{{Castro-Tirado}}, \binits{A.J.}},
\bauthor{\bsnm{{Fender}}, \binits{R.}},
\bauthor{\bsnm{{Fremling}}, \binits{C.}},
\bauthor{\bsnm{{Gal-Yam}}, \binits{A.}},
\bauthor{\bsnm{{Ho}}, \binits{A.Y.Q.}},
\bauthor{\bsnm{{Horesh}}, \binits{A.}},
\bauthor{\bsnm{{Hosseinzadeh}}, \binits{G.}},
\bauthor{\bsnm{{Kasliwal}}, \binits{M.M.}},
\bauthor{\bsnm{{Kong}}, \binits{A.K.H.}},
\bauthor{\bsnm{{Laher}}, \binits{R.R.}},
\bauthor{\bsnm{{Leloudas}}, \binits{G.}},
\bauthor{\bsnm{{Lunnan}}, \binits{R.}},
\bauthor{\bsnm{{Masci}}, \binits{F.J.}},
\bauthor{\bsnm{{Mooley}}, \binits{K.}},
\bauthor{\bsnm{{Neill}}, \binits{J.D.}},
\bauthor{\bsnm{{Nugent}}, \binits{P.}},
\bauthor{\bsnm{{Powell}}, \binits{M.}},
\bauthor{\bsnm{{Valeev}}, \binits{A.F.}},
\bauthor{\bsnm{{Vreeswijk}}, \binits{P.M.}},
\bauthor{\bsnm{{Walters}}, \binits{R.}},
\bauthor{\bsnm{{Wozniak}}, \binits{P.}}:
\batitle{{iPTF16fnl: A Faint and Fast Tidal Disruption Event in an E+A
  Galaxy}}.
\bjtitle{\apj}
\bvolume{844}(\bissue{1}),
\bfpage{46}
(\byear{2017})
{\href{https://arxiv.org/abs/1703.00965}{{arXiv:1703.00965}}}
{[astro-ph.HE]}.
\doiurl{10.3847/1538-4357/aa7579}
\end{barticle}
\endbibitem

\bibitem{Evans07}
\begin{barticle}
\bauthor{\bsnm{{Evans}}, \binits{P.A.}},
\bauthor{\bsnm{{Beardmore}}, \binits{A.P.}},
\bauthor{\bsnm{{Page}}, \binits{K.L.}},
\bauthor{\bsnm{{Tyler}}, \binits{L.G.}},
\bauthor{\bsnm{{Osborne}}, \binits{J.P.}},
\bauthor{\bsnm{{Goad}}, \binits{M.R.}},
\bauthor{\bsnm{{O'Brien}}, \binits{P.T.}},
\bauthor{\bsnm{{Vetere}}, \binits{L.}},
\bauthor{\bsnm{{Racusin}}, \binits{J.}},
\bauthor{\bsnm{{Morris}}, \binits{D.}},
\bauthor{\bsnm{{Burrows}}, \binits{D.N.}},
\bauthor{\bsnm{{Capalbi}}, \binits{M.}},
\bauthor{\bsnm{{Perri}}, \binits{M.}},
\bauthor{\bsnm{{Gehrels}}, \binits{N.}},
\bauthor{\bsnm{{Romano}}, \binits{P.}}:
\batitle{{An online repository of Swift/XRT light curves of
  {\ensuremath{\gamma}}-ray bursts}}.
\bjtitle{\aap}
\bvolume{469}(\bissue{1}),
\bfpage{379}--\blpage{385}
(\byear{2007})
{\href{https://arxiv.org/abs/0704.0128}{{arXiv:0704.0128}}}
{[astro-ph]}.
\doiurl{10.1051/0004-6361:20077530}
\end{barticle}
\endbibitem

\bibitem{Evans09}
\begin{barticle}
\bauthor{\bsnm{{Evans}}, \binits{P.A.}},
\bauthor{\bsnm{{Beardmore}}, \binits{A.P.}},
\bauthor{\bsnm{{Page}}, \binits{K.L.}},
\bauthor{\bsnm{{Osborne}}, \binits{J.P.}},
\bauthor{\bsnm{{O'Brien}}, \binits{P.T.}},
\bauthor{\bsnm{{Willingale}}, \binits{R.}},
\bauthor{\bsnm{{Starling}}, \binits{R.L.C.}},
\bauthor{\bsnm{{Burrows}}, \binits{D.N.}},
\bauthor{\bsnm{{Godet}}, \binits{O.}},
\bauthor{\bsnm{{Vetere}}, \binits{L.}},
\bauthor{\bsnm{{Racusin}}, \binits{J.}},
\bauthor{\bsnm{{Goad}}, \binits{M.R.}},
\bauthor{\bsnm{{Wiersema}}, \binits{K.}},
\bauthor{\bsnm{{Angelini}}, \binits{L.}},
\bauthor{\bsnm{{Capalbi}}, \binits{M.}},
\bauthor{\bsnm{{Chincarini}}, \binits{G.}},
\bauthor{\bsnm{{Gehrels}}, \binits{N.}},
\bauthor{\bsnm{{Kennea}}, \binits{J.A.}},
\bauthor{\bsnm{{Margutti}}, \binits{R.}},
\bauthor{\bsnm{{Morris}}, \binits{D.C.}},
\bauthor{\bsnm{{Mountford}}, \binits{C.J.}},
\bauthor{\bsnm{{Pagani}}, \binits{C.}},
\bauthor{\bsnm{{Perri}}, \binits{M.}},
\bauthor{\bsnm{{Romano}}, \binits{P.}},
\bauthor{\bsnm{{Tanvir}}, \binits{N.}}:
\batitle{{Methods and results of an automatic analysis of a complete sample of
  Swift-XRT observations of GRBs}}.
\bjtitle{\mnras}
\bvolume{397}(\bissue{3}),
\bfpage{1177}--\blpage{1201}
(\byear{2009})
{\href{https://arxiv.org/abs/0812.3662}{{arXiv:0812.3662}}}
{[astro-ph]}.
\doiurl{10.1111/j.1365-2966.2009.14913.x}
\end{barticle}
\endbibitem

\bibitem{HEAsoft}
\begin{botherref}
\oauthor{\bsnm{{Nasa High Energy Astrophysics Science Archive Research Center
  (Heasarc)}}}:
{HEAsoft: Unified Release of FTOOLS and XANADU}
(2014)
\end{botherref}
\endbibitem

\bibitem{SciPy}
\begin{barticle}
\bauthor{\bsnm{Virtanen}, \binits{P.}},
\bauthor{\bsnm{Gommers}, \binits{R.}},
\bauthor{\bsnm{Oliphant}, \binits{T.E.}},
\bauthor{\bsnm{Haberland}, \binits{M.}},
\bauthor{\bsnm{Reddy}, \binits{T.}},
\bauthor{\bsnm{Cournapeau}, \binits{D.}},
\bauthor{\bsnm{Burovski}, \binits{E.}},
\bauthor{\bsnm{Peterson}, \binits{P.}},
\bauthor{\bsnm{Weckesser}, \binits{W.}},
\bauthor{\bsnm{Bright}, \binits{J.}},
\bauthor{\bsnm{{van der Walt}}, \binits{S.J.}},
\bauthor{\bsnm{Brett}, \binits{M.}},
\bauthor{\bsnm{Wilson}, \binits{J.}},
\bauthor{\bsnm{Millman}, \binits{K.J.}},
\bauthor{\bsnm{Mayorov}, \binits{N.}},
\bauthor{\bsnm{Nelson}, \binits{A.R.J.}},
\bauthor{\bsnm{Jones}, \binits{E.}},
\bauthor{\bsnm{Kern}, \binits{R.}},
\bauthor{\bsnm{Larson}, \binits{E.}},
\bauthor{\bsnm{Carey}, \binits{C.J.}},
\bauthor{\bsnm{Polat}, \binits{{\. I}.}},
\bauthor{\bsnm{Feng}, \binits{Y.}},
\bauthor{\bsnm{Moore}, \binits{E.W.}},
\bauthor{\bsnm{{VanderPlas}}, \binits{J.}},
\bauthor{\bsnm{Laxalde}, \binits{D.}},
\bauthor{\bsnm{Perktold}, \binits{J.}},
\bauthor{\bsnm{Cimrman}, \binits{R.}},
\bauthor{\bsnm{Henriksen}, \binits{I.}},
\bauthor{\bsnm{Quintero}, \binits{E.A.}},
\bauthor{\bsnm{Harris}, \binits{C.R.}},
\bauthor{\bsnm{Archibald}, \binits{A.M.}},
\bauthor{\bsnm{Ribeiro}, \binits{A.H.}},
\bauthor{\bsnm{Pedregosa}, \binits{F.}},
\bauthor{\bsnm{{van Mulbregt}}, \binits{P.}},
\bauthor{\bsnm{{SciPy 1.0 Contributors}}}:
\batitle{{{SciPy} 1.0: Fundamental Algorithms for Scientific Computing in
  Python}}.
\bjtitle{Nature Methods}
\bvolume{17},
\bfpage{261}--\blpage{272}
(\byear{2020}).
\doiurl{10.1038/s41592-019-0686-2}
\end{barticle}
\endbibitem

\bibitem{Tonry79}
\begin{barticle}
\bauthor{\bsnm{{Tonry}}, \binits{J.}},
\bauthor{\bsnm{{Davis}}, \binits{M.}}:
\batitle{{A survey of galaxy redshifts. I. Data reduction techniques.}}
\bjtitle{\aj}
\bvolume{84},
\bfpage{1511}--\blpage{1525}
(\byear{1979}).
\doiurl{10.1086/112569}
\end{barticle}
\endbibitem

\bibitem{Barthelmy05}
\begin{barticle}
\bauthor{\bsnm{{Barthelmy}}, \binits{S.D.}},
\bauthor{\bsnm{{Chincarini}}, \binits{G.}},
\bauthor{\bsnm{{Burrows}}, \binits{D.N.}},
\bauthor{\bsnm{{Gehrels}}, \binits{N.}},
\bauthor{\bsnm{{Covino}}, \binits{S.}},
\bauthor{\bsnm{{Moretti}}, \binits{A.}},
\bauthor{\bsnm{{Romano}}, \binits{P.}},
\bauthor{\bsnm{{O'Brien}}, \binits{P.T.}},
\bauthor{\bsnm{{Sarazin}}, \binits{C.L.}},
\bauthor{\bsnm{{Kouveliotou}}, \binits{C.}},
\bauthor{\bsnm{{Goad}}, \binits{M.}},
\bauthor{\bsnm{{Vaughan}}, \binits{S.}},
\bauthor{\bsnm{{Tagliaferri}}, \binits{G.}},
\bauthor{\bsnm{{Zhang}}, \binits{B.}},
\bauthor{\bsnm{{Antonelli}}, \binits{L.A.}},
\bauthor{\bsnm{{Campana}}, \binits{S.}},
\bauthor{\bsnm{{Cummings}}, \binits{J.R.}},
\bauthor{\bsnm{{D'Avanzo}}, \binits{P.}},
\bauthor{\bsnm{{Davies}}, \binits{M.B.}},
\bauthor{\bsnm{{Giommi}}, \binits{P.}},
\bauthor{\bsnm{{Grupe}}, \binits{D.}},
\bauthor{\bsnm{{Kaneko}}, \binits{Y.}},
\bauthor{\bsnm{{Kennea}}, \binits{J.A.}},
\bauthor{\bsnm{{King}}, \binits{A.}},
\bauthor{\bsnm{{Kobayashi}}, \binits{S.}},
\bauthor{\bsnm{{Melandri}}, \binits{A.}},
\bauthor{\bsnm{{Meszaros}}, \binits{P.}},
\bauthor{\bsnm{{Nousek}}, \binits{J.A.}},
\bauthor{\bsnm{{Patel}}, \binits{S.}},
\bauthor{\bsnm{{Sakamoto}}, \binits{T.}},
\bauthor{\bsnm{{Wijers}}, \binits{R.A.M.J.}}:
\batitle{{An origin for short {\ensuremath{\gamma}}-ray bursts unassociated
  with current star formation}}.
\bjtitle{\nat}
\bvolume{438}(\bissue{7070}),
\bfpage{994}--\blpage{996}
(\byear{2005})
{\href{https://arxiv.org/abs/astro-ph/0511579}{{arXiv:astro-ph/0511579}}}
{[astro-ph]}.
\doiurl{10.1038/nature04392}
\end{barticle}
\endbibitem

\bibitem{Wilms00}
\begin{barticle}
\bauthor{\bsnm{{Wilms}}, \binits{J.}},
\bauthor{\bsnm{{Allen}}, \binits{A.}},
\bauthor{\bsnm{{McCray}}, \binits{R.}}:
\batitle{{On the Absorption of X-Rays in the Interstellar Medium}}.
\bjtitle{\apj}
\bvolume{542}(\bissue{2}),
\bfpage{914}--\blpage{924}
(\byear{2000})
{\href{https://arxiv.org/abs/astro-ph/0008425}{{arXiv:astro-ph/0008425}}}
{[astro-ph]}.
\doiurl{10.1086/317016}
\end{barticle}
\endbibitem

\bibitem{Willingale13}
\begin{barticle}
\bauthor{\bsnm{{Willingale}}, \binits{R.}},
\bauthor{\bsnm{{Starling}}, \binits{R.L.C.}},
\bauthor{\bsnm{{Beardmore}}, \binits{A.P.}},
\bauthor{\bsnm{{Tanvir}}, \binits{N.R.}},
\bauthor{\bsnm{{O'Brien}}, \binits{P.T.}}:
\batitle{{Calibration of X-ray absorption in our Galaxy}}.
\bjtitle{\mnras}
\bvolume{431}(\bissue{1}),
\bfpage{394}--\blpage{404}
(\byear{2013})
{\href{https://arxiv.org/abs/1303.0843}{{arXiv:1303.0843}}}
{[astro-ph.HE]}.
\doiurl{10.1093/mnras/stt175}
\end{barticle}
\endbibitem

\bibitem{alard98}
\begin{barticle}
\bauthor{\bsnm{{Alard}}, \binits{C.}},
\bauthor{\bsnm{{Lupton}}, \binits{R.H.}}:
\batitle{{A Method for Optimal Image Subtraction}}.
\bjtitle{\apj}
\bvolume{503}(\bissue{1}),
\bfpage{325}--\blpage{331}
(\byear{1998})
{\href{https://arxiv.org/abs/astro-ph/9712287}{{arXiv:astro-ph/9712287}}}
{[astro-ph]}.
\doiurl{10.1086/305984}
\end{barticle}
\endbibitem

\bibitem{Bloom+02}
\begin{barticle}
\bauthor{\bsnm{{Bloom}}, \binits{J.S.}},
\bauthor{\bsnm{{Kulkarni}}, \binits{S.R.}},
\bauthor{\bsnm{{Djorgovski}}, \binits{S.G.}}:
\batitle{{The Observed Offset Distribution of Gamma-Ray Bursts from Their Host
  Galaxies: A Robust Clue to the Nature of the Progenitors}}.
\bjtitle{\aj}
\bvolume{123}(\bissue{3}),
\bfpage{1111}--\blpage{1148}
(\byear{2002})
{\href{https://arxiv.org/abs/astro-ph/0010176}{{arXiv:astro-ph/0010176}}}
{[astro-ph]}.
\doiurl{10.1086/338893}
\end{barticle}
\endbibitem

\bibitem{Evans+09}
\begin{barticle}
\bauthor{\bsnm{{Evans}}, \binits{P.A.}},
\bauthor{\bsnm{{Beardmore}}, \binits{A.P.}},
\bauthor{\bsnm{{Page}}, \binits{K.L.}},
\bauthor{\bsnm{{Osborne}}, \binits{J.P.}},
\bauthor{\bsnm{{O'Brien}}, \binits{P.T.}},
\bauthor{\bsnm{{Willingale}}, \binits{R.}},
\bauthor{\bsnm{{Starling}}, \binits{R.L.C.}},
\bauthor{\bsnm{{Burrows}}, \binits{D.N.}},
\bauthor{\bsnm{{Godet}}, \binits{O.}},
\bauthor{\bsnm{{Vetere}}, \binits{L.}},
\bauthor{\bsnm{{Racusin}}, \binits{J.}},
\bauthor{\bsnm{{Goad}}, \binits{M.R.}},
\bauthor{\bsnm{{Wiersema}}, \binits{K.}},
\bauthor{\bsnm{{Angelini}}, \binits{L.}},
\bauthor{\bsnm{{Capalbi}}, \binits{M.}},
\bauthor{\bsnm{{Chincarini}}, \binits{G.}},
\bauthor{\bsnm{{Gehrels}}, \binits{N.}},
\bauthor{\bsnm{{Kennea}}, \binits{J.A.}},
\bauthor{\bsnm{{Margutti}}, \binits{R.}},
\bauthor{\bsnm{{Morris}}, \binits{D.C.}},
\bauthor{\bsnm{{Mountford}}, \binits{C.J.}},
\bauthor{\bsnm{{Pagani}}, \binits{C.}},
\bauthor{\bsnm{{Perri}}, \binits{M.}},
\bauthor{\bsnm{{Romano}}, \binits{P.}},
\bauthor{\bsnm{{Tanvir}}, \binits{N.}}:
\batitle{{Methods and results of an automatic analysis of a complete sample of
  Swift-XRT observations of GRBs}}.
\bjtitle{\mnras}
\bvolume{397}(\bissue{3}),
\bfpage{1177}--\blpage{1201}
(\byear{2009})
{\href{https://arxiv.org/abs/0812.3662}{{arXiv:0812.3662}}}
{[astro-ph]}.
\doiurl{10.1111/j.1365-2966.2009.14913.x}
\end{barticle}
\endbibitem

\bibitem{Schulze+14}
\begin{barticle}
\bauthor{\bsnm{{Schulze}}, \binits{S.}},
\bauthor{\bsnm{{Malesani}}, \binits{D.}},
\bauthor{\bsnm{{Cucchiara}}, \binits{A.}},
\bauthor{\bsnm{{Tanvir}}, \binits{N.R.}},
\bauthor{\bsnm{{Kr{\"u}hler}}, \binits{T.}},
\bauthor{\bsnm{{de Ugarte Postigo}}, \binits{A.}},
\bauthor{\bsnm{{Leloudas}}, \binits{G.}},
\bauthor{\bsnm{{Lyman}}, \binits{J.}},
\bauthor{\bsnm{{Bersier}}, \binits{D.}},
\bauthor{\bsnm{{Wiersema}}, \binits{K.}},
\bauthor{\bsnm{{Perley}}, \binits{D.A.}},
\bauthor{\bsnm{{Schady}}, \binits{P.}},
\bauthor{\bsnm{{Gorosabel}}, \binits{J.}},
\bauthor{\bsnm{{Anderson}}, \binits{J.P.}},
\bauthor{\bsnm{{Castro-Tirado}}, \binits{A.J.}},
\bauthor{\bsnm{{Cenko}}, \binits{S.B.}},
\bauthor{\bsnm{{De Cia}}, \binits{A.}},
\bauthor{\bsnm{{Ellerbroek}}, \binits{L.E.}},
\bauthor{\bsnm{{Fynbo}}, \binits{J.P.U.}},
\bauthor{\bsnm{{Greiner}}, \binits{J.}},
\bauthor{\bsnm{{Hjorth}}, \binits{J.}},
\bauthor{\bsnm{{Kann}}, \binits{D.A.}},
\bauthor{\bsnm{{Kaper}}, \binits{L.}},
\bauthor{\bsnm{{Klose}}, \binits{S.}},
\bauthor{\bsnm{{Levan}}, \binits{A.J.}},
\bauthor{\bsnm{{Mart{\'\i}n}}, \binits{S.}},
\bauthor{\bsnm{{O'Brien}}, \binits{P.T.}},
\bauthor{\bsnm{{Page}}, \binits{K.L.}},
\bauthor{\bsnm{{Pignata}}, \binits{G.}},
\bauthor{\bsnm{{Rapaport}}, \binits{S.}},
\bauthor{\bsnm{{S{\'a}nchez-Ram{\'\i}rez}}, \binits{R.}},
\bauthor{\bsnm{{Sollerman}}, \binits{J.}},
\bauthor{\bsnm{{Smith}}, \binits{I.A.}},
\bauthor{\bsnm{{Sparre}}, \binits{M.}},
\bauthor{\bsnm{{Th{\"o}ne}}, \binits{C.C.}},
\bauthor{\bsnm{{Watson}}, \binits{D.J.}},
\bauthor{\bsnm{{Xu}}, \binits{D.}},
\bauthor{\bsnm{{Bauer}}, \binits{F.E.}},
\bauthor{\bsnm{{Bayliss}}, \binits{M.}},
\bauthor{\bsnm{{Bj{\"o}rnsson}}, \binits{G.}},
\bauthor{\bsnm{{Bremer}}, \binits{M.}},
\bauthor{\bsnm{{Cano}}, \binits{Z.}},
\bauthor{\bsnm{{Covino}}, \binits{S.}},
\bauthor{\bsnm{{D'Elia}}, \binits{V.}},
\bauthor{\bsnm{{Frail}}, \binits{D.A.}},
\bauthor{\bsnm{{Geier}}, \binits{S.}},
\bauthor{\bsnm{{Goldoni}}, \binits{P.}},
\bauthor{\bsnm{{Hartoog}}, \binits{O.E.}},
\bauthor{\bsnm{{Jakobsson}}, \binits{P.}},
\bauthor{\bsnm{{Korhonen}}, \binits{H.}},
\bauthor{\bsnm{{Lee}}, \binits{K.Y.}},
\bauthor{\bsnm{{Milvang-Jensen}}, \binits{B.}},
\bauthor{\bsnm{{Nardini}}, \binits{M.}},
\bauthor{\bsnm{{Nicuesa Guelbenzu}}, \binits{A.}},
\bauthor{\bsnm{{Oguri}}, \binits{M.}},
\bauthor{\bsnm{{Pandey}}, \binits{S.B.}},
\bauthor{\bsnm{{Petitpas}}, \binits{G.}},
\bauthor{\bsnm{{Rossi}}, \binits{A.}},
\bauthor{\bsnm{{Sandberg}}, \binits{A.}},
\bauthor{\bsnm{{Schmidl}}, \binits{S.}},
\bauthor{\bsnm{{Tagliaferri}}, \binits{G.}},
\bauthor{\bsnm{{Tilanus}}, \binits{R.P.J.}},
\bauthor{\bsnm{{Winters}}, \binits{J.M.}},
\bauthor{\bsnm{{Wright}}, \binits{D.}},
\bauthor{\bsnm{{Wuyts}}, \binits{E.}}:
\batitle{{GRB 120422A/SN 2012bz: Bridging the gap between low- and
  high-luminosity gamma-ray bursts}}.
\bjtitle{\aap}
\bvolume{566},
\bfpage{102}
(\byear{2014})
{\href{https://arxiv.org/abs/1401.3774}{{arXiv:1401.3774}}}
{[astro-ph.HE]}.
\doiurl{10.1051/0004-6361/201423387}
\end{barticle}
\endbibitem

\bibitem{Gompertz+20}
\begin{barticle}
\bauthor{\bsnm{{Gompertz}}, \binits{B.P.}},
\bauthor{\bsnm{{Levan}}, \binits{A.J.}},
\bauthor{\bsnm{{Tanvir}}, \binits{N.R.}}:
\batitle{{A Search for Neutron Star-Black Hole Binary Mergers in the Short
  Gamma-Ray Burst Population}}.
\bjtitle{\apj}
\bvolume{895}(\bissue{1}),
\bfpage{58}
(\byear{2020})
{\href{https://arxiv.org/abs/2001.08706}{{arXiv:2001.08706}}}
{[astro-ph.HE]}.
\doiurl{10.3847/1538-4357/ab8d24}
\end{barticle}
\endbibitem

\bibitem{DAi19}
\begin{barticle}
\bauthor{\bsnm{{D'Ai}}, \binits{A.}},
\bauthor{\bsnm{{Melandri}}, \binits{A.}},
\bauthor{\bsnm{{D'Avanzo}}, \binits{P.}},
\bauthor{\bsnm{{Gropp}}, \binits{J.D.}},
\bauthor{\bsnm{{Tohuvavohu}}, \binits{A.}},
\bauthor{\bsnm{{Kennea}}, \binits{J.A.}},
\bauthor{\bsnm{{Evans}}, \binits{P.A.}},
\bauthor{\bsnm{{Osborne}}, \binits{J.P.}},
\bauthor{\bsnm{{Page}}, \binits{K.L.}},
\bauthor{\bsnm{{Simpson}}, \binits{K.K.}},
\bauthor{\bsnm{{Swift-XRT Team}}}:
\batitle{{GRB 191019A: Swift-XRT refined Analysis}}.
\bjtitle{GRB Coordinates Network}
\bvolume{26045},
\bfpage{1}
(\byear{2019})
\end{barticle}
\endbibitem

\bibitem{conselice03}
\begin{barticle}
\bauthor{\bsnm{{Conselice}}, \binits{C.J.}}:
\batitle{{The Relationship between Stellar Light Distributions of Galaxies and
  Their Formation Histories}}.
\bjtitle{\apjs}
\bvolume{147}(\bissue{1}),
\bfpage{1}--\blpage{28}
(\byear{2003})
{\href{https://arxiv.org/abs/astro-ph/0303065}{{arXiv:astro-ph/0303065}}}
{[astro-ph]}.
\doiurl{10.1086/375001}
\end{barticle}
\endbibitem

\bibitem{Leja+17}
\begin{barticle}
\bauthor{\bsnm{{Leja}}, \binits{J.}},
\bauthor{\bsnm{{Johnson}}, \binits{B.D.}},
\bauthor{\bsnm{{Conroy}}, \binits{C.}},
\bauthor{\bsnm{{van Dokkum}}, \binits{P.G.}},
\bauthor{\bsnm{{Byler}}, \binits{N.}}:
\batitle{{Deriving Physical Properties from Broadband Photometry with
  Prospector: Description of the Model and a Demonstration of its Accuracy
  Using 129 Galaxies in the Local Universe}}.
\bjtitle{\apj}
\bvolume{837}(\bissue{2}),
\bfpage{170}
(\byear{2017})
{\href{https://arxiv.org/abs/1609.09073}{{arXiv:1609.09073}}}
{[astro-ph.GA]}.
\doiurl{10.3847/1538-4357/aa5ffe}
\end{barticle}
\endbibitem

\bibitem{jlc+2021}
\begin{barticle}
\bauthor{\bsnm{{Johnson}}, \binits{B.D.}},
\bauthor{\bsnm{{Leja}}, \binits{J.}},
\bauthor{\bsnm{{Conroy}}, \binits{C.}},
\bauthor{\bsnm{{Speagle}}, \binits{J.S.}}:
\batitle{{Stellar Population Inference with Prospector}}.
\bjtitle{\apjs}
\bvolume{254}(\bissue{2}),
\bfpage{22}
(\byear{2021})
{\href{https://arxiv.org/abs/2012.01426}{{arXiv:2012.01426}}}
{[astro-ph.GA]}.
\doiurl{10.3847/1538-4365/abef67}
\end{barticle}
\endbibitem

\bibitem{Dynesty}
\begin{barticle}
\bauthor{\bsnm{{Speagle}}, \binits{J.S.}}:
\batitle{{dynesty: A Dynamic Nested Sampling Package for Estimating Bayesian
  Posteriors and Evidences}}.
\bjtitle{\mnras}
(\byear{2020})
{\href{https://arxiv.org/abs/1904.02180}{{arXiv:1904.02180}}}
{[astro-ph.IM]}.
\doiurl{10.1093/mnras/staa278}
\end{barticle}
\endbibitem

\bibitem{FSPS_2009}
\begin{barticle}
\bauthor{\bsnm{{Conroy}}, \binits{C.}},
\bauthor{\bsnm{{Gunn}}, \binits{J.E.}},
\bauthor{\bsnm{{White}}, \binits{M.}}:
\batitle{{The Propagation of Uncertainties in Stellar Population Synthesis
  Modeling. I. The Relevance of Uncertain Aspects of Stellar Evolution and the
  Initial Mass Function to the Derived Physical Properties of Galaxies}}.
\bjtitle{\apj}
\bvolume{699}(\bissue{1}),
\bfpage{486}--\blpage{506}
(\byear{2009})
{\href{https://arxiv.org/abs/0809.4261}{{arXiv:0809.4261}}}
{[astro-ph]}.
\doiurl{10.1088/0004-637X/699/1/486}
\end{barticle}
\endbibitem

\bibitem{FSPS_2010}
\begin{barticle}
\bauthor{\bsnm{{Conroy}}, \binits{C.}},
\bauthor{\bsnm{{Gunn}}, \binits{J.E.}}:
\batitle{{The Propagation of Uncertainties in Stellar Population Synthesis
  Modeling. III. Model Calibration, Comparison, and Evaluation}}.
\bjtitle{\apj}
\bvolume{712}(\bissue{2}),
\bfpage{833}--\blpage{857}
(\byear{2010})
{\href{https://arxiv.org/abs/0911.3151}{{arXiv:0911.3151}}}
{[astro-ph.CO]}.
\doiurl{10.1088/0004-637X/712/2/833}
\end{barticle}
\endbibitem

\bibitem{MilkyWay}
\begin{barticle}
\bauthor{\bsnm{{Cardelli}}, \binits{J.A.}},
\bauthor{\bsnm{{Clayton}}, \binits{G.C.}},
\bauthor{\bsnm{{Mathis}}, \binits{J.S.}}:
\batitle{{The Relationship between Infrared, Optical, and Ultraviolet
  Extinction}}.
\bjtitle{\apj}
\bvolume{345},
\bfpage{245}
(\byear{1989}).
\doiurl{10.1086/167900}
\end{barticle}
\endbibitem

\bibitem{Chabrier2003}
\begin{barticle}
\bauthor{\bsnm{{Chabrier}}, \binits{G.}}:
\batitle{{Galactic Stellar and Substellar Initial Mass Function}}.
\bjtitle{\pasp}
\bvolume{115}(\bissue{809}),
\bfpage{763}--\blpage{795}
(\byear{2003})
{\href{https://arxiv.org/abs/astro-ph/0304382}{{arXiv:astro-ph/0304382}}}
{[astro-ph]}.
\doiurl{10.1086/376392}
\end{barticle}
\endbibitem

\bibitem{gcb+05}
\begin{barticle}
\bauthor{\bsnm{{Gallazzi}}, \binits{A.}},
\bauthor{\bsnm{{Charlot}}, \binits{S.}},
\bauthor{\bsnm{{Brinchmann}}, \binits{J.}},
\bauthor{\bsnm{{White}}, \binits{S.D.M.}},
\bauthor{\bsnm{{Tremonti}}, \binits{C.A.}}:
\batitle{{The ages and metallicities of galaxies in the local universe}}.
\bjtitle{\mnras}
\bvolume{362}(\bissue{1}),
\bfpage{41}--\blpage{58}
(\byear{2005})
{\href{https://arxiv.org/abs/astro-ph/0506539}{{arXiv:astro-ph/0506539}}}
{[astro-ph]}.
\doiurl{10.1111/j.1365-2966.2005.09321.x}
\end{barticle}
\endbibitem

\bibitem{pkb+14}
\begin{barticle}
\bauthor{\bsnm{{Price}}, \binits{S.H.}},
\bauthor{\bsnm{{Kriek}}, \binits{M.}},
\bauthor{\bsnm{{Brammer}}, \binits{G.B.}},
\bauthor{\bsnm{{Conroy}}, \binits{C.}},
\bauthor{\bsnm{{F{\"o}rster Schreiber}}, \binits{N.M.}},
\bauthor{\bsnm{{Franx}}, \binits{M.}},
\bauthor{\bsnm{{Fumagalli}}, \binits{M.}},
\bauthor{\bsnm{{Lundgren}}, \binits{B.}},
\bauthor{\bsnm{{Momcheva}}, \binits{I.}},
\bauthor{\bsnm{{Nelson}}, \binits{E.J.}},
\bauthor{\bsnm{{Skelton}}, \binits{R.E.}},
\bauthor{\bsnm{{van Dokkum}}, \binits{P.G.}},
\bauthor{\bsnm{{Whitaker}}, \binits{K.E.}},
\bauthor{\bsnm{{Wuyts}}, \binits{S.}}:
\batitle{{Direct Measurements of Dust Attenuation in z
  \raisebox{-0.5ex}\textasciitilde 1.5 Star-forming Galaxies from 3D-HST:
  Implications for Dust Geometry and Star Formation Rates}}.
\bjtitle{\apj}
\bvolume{788}(\bissue{1}),
\bfpage{86}
(\byear{2014})
{\href{https://arxiv.org/abs/1310.4177}{{arXiv:1310.4177}}}
{[astro-ph.CO]}.
\doiurl{10.1088/0004-637X/788/1/86}
\end{barticle}
\endbibitem

\bibitem{Leja2019}
\begin{barticle}
\bauthor{\bsnm{{Leja}}, \binits{J.}},
\bauthor{\bsnm{{Johnson}}, \binits{B.D.}},
\bauthor{\bsnm{{Conroy}}, \binits{C.}},
\bauthor{\bsnm{{van Dokkum}}, \binits{P.}},
\bauthor{\bsnm{{Speagle}}, \binits{J.S.}},
\bauthor{\bsnm{{Brammer}}, \binits{G.}},
\bauthor{\bsnm{{Momcheva}}, \binits{I.}},
\bauthor{\bsnm{{Skelton}}, \binits{R.}},
\bauthor{\bsnm{{Whitaker}}, \binits{K.E.}},
\bauthor{\bsnm{{Franx}}, \binits{M.}},
\bauthor{\bsnm{{Nelson}}, \binits{E.J.}}:
\batitle{{An Older, More Quiescent Universe from Panchromatic SED Fitting of
  the 3D-HST Survey}}.
\bjtitle{\apj}
\bvolume{877}(\bissue{2}),
\bfpage{140}
(\byear{2019})
{\href{https://arxiv.org/abs/1812.05608}{{arXiv:1812.05608}}}
{[astro-ph.GA]}.
\doiurl{10.3847/1538-4357/ab1d5a}
\end{barticle}
\endbibitem

\bibitem{cab+20}
\begin{barticle}
\bauthor{\bsnm{{Calzetti}}, \binits{D.}},
\bauthor{\bsnm{{Armus}}, \binits{L.}},
\bauthor{\bsnm{{Bohlin}}, \binits{R.C.}},
\bauthor{\bsnm{{Kinney}}, \binits{A.L.}},
\bauthor{\bsnm{{Koornneef}}, \binits{J.}},
\bauthor{\bsnm{{Storchi-Bergmann}}, \binits{T.}}:
\batitle{{The Dust Content and Opacity of Actively Star-forming Galaxies}}.
\bjtitle{\apj}
\bvolume{533}(\bissue{2}),
\bfpage{682}--\blpage{695}
(\byear{2000})
{\href{https://arxiv.org/abs/astro-ph/9911459}{{arXiv:astro-ph/9911459}}}
{[astro-ph]}.
\doiurl{10.1086/308692}
\end{barticle}
\endbibitem

\bibitem{Cappellari2022}
\begin{barticle}
\bauthor{\bsnm{{Cappellari}}, \binits{M.}}:
\batitle{{Full spectrum fitting with photometry in ppxf: non-parametric star
  formation history, metallicity and the quenching boundary from 3200 LEGA-C
  galaxies at redshift $z\approx0.8$}}.
\bjtitle{MNRAS submitted}
(\byear{2022})
{\href{https://arxiv.org/abs/2208.14974}{{2208.14974}}}.
\doiurl{10.48550/arXiv.2208.14974}
\end{barticle}
\endbibitem

\bibitem{inkenhaag22}
\begin{barticle}
\bauthor{\bsnm{{Inkenhaag}}, \binits{A.}},
\bauthor{\bsnm{{Jonker}}, \binits{P.G.}},
\bauthor{\bsnm{{Cannizzaro}}, \binits{G.}},
\bauthor{\bsnm{{Mata S{\'a}nchez}}, \binits{D.}},
\bauthor{\bsnm{{Saxton}}, \binits{R.D.}}:
\batitle{{Host galaxy line diagnostics for the candidate tidal disruption
  events XMMSL1 J111527.3+180638 and PTF09axc}}.
\bjtitle{\mnras}
\bvolume{507}(\bissue{4}),
\bfpage{6196}--\blpage{6204}
(\byear{2021})
{\href{https://arxiv.org/abs/2109.01092}{{arXiv:2109.01092}}}
{[astro-ph.HE]}.
\doiurl{10.1093/mnras/stab2541}
\end{barticle}
\endbibitem

\bibitem{Perley2013}
\begin{barticle}
\bauthor{\bsnm{{Perley}}, \binits{D.A.}},
\bauthor{\bsnm{{Levan}}, \binits{A.J.}},
\bauthor{\bsnm{{Tanvir}}, \binits{N.R.}},
\bauthor{\bsnm{{Cenko}}, \binits{S.B.}},
\bauthor{\bsnm{{Bloom}}, \binits{J.S.}},
\bauthor{\bsnm{{Hjorth}}, \binits{J.}},
\bauthor{\bsnm{{Kr{\"u}hler}}, \binits{T.}},
\bauthor{\bsnm{{Filippenko}}, \binits{A.V.}},
\bauthor{\bsnm{{Fruchter}}, \binits{A.}},
\bauthor{\bsnm{{Fynbo}}, \binits{J.P.U.}},
\bauthor{\bsnm{{Jakobsson}}, \binits{P.}},
\bauthor{\bsnm{{Kalirai}}, \binits{J.}},
\bauthor{\bsnm{{Milvang-Jensen}}, \binits{B.}},
\bauthor{\bsnm{{Morgan}}, \binits{A.N.}},
\bauthor{\bsnm{{Prochaska}}, \binits{J.X.}},
\bauthor{\bsnm{{Silverman}}, \binits{J.M.}}:
\batitle{{A Population of Massive, Luminous Galaxies Hosting Heavily
  Dust-obscured Gamma-Ray Bursts: Implications for the Use of GRBs as Tracers
  of Cosmic Star Formation}}.
\bjtitle{\apj}
\bvolume{778}(\bissue{2}),
\bfpage{128}
(\byear{2013})
{\href{https://arxiv.org/abs/1301.5903}{{arXiv:1301.5903}}}
{[astro-ph.CO]}.
\doiurl{10.1088/0004-637X/778/2/128}
\end{barticle}
\endbibitem

\bibitem{nugent22}
\begin{botherref}
\oauthor{\bsnm{{Nugent}}, \binits{A.E.}},
\oauthor{\bsnm{{Fong}}, \binits{W.-f.}},
\oauthor{\bsnm{{Dong}}, \binits{Y.}},
\oauthor{\bsnm{{Leja}}, \binits{J.}},
\oauthor{\bsnm{{Berger}}, \binits{E.}},
\oauthor{\bsnm{{Zevin}}, \binits{M.}},
\oauthor{\bsnm{{Chornock}}, \binits{R.}},
\oauthor{\bsnm{{Cobb}}, \binits{B.E.}},
\oauthor{\bsnm{{Kelley}}, \binits{L.Z.}},
\oauthor{\bsnm{{Kilpatrick}}, \binits{C.D.}},
\oauthor{\bsnm{{Levan}}, \binits{A.}},
\oauthor{\bsnm{{Margutti}}, \binits{R.}},
\oauthor{\bsnm{{Paterson}}, \binits{K.}},
\oauthor{\bsnm{{Perley}}, \binits{D.}},
\oauthor{\bsnm{{Rouco Escorial}}, \binits{A.}},
\oauthor{\bsnm{{Smith}}, \binits{N.}},
\oauthor{\bsnm{{Tanvir}}, \binits{N.}}:
{Short GRB Host Galaxies II: A Legacy Sample of Redshifts, Stellar Population
  Properties, and Implications for their Neutron Star Merger Origins}.
arXiv e-prints,
2206--01764
(2022)
{\href{https://arxiv.org/abs/2206.01764}{{arXiv:2206.01764}}}
{[astro-ph.GA]}
\end{botherref}
\endbibitem

\bibitem{french20}
\begin{barticle}
\bauthor{\bsnm{{French}}, \binits{K.D.}},
\bauthor{\bsnm{{Arcavi}}, \binits{I.}},
\bauthor{\bsnm{{Zabludoff}}, \binits{A.I.}},
\bauthor{\bsnm{{Stone}}, \binits{N.}},
\bauthor{\bsnm{{Hiramatsu}}, \binits{D.}},
\bauthor{\bsnm{{van Velzen}}, \binits{S.}},
\bauthor{\bsnm{{McCully}}, \binits{C.}},
\bauthor{\bsnm{{Jiang}}, \binits{N.}}:
\batitle{{The Structure of Tidal Disruption Event Host Galaxies on Scales of
  Tens to Thousands of Parsecs}}.
\bjtitle{\apj}
\bvolume{891}(\bissue{1}),
\bfpage{93}
(\byear{2020})
{\href{https://arxiv.org/abs/2002.02498}{{arXiv:2002.02498}}}
{[astro-ph.HE]}.
\doiurl{10.3847/1538-4357/ab7450}
\end{barticle}
\endbibitem

\bibitem{DellaValle+06}
\begin{barticle}
\bauthor{\bsnm{{Della Valle}}, \binits{M.}},
\bauthor{\bsnm{{Chincarini}}, \binits{G.}},
\bauthor{\bsnm{{Panagia}}, \binits{N.}},
\bauthor{\bsnm{{Tagliaferri}}, \binits{G.}},
\bauthor{\bsnm{{Malesani}}, \binits{D.}},
\bauthor{\bsnm{{Testa}}, \binits{V.}},
\bauthor{\bsnm{{Fugazza}}, \binits{D.}},
\bauthor{\bsnm{{Campana}}, \binits{S.}},
\bauthor{\bsnm{{Covino}}, \binits{S.}},
\bauthor{\bsnm{{Mangano}}, \binits{V.}},
\bauthor{\bsnm{{Antonelli}}, \binits{L.A.}},
\bauthor{\bsnm{{D'Avanzo}}, \binits{P.}},
\bauthor{\bsnm{{Hurley}}, \binits{K.}},
\bauthor{\bsnm{{Mirabel}}, \binits{I.F.}},
\bauthor{\bsnm{{Pellizza}}, \binits{L.J.}},
\bauthor{\bsnm{{Piranomonte}}, \binits{S.}},
\bauthor{\bsnm{{Stella}}, \binits{L.}}:
\batitle{{An enigmatic long-lasting {\ensuremath{\gamma}}-ray burst not
  accompanied by a bright supernova}}.
\bjtitle{\nat}
\bvolume{444}(\bissue{7122}),
\bfpage{1050}--\blpage{1052}
(\byear{2006})
{\href{https://arxiv.org/abs/astro-ph/0608322}{{arXiv:astro-ph/0608322}}}
{[astro-ph]}.
\doiurl{10.1038/nature05374}
\end{barticle}
\endbibitem

\bibitem{pg04}
\begin{barticle}
\bauthor{\bsnm{{Portegies Zwart}}, \binits{S.F.}},
\bauthor{\bsnm{{Baumgardt}}, \binits{H.}},
\bauthor{\bsnm{{Hut}}, \binits{P.}},
\bauthor{\bsnm{{Makino}}, \binits{J.}},
\bauthor{\bsnm{{McMillan}}, \binits{S.L.W.}}:
\batitle{{Formation of massive black holes through runaway collisions in dense
  young star clusters}}.
\bjtitle{\nat}
\bvolume{428}(\bissue{6984}),
\bfpage{724}--\blpage{726}
(\byear{2004})
{\href{https://arxiv.org/abs/astro-ph/0402622}{{arXiv:astro-ph/0402622}}}
{[astro-ph]}.
\doiurl{10.1038/nature02448}
\end{barticle}
\endbibitem

\bibitem{bloom11}
\begin{barticle}
\bauthor{\bsnm{{Bloom}}, \binits{J.S.}},
\bauthor{\bsnm{{Giannios}}, \binits{D.}},
\bauthor{\bsnm{{Metzger}}, \binits{B.D.}},
\bauthor{\bsnm{{Cenko}}, \binits{S.B.}},
\bauthor{\bsnm{{Perley}}, \binits{D.A.}},
\bauthor{\bsnm{{Butler}}, \binits{N.R.}},
\bauthor{\bsnm{{Tanvir}}, \binits{N.R.}},
\bauthor{\bsnm{{Levan}}, \binits{A.J.}},
\bauthor{\bsnm{{O'Brien}}, \binits{P.T.}},
\bauthor{\bsnm{{Strubbe}}, \binits{L.E.}},
\bauthor{\bsnm{{De Colle}}, \binits{F.}},
\bauthor{\bsnm{{Ramirez-Ruiz}}, \binits{E.}},
\bauthor{\bsnm{{Lee}}, \binits{W.H.}},
\bauthor{\bsnm{{Nayakshin}}, \binits{S.}},
\bauthor{\bsnm{{Quataert}}, \binits{E.}},
\bauthor{\bsnm{{King}}, \binits{A.R.}},
\bauthor{\bsnm{{Cucchiara}}, \binits{A.}},
\bauthor{\bsnm{{Guillochon}}, \binits{J.}},
\bauthor{\bsnm{{Bower}}, \binits{G.C.}},
\bauthor{\bsnm{{Fruchter}}, \binits{A.S.}},
\bauthor{\bsnm{{Morgan}}, \binits{A.N.}},
\bauthor{\bsnm{{van der Horst}}, \binits{A.J.}}:
\batitle{{A Possible Relativistic Jetted Outburst from a Massive Black Hole Fed
  by a Tidally Disrupted Star}}.
\bjtitle{Science}
\bvolume{333},
\bfpage{203}
(\byear{2011})
{\href{https://arxiv.org/abs/1104.3257}{{arXiv:1104.3257}}}
{[astro-ph.HE]}.
\doiurl{10.1126/science.1207150}
\end{barticle}
\endbibitem

\bibitem{burrows11}
\begin{barticle}
\bauthor{\bsnm{{Burrows}}, \binits{D.N.}},
\bauthor{\bsnm{{Kennea}}, \binits{J.A.}},
\bauthor{\bsnm{{Ghisellini}}, \binits{G.}},
\bauthor{\bsnm{{Mangano}}, \binits{V.}},
\bauthor{\bsnm{{Zhang}}, \binits{B.}},
\bauthor{\bsnm{{Page}}, \binits{K.L.}},
\bauthor{\bsnm{{Eracleous}}, \binits{M.}},
\bauthor{\bsnm{{Romano}}, \binits{P.}},
\bauthor{\bsnm{{Sakamoto}}, \binits{T.}},
\bauthor{\bsnm{{Falcone}}, \binits{A.D.}},
\bauthor{\bsnm{{Osborne}}, \binits{J.P.}},
\bauthor{\bsnm{{Campana}}, \binits{S.}},
\bauthor{\bsnm{{Beardmore}}, \binits{A.P.}},
\bauthor{\bsnm{{Breeveld}}, \binits{A.A.}},
\bauthor{\bsnm{{Chester}}, \binits{M.M.}},
\bauthor{\bsnm{{Corbet}}, \binits{R.}},
\bauthor{\bsnm{{Covino}}, \binits{S.}},
\bauthor{\bsnm{{Cummings}}, \binits{J.R.}},
\bauthor{\bsnm{{D'Avanzo}}, \binits{P.}},
\bauthor{\bsnm{{D'Elia}}, \binits{V.}},
\bauthor{\bsnm{{Esposito}}, \binits{P.}},
\bauthor{\bsnm{{Evans}}, \binits{P.A.}},
\bauthor{\bsnm{{Fugazza}}, \binits{D.}},
\bauthor{\bsnm{{Gelbord}}, \binits{J.M.}},
\bauthor{\bsnm{{Hiroi}}, \binits{K.}},
\bauthor{\bsnm{{Holland}}, \binits{S.T.}},
\bauthor{\bsnm{{Huang}}, \binits{K.Y.}},
\bauthor{\bsnm{{Im}}, \binits{M.}},
\bauthor{\bsnm{{Israel}}, \binits{G.}},
\bauthor{\bsnm{{Jeon}}, \binits{Y.}},
\bauthor{\bsnm{{Jeon}}, \binits{Y.-B.}},
\bauthor{\bsnm{{Jun}}, \binits{H.D.}},
\bauthor{\bsnm{{Kawai}}, \binits{N.}},
\bauthor{\bsnm{{Kim}}, \binits{J.H.}},
\bauthor{\bsnm{{Krimm}}, \binits{H.A.}},
\bauthor{\bsnm{{Marshall}}, \binits{F.E.}},
\bauthor{\bsnm{{P.~M{\'e}sz{\'a}ros}}},
\bauthor{\bsnm{{Negoro}}, \binits{H.}},
\bauthor{\bsnm{{Omodei}}, \binits{N.}},
\bauthor{\bsnm{{Park}}, \binits{W.-K.}},
\bauthor{\bsnm{{Perkins}}, \binits{J.S.}},
\bauthor{\bsnm{{Sugizaki}}, \binits{M.}},
\bauthor{\bsnm{{Sung}}, \binits{H.-I.}},
\bauthor{\bsnm{{Tagliaferri}}, \binits{G.}},
\bauthor{\bsnm{{Troja}}, \binits{E.}},
\bauthor{\bsnm{{Ueda}}, \binits{Y.}},
\bauthor{\bsnm{{Urata}}, \binits{Y.}},
\bauthor{\bsnm{{Usui}}, \binits{R.}},
\bauthor{\bsnm{{Antonelli}}, \binits{L.A.}},
\bauthor{\bsnm{{Barthelmy}}, \binits{S.D.}},
\bauthor{\bsnm{{Cusumano}}, \binits{G.}},
\bauthor{\bsnm{{Giommi}}, \binits{P.}},
\bauthor{\bsnm{{Melandri}}, \binits{A.}},
\bauthor{\bsnm{{Perri}}, \binits{M.}},
\bauthor{\bsnm{{Racusin}}, \binits{J.L.}},
\bauthor{\bsnm{{Sbarufatti}}, \binits{B.}},
\bauthor{\bsnm{{Siegel}}, \binits{M.H.}},
\bauthor{\bsnm{{Gehrels}}, \binits{N.}}:
\batitle{{Relativistic jet activity from the tidal disruption of a star by a
  massive black hole}}.
\bjtitle{\nat}
\bvolume{476},
\bfpage{421}--\blpage{424}
(\byear{2011})
{\href{https://arxiv.org/abs/1104.4787}{{arXiv:1104.4787}}}
{[astro-ph.HE]}.
\doiurl{10.1038/nature10374}
\end{barticle}
\endbibitem

\bibitem{levan11}
\begin{barticle}
\bauthor{\bsnm{{Levan}}, \binits{A.J.}},
\bauthor{\bsnm{{Tanvir}}, \binits{N.R.}},
\bauthor{\bsnm{{Cenko}}, \binits{S.B.}},
\bauthor{\bsnm{{Perley}}, \binits{D.A.}},
\bauthor{\bsnm{{Wiersema}}, \binits{K.}},
\bauthor{\bsnm{{Bloom}}, \binits{J.S.}},
\bauthor{\bsnm{{Fruchter}}, \binits{A.S.}},
\bauthor{\bsnm{{Postigo}}, \binits{A.d.U.}},
\bauthor{\bsnm{{O'Brien}}, \binits{P.T.}},
\bauthor{\bsnm{{Butler}}, \binits{N.}},
\bauthor{\bsnm{{van der Horst}}, \binits{A.J.}},
\bauthor{\bsnm{{Leloudas}}, \binits{G.}},
\bauthor{\bsnm{{Morgan}}, \binits{A.N.}},
\bauthor{\bsnm{{Misra}}, \binits{K.}},
\bauthor{\bsnm{{Bower}}, \binits{G.C.}},
\bauthor{\bsnm{{Farihi}}, \binits{J.}},
\bauthor{\bsnm{{Tunnicliffe}}, \binits{R.L.}},
\bauthor{\bsnm{{Modjaz}}, \binits{M.}},
\bauthor{\bsnm{{Silverman}}, \binits{J.M.}},
\bauthor{\bsnm{{Hjorth}}, \binits{J.}},
\bauthor{\bsnm{{Th{\"o}ne}}, \binits{C.}},
\bauthor{\bsnm{{Cucchiara}}, \binits{A.}},
\bauthor{\bsnm{{Cer{\'o}n}}, \binits{J.M.C.}},
\bauthor{\bsnm{{Castro-Tirado}}, \binits{A.J.}},
\bauthor{\bsnm{{Arnold}}, \binits{J.A.}},
\bauthor{\bsnm{{Bremer}}, \binits{M.}},
\bauthor{\bsnm{{Brodie}}, \binits{J.P.}},
\bauthor{\bsnm{{Carroll}}, \binits{T.}},
\bauthor{\bsnm{{Cooper}}, \binits{M.C.}},
\bauthor{\bsnm{{Curran}}, \binits{P.A.}},
\bauthor{\bsnm{{Cutri}}, \binits{R.M.}},
\bauthor{\bsnm{{Ehle}}, \binits{J.}},
\bauthor{\bsnm{{Forbes}}, \binits{D.}},
\bauthor{\bsnm{{Fynbo}}, \binits{J.}},
\bauthor{\bsnm{{Gorosabel}}, \binits{J.}},
\bauthor{\bsnm{{Graham}}, \binits{J.}},
\bauthor{\bsnm{{Hoffman}}, \binits{D.I.}},
\bauthor{\bsnm{{Guziy}}, \binits{S.}},
\bauthor{\bsnm{{Jakobsson}}, \binits{P.}},
\bauthor{\bsnm{{Kamble}}, \binits{A.}},
\bauthor{\bsnm{{Kerr}}, \binits{T.}},
\bauthor{\bsnm{{Kasliwal}}, \binits{M.M.}},
\bauthor{\bsnm{{Kouveliotou}}, \binits{C.}},
\bauthor{\bsnm{{Kocevski}}, \binits{D.}},
\bauthor{\bsnm{{Law}}, \binits{N.M.}},
\bauthor{\bsnm{{Nugent}}, \binits{P.E.}},
\bauthor{\bsnm{{Ofek}}, \binits{E.O.}},
\bauthor{\bsnm{{Poznanski}}, \binits{D.}},
\bauthor{\bsnm{{Quimby}}, \binits{R.M.}},
\bauthor{\bsnm{{Rol}}, \binits{E.}},
\bauthor{\bsnm{{Romanowsky}}, \binits{A.J.}},
\bauthor{\bsnm{{S{\'a}nchez-Ram{\'{\i}}rez}}, \binits{R.}},
\bauthor{\bsnm{{Schulze}}, \binits{S.}},
\bauthor{\bsnm{{Singh}}, \binits{N.}},
\bauthor{\bsnm{{van Spaandonk}}, \binits{L.}},
\bauthor{\bsnm{{Starling}}, \binits{R.L.C.}},
\bauthor{\bsnm{{Strom}}, \binits{R.G.}},
\bauthor{\bsnm{{Tello}}, \binits{J.C.}},
\bauthor{\bsnm{{Vaduvescu}}, \binits{O.}},
\bauthor{\bsnm{{Wheatley}}, \binits{P.J.}},
\bauthor{\bsnm{{Wijers}}, \binits{R.A.M.J.}},
\bauthor{\bsnm{{Winters}}, \binits{J.M.}},
\bauthor{\bsnm{{Xu}}, \binits{D.}}:
\batitle{{An Extremely Luminous Panchromatic Outburst from the Nucleus of a
  Distant Galaxy}}.
\bjtitle{Science}
\bvolume{333},
\bfpage{199}
(\byear{2011})
{\href{https://arxiv.org/abs/1104.3356}{{arXiv:1104.3356}}}
{[astro-ph.HE]}.
\doiurl{10.1126/science.1207143}
\end{barticle}
\endbibitem

\bibitem{cenko12}
\begin{barticle}
\bauthor{\bsnm{{Cenko}}, \binits{S.B.}},
\bauthor{\bsnm{{Krimm}}, \binits{H.A.}},
\bauthor{\bsnm{{Horesh}}, \binits{A.}},
\bauthor{\bsnm{{Rau}}, \binits{A.}},
\bauthor{\bsnm{{Frail}}, \binits{D.A.}},
\bauthor{\bsnm{{Kennea}}, \binits{J.A.}},
\bauthor{\bsnm{{Levan}}, \binits{A.J.}},
\bauthor{\bsnm{{Holland}}, \binits{S.T.}},
\bauthor{\bsnm{{Butler}}, \binits{N.R.}},
\bauthor{\bsnm{{Quimby}}, \binits{R.M.}},
\bauthor{\bsnm{{Bloom}}, \binits{J.S.}},
\bauthor{\bsnm{{Filippenko}}, \binits{A.V.}},
\bauthor{\bsnm{{Gal-Yam}}, \binits{A.}},
\bauthor{\bsnm{{Greiner}}, \binits{J.}},
\bauthor{\bsnm{{Kulkarni}}, \binits{S.R.}},
\bauthor{\bsnm{{Ofek}}, \binits{E.O.}},
\bauthor{\bsnm{{Olivares E.}}, \binits{F.}},
\bauthor{\bsnm{{Schady}}, \binits{P.}},
\bauthor{\bsnm{{Silverman}}, \binits{J.M.}},
\bauthor{\bsnm{{Tanvir}}, \binits{N.R.}},
\bauthor{\bsnm{{Xu}}, \binits{D.}}:
\batitle{{Swift J2058.4+0516: Discovery of a Possible Second Relativistic Tidal
  Disruption Flare?}}
\bjtitle{\apj}
\bvolume{753},
\bfpage{77}
(\byear{2012})
{\href{https://arxiv.org/abs/1107.5307}{{arXiv:1107.5307}}}
{[astro-ph.HE]}.
\doiurl{10.1088/0004-637X/753/1/77}
\end{barticle}
\endbibitem

\bibitem{brown15}
\begin{barticle}
\bauthor{\bsnm{{Brown}}, \binits{G.C.}},
\bauthor{\bsnm{{Levan}}, \binits{A.J.}},
\bauthor{\bsnm{{Stanway}}, \binits{E.R.}},
\bauthor{\bsnm{{Tanvir}}, \binits{N.R.}},
\bauthor{\bsnm{{Cenko}}, \binits{S.B.}},
\bauthor{\bsnm{{Berger}}, \binits{E.}},
\bauthor{\bsnm{{Chornock}}, \binits{R.}},
\bauthor{\bsnm{{Cucchiaria}}, \binits{A.}}:
\batitle{{Swift J1112.2-8238: a candidate relativistic tidal disruption
  flare}}.
\bjtitle{\mnras}
\bvolume{452},
\bfpage{4297}--\blpage{4306}
(\byear{2015})
{\href{https://arxiv.org/abs/1507.03582}{{arXiv:1507.03582}}}
{[astro-ph.HE]}.
\doiurl{10.1093/mnras/stv1520}
\end{barticle}
\endbibitem

\bibitem{auchettl17}
\begin{barticle}
\bauthor{\bsnm{{Auchettl}}, \binits{K.}},
\bauthor{\bsnm{{Guillochon}}, \binits{J.}},
\bauthor{\bsnm{{Ramirez-Ruiz}}, \binits{E.}}:
\batitle{{New Physical Insights about Tidal Disruption Events from a
  Comprehensive Observational Inventory at X-Ray Wavelengths}}.
\bjtitle{\apj}
\bvolume{838}(\bissue{2}),
\bfpage{149}
(\byear{2017})
{\href{https://arxiv.org/abs/1611.02291}{{arXiv:1611.02291}}}
{[astro-ph.HE]}.
\doiurl{10.3847/1538-4357/aa633b}
\end{barticle}
\endbibitem

\bibitem{saxton20}
\begin{barticle}
\bauthor{\bsnm{{Saxton}}, \binits{R.}},
\bauthor{\bsnm{{Komossa}}, \binits{S.}},
\bauthor{\bsnm{{Auchettl}}, \binits{K.}},
\bauthor{\bsnm{{Jonker}}, \binits{P.G.}}:
\batitle{{X-Ray Properties of TDEs}}.
\bjtitle{\ssr}
\bvolume{216}(\bissue{5}),
\bfpage{85}
(\byear{2020}).
\doiurl{10.1007/s11214-020-00708-4}
\end{barticle}
\endbibitem

\bibitem{maguire20}
\begin{barticle}
\bauthor{\bsnm{{Maguire}}, \binits{K.}},
\bauthor{\bsnm{{Eracleous}}, \binits{M.}},
\bauthor{\bsnm{{Jonker}}, \binits{P.G.}},
\bauthor{\bsnm{{MacLeod}}, \binits{M.}},
\bauthor{\bsnm{{Rosswog}}, \binits{S.}}:
\batitle{{Tidal Disruptions of White Dwarfs: Theoretical Models and
  Observational Prospects}}.
\bjtitle{\ssr}
\bvolume{216}(\bissue{3}),
\bfpage{39}
(\byear{2020})
{\href{https://arxiv.org/abs/2004.00146}{{arXiv:2004.00146}}}
{[astro-ph.HE]}.
\doiurl{10.1007/s11214-020-00661-2}
\end{barticle}
\endbibitem

\bibitem{levan14}
\begin{barticle}
\bauthor{\bsnm{{Levan}}, \binits{A.J.}},
\bauthor{\bsnm{{Tanvir}}, \binits{N.R.}},
\bauthor{\bsnm{{Starling}}, \binits{R.L.C.}},
\bauthor{\bsnm{{Wiersema}}, \binits{K.}},
\bauthor{\bsnm{{Page}}, \binits{K.L.}},
\bauthor{\bsnm{{Perley}}, \binits{D.A.}},
\bauthor{\bsnm{{Schulze}}, \binits{S.}},
\bauthor{\bsnm{{Wynn}}, \binits{G.A.}},
\bauthor{\bsnm{{Chornock}}, \binits{R.}},
\bauthor{\bsnm{{Hjorth}}, \binits{J.}},
\bauthor{\bsnm{{Cenko}}, \binits{S.B.}},
\bauthor{\bsnm{{Fruchter}}, \binits{A.S.}},
\bauthor{\bsnm{{O'Brien}}, \binits{P.T.}},
\bauthor{\bsnm{{Brown}}, \binits{G.C.}},
\bauthor{\bsnm{{Tunnicliffe}}, \binits{R.L.}},
\bauthor{\bsnm{{Malesani}}, \binits{D.}},
\bauthor{\bsnm{{Jakobsson}}, \binits{P.}},
\bauthor{\bsnm{{Watson}}, \binits{D.}},
\bauthor{\bsnm{{Berger}}, \binits{E.}},
\bauthor{\bsnm{{Bersier}}, \binits{D.}},
\bauthor{\bsnm{{Cobb}}, \binits{B.E.}},
\bauthor{\bsnm{{Covino}}, \binits{S.}},
\bauthor{\bsnm{{Cucchiara}}, \binits{A.}},
\bauthor{\bsnm{{de Ugarte Postigo}}, \binits{A.}},
\bauthor{\bsnm{{Fox}}, \binits{D.B.}},
\bauthor{\bsnm{{Gal-Yam}}, \binits{A.}},
\bauthor{\bsnm{{Goldoni}}, \binits{P.}},
\bauthor{\bsnm{{Gorosabel}}, \binits{J.}},
\bauthor{\bsnm{{Kaper}}, \binits{L.}},
\bauthor{\bsnm{{Kr{\"u}hler}}, \binits{T.}},
\bauthor{\bsnm{{Karjalainen}}, \binits{R.}},
\bauthor{\bsnm{{Osborne}}, \binits{J.P.}},
\bauthor{\bsnm{{Pian}}, \binits{E.}},
\bauthor{\bsnm{{S{\'a}nchez-Ram{\'{\i}}rez}}, \binits{R.}},
\bauthor{\bsnm{{Schmidt}}, \binits{B.}},
\bauthor{\bsnm{{Skillen}}, \binits{I.}},
\bauthor{\bsnm{{Tagliaferri}}, \binits{G.}},
\bauthor{\bsnm{{Th{\"o}ne}}, \binits{C.}},
\bauthor{\bsnm{{Vaduvescu}}, \binits{O.}},
\bauthor{\bsnm{{Wijers}}, \binits{R.A.M.J.}},
\bauthor{\bsnm{{Zauderer}}, \binits{B.A.}}:
\batitle{{A New Population of Ultra-long Duration Gamma-Ray Bursts}}.
\bjtitle{\apj}
\bvolume{781},
\bfpage{13}
(\byear{2014})
{\href{https://arxiv.org/abs/1302.2352}{{arXiv:1302.2352}}}
{[astro-ph.HE]}.
\doiurl{10.1088/0004-637X/781/1/13}
\end{barticle}
\endbibitem

\bibitem{macleod14}
\begin{barticle}
\bauthor{\bsnm{{MacLeod}}, \binits{M.}},
\bauthor{\bsnm{{Goldstein}}, \binits{J.}},
\bauthor{\bsnm{{Ramirez-Ruiz}}, \binits{E.}},
\bauthor{\bsnm{{Guillochon}}, \binits{J.}},
\bauthor{\bsnm{{Samsing}}, \binits{J.}}:
\batitle{{Illuminating Massive Black Holes with White Dwarfs: Orbital Dynamics
  and High-energy Transients from Tidal Interactions}}.
\bjtitle{\apj}
\bvolume{794}(\bissue{1}),
\bfpage{9}
(\byear{2014})
{\href{https://arxiv.org/abs/1405.1426}{{arXiv:1405.1426}}}
{[astro-ph.HE]}.
\doiurl{10.1088/0004-637X/794/1/9}
\end{barticle}
\endbibitem

\bibitem{macleod16}
\begin{barticle}
\bauthor{\bsnm{{MacLeod}}, \binits{M.}},
\bauthor{\bsnm{{Guillochon}}, \binits{J.}},
\bauthor{\bsnm{{Ramirez-Ruiz}}, \binits{E.}},
\bauthor{\bsnm{{Kasen}}, \binits{D.}},
\bauthor{\bsnm{{Rosswog}}, \binits{S.}}:
\batitle{{Optical Thermonuclear Transients from Tidal Compression of White
  Dwarfs as Tracers of the Low End of the Massive Black Hole Mass Function}}.
\bjtitle{\apj}
\bvolume{819}(\bissue{1}),
\bfpage{3}
(\byear{2016})
{\href{https://arxiv.org/abs/1508.02399}{{arXiv:1508.02399}}}
{[astro-ph.HE]}.
\doiurl{10.3847/0004-637X/819/1/3}
\end{barticle}
\endbibitem

\bibitem{Perets16}
\begin{barticle}
\bauthor{\bsnm{{Perets}}, \binits{H.B.}},
\bauthor{\bsnm{{Li}}, \binits{Z.}},
\bauthor{\bsnm{{Lombardi}}, \binits{J.} \bsuffix{James~C.}},
\bauthor{\bsnm{{Milcarek}}, \binits{J.} \bsuffix{Stephen~R.}}:
\batitle{{Micro-tidal Disruption Events by Stellar Compact Objects and the
  Production of Ultra-long GRBs}}.
\bjtitle{\apj}
\bvolume{823}(\bissue{2}),
\bfpage{113}
(\byear{2016})
{\href{https://arxiv.org/abs/1602.07698}{{arXiv:1602.07698}}}
{[astro-ph.HE]}.
\doiurl{10.3847/0004-637X/823/2/113}
\end{barticle}
\endbibitem

\bibitem{NorrisBonnell06}
\begin{barticle}
\bauthor{\bsnm{{Norris}}, \binits{J.P.}},
\bauthor{\bsnm{{Bonnell}}, \binits{J.T.}}:
\batitle{{Short Gamma-Ray Bursts with Extended Emission}}.
\bjtitle{\apj}
\bvolume{643}(\bissue{1}),
\bfpage{266}--\blpage{275}
(\byear{2006})
{\href{https://arxiv.org/abs/astro-ph/0601190}{{arXiv:astro-ph/0601190}}}
{[astro-ph]}.
\doiurl{10.1086/502796}
\end{barticle}
\endbibitem

\bibitem{xue19}
\begin{barticle}
\bauthor{\bsnm{{Xue}}, \binits{Y.Q.}},
\bauthor{\bsnm{{Zheng}}, \binits{X.C.}},
\bauthor{\bsnm{{Li}}, \binits{Y.}},
\bauthor{\bsnm{{Brandt}}, \binits{W.N.}},
\bauthor{\bsnm{{Zhang}}, \binits{B.}},
\bauthor{\bsnm{{Luo}}, \binits{B.}},
\bauthor{\bsnm{{Zhang}}, \binits{B.-B.}},
\bauthor{\bsnm{{Bauer}}, \binits{F.E.}},
\bauthor{\bsnm{{Sun}}, \binits{H.}},
\bauthor{\bsnm{{Lehmer}}, \binits{B.D.}},
\bauthor{\bsnm{{Wu}}, \binits{X.-F.}},
\bauthor{\bsnm{{Yang}}, \binits{G.}},
\bauthor{\bsnm{{Kong}}, \binits{X.}},
\bauthor{\bsnm{{Li}}, \binits{J.Y.}},
\bauthor{\bsnm{{Sun}}, \binits{M.Y.}},
\bauthor{\bsnm{{Wang}}, \binits{J.-X.}},
\bauthor{\bsnm{{Vito}}, \binits{F.}}:
\batitle{{A magnetar-powered X-ray transient as the aftermath of a binary
  neutron-star merger}}.
\bjtitle{\nat}
\bvolume{568}(\bissue{7751}),
\bfpage{198}--\blpage{201}
(\byear{2019})
{\href{https://arxiv.org/abs/1904.05368}{{arXiv:1904.05368}}}
{[astro-ph.HE]}.
\doiurl{10.1038/s41586-019-1079-5}
\end{barticle}
\endbibitem

\bibitem{qv22}
\begin{barticle}
\bauthor{\bsnm{{Quirola-V{\'a}squez}}, \binits{J.}},
\bauthor{\bsnm{{Bauer}}, \binits{F.E.}},
\bauthor{\bsnm{{Jonker}}, \binits{P.G.}},
\bauthor{\bsnm{{Brandt}}, \binits{W.N.}},
\bauthor{\bsnm{{Yang}}, \binits{G.}},
\bauthor{\bsnm{{Levan}}, \binits{A.J.}},
\bauthor{\bsnm{{Xue}}, \binits{Y.Q.}},
\bauthor{\bsnm{{Eappachen}}, \binits{D.}},
\bauthor{\bsnm{{Zheng}}, \binits{X.C.}},
\bauthor{\bsnm{{Luo}}, \binits{B.}}:
\batitle{{Extragalactic fast X-ray transient candidates discovered by Chandra
  (2000-2014)}}.
\bjtitle{\aap}
\bvolume{663},
\bfpage{168}
(\byear{2022})
{\href{https://arxiv.org/abs/2201.07773}{{arXiv:2201.07773}}}
{[astro-ph.HE]}.
\doiurl{10.1051/0004-6361/202243047}
\end{barticle}
\endbibitem

\bibitem{heggie03}
\begin{bbook}
\bauthor{\bsnm{{Heggie}}, \binits{D.}},
\bauthor{\bsnm{{Hut}}, \binits{P.}}:
\bbtitle{The Gravitational Million-Body Problem: A Multidisciplinary Approach
  to Star Cluster Dynamics},
(\byear{2003})
\end{bbook}
\endbibitem

\bibitem{hoang18}
\begin{barticle}
\bauthor{\bsnm{{Hoang}}, \binits{B.-M.}},
\bauthor{\bsnm{{Naoz}}, \binits{S.}},
\bauthor{\bsnm{{Kocsis}}, \binits{B.}},
\bauthor{\bsnm{{Rasio}}, \binits{F.A.}},
\bauthor{\bsnm{{Dosopoulou}}, \binits{F.}}:
\batitle{{Black Hole Mergers in Galactic Nuclei Induced by the Eccentric
  Kozai-Lidov Effect}}.
\bjtitle{\apj}
\bvolume{856}(\bissue{2}),
\bfpage{140}
(\byear{2018})
{\href{https://arxiv.org/abs/1706.09896}{{arXiv:1706.09896}}}
{[astro-ph.HE]}.
\doiurl{10.3847/1538-4357/aaafce}
\end{barticle}
\endbibitem

\bibitem{graham20}
\begin{barticle}
\bauthor{\bsnm{{Graham}}, \binits{M.J.}},
\bauthor{\bsnm{{Ford}}, \binits{K.E.S.}},
\bauthor{\bsnm{{McKernan}}, \binits{B.}},
\bauthor{\bsnm{{Ross}}, \binits{N.P.}},
\bauthor{\bsnm{{Stern}}, \binits{D.}},
\bauthor{\bsnm{{Burdge}}, \binits{K.}},
\bauthor{\bsnm{{Coughlin}}, \binits{M.}},
\bauthor{\bsnm{{Djorgovski}}, \binits{S.G.}},
\bauthor{\bsnm{{Drake}}, \binits{A.J.}},
\bauthor{\bsnm{{Duev}}, \binits{D.}},
\bauthor{\bsnm{{Kasliwal}}, \binits{M.}},
\bauthor{\bsnm{{Mahabal}}, \binits{A.A.}},
\bauthor{\bsnm{{van Velzen}}, \binits{S.}},
\bauthor{\bsnm{{Belecki}}, \binits{J.}},
\bauthor{\bsnm{{Bellm}}, \binits{E.C.}},
\bauthor{\bsnm{{Burruss}}, \binits{R.}},
\bauthor{\bsnm{{Cenko}}, \binits{S.B.}},
\bauthor{\bsnm{{Cunningham}}, \binits{V.}},
\bauthor{\bsnm{{Helou}}, \binits{G.}},
\bauthor{\bsnm{{Kulkarni}}, \binits{S.R.}},
\bauthor{\bsnm{{Masci}}, \binits{F.J.}},
\bauthor{\bsnm{{Prince}}, \binits{T.}},
\bauthor{\bsnm{{Reiley}}, \binits{D.}},
\bauthor{\bsnm{{Rodriguez}}, \binits{H.}},
\bauthor{\bsnm{{Rusholme}}, \binits{B.}},
\bauthor{\bsnm{{Smith}}, \binits{R.M.}},
\bauthor{\bsnm{{Soumagnac}}, \binits{M.T.}}:
\batitle{{Candidate Electromagnetic Counterpart to the Binary Black Hole Merger
  Gravitational-Wave Event S190521g$^{*}$}}.
\bjtitle{\prl}
\bvolume{124}(\bissue{25}),
\bfpage{251102}
(\byear{2020})
{\href{https://arxiv.org/abs/2006.14122}{{arXiv:2006.14122}}}
{[astro-ph.HE]}.
\doiurl{10.1103/PhysRevLett.124.251102}
\end{barticle}
\endbibitem

\bibitem{graham22}
\begin{botherref}
\oauthor{\bsnm{{Graham}}, \binits{M.J.}},
\oauthor{\bsnm{{McKernan}}, \binits{B.}},
\oauthor{\bsnm{{Ford}}, \binits{K.E.S.}},
\oauthor{\bsnm{{Stern}}, \binits{D.}},
\oauthor{\bsnm{{Djorgovski}}, \binits{S.G.}},
\oauthor{\bsnm{{Coughlin}}, \binits{M.}},
\oauthor{\bsnm{{Burdge}}, \binits{K.B.}},
\oauthor{\bsnm{{Bellm}}, \binits{E.C.}},
\oauthor{\bsnm{{Helou}}, \binits{G.}},
\oauthor{\bsnm{{Mahabal}}, \binits{A.A.}},
\oauthor{\bsnm{{Masci}}, \binits{F.J.}},
\oauthor{\bsnm{{Purdum}}, \binits{J.}},
\oauthor{\bsnm{{Rosnet}}, \binits{P.}},
\oauthor{\bsnm{{Rusholme}}, \binits{B.}}:
{A light in the dark: searching for electromagnetic counterparts to black
  hole-black hole mergers in LIGO/Virgo O3 with the Zwicky Transient Facility}.
arXiv e-prints,
2209--13004
(2022)
{\href{https://arxiv.org/abs/2209.13004}{{arXiv:2209.13004}}}
{[astro-ph.HE]}
\end{botherref}
\endbibitem

\bibitem{mandel22}
\begin{barticle}
\bauthor{\bsnm{{Mandel}}, \binits{I.}},
\bauthor{\bsnm{{Broekgaarden}}, \binits{F.S.}}:
\batitle{{Rates of compact object coalescences}}.
\bjtitle{Living Reviews in Relativity}
\bvolume{25}(\bissue{1}),
\bfpage{1}
(\byear{2022})
{\href{https://arxiv.org/abs/2107.14239}{{arXiv:2107.14239}}}
{[astro-ph.HE]}.
\doiurl{10.1007/s41114-021-00034-3}
\end{barticle}
\endbibitem

\bibitem{Belczynski06}
\begin{barticle}
\bauthor{\bsnm{{Belczynski}}, \binits{K.}},
\bauthor{\bsnm{{Perna}}, \binits{R.}},
\bauthor{\bsnm{{Bulik}}, \binits{T.}},
\bauthor{\bsnm{{Kalogera}}, \binits{V.}},
\bauthor{\bsnm{{Ivanova}}, \binits{N.}},
\bauthor{\bsnm{{Lamb}}, \binits{D.Q.}}:
\batitle{{A Study of Compact Object Mergers as Short Gamma-Ray Burst
  Progenitors}}.
\bjtitle{\apj}
\bvolume{648}(\bissue{2}),
\bfpage{1110}--\blpage{1116}
(\byear{2006})
{\href{https://arxiv.org/abs/astro-ph/0601458}{{arXiv:astro-ph/0601458}}}
{[astro-ph]}.
\doiurl{10.1086/505169}
\end{barticle}
\endbibitem

\bibitem{Wiggins18}
\begin{barticle}
\bauthor{\bsnm{{Wiggins}}, \binits{B.K.}},
\bauthor{\bsnm{{Fryer}}, \binits{C.L.}},
\bauthor{\bsnm{{Smidt}}, \binits{J.M.}},
\bauthor{\bsnm{{Hartmann}}, \binits{D.}},
\bauthor{\bsnm{{Lloyd-Ronning}}, \binits{N.}},
\bauthor{\bsnm{{Belcynski}}, \binits{C.}}:
\batitle{{The Location and Environments of Neutron Star Mergers in an Evolving
  Universe}}.
\bjtitle{\apj}
\bvolume{865}(\bissue{1}),
\bfpage{27}
(\byear{2018})
{\href{https://arxiv.org/abs/1807.02853}{{arXiv:1807.02853}}}
{[astro-ph.HE]}.
\doiurl{10.3847/1538-4357/aad2d4}
\end{barticle}
\endbibitem

\bibitem{Chambers+16}
\begin{botherref}
\oauthor{\bsnm{{Chambers}}, \binits{K.C.}},
\oauthor{\bsnm{{Magnier}}, \binits{E.A.}},
\oauthor{\bsnm{{Metcalfe}}, \binits{N.}},
\oauthor{\bsnm{{Flewelling}}, \binits{H.A.}},
\oauthor{\bsnm{{Huber}}, \binits{M.E.}},
\oauthor{\bsnm{{Waters}}, \binits{C.Z.}},
\oauthor{\bsnm{{Denneau}}, \binits{L.}},
\oauthor{\bsnm{{Draper}}, \binits{P.W.}},
\oauthor{\bsnm{{Farrow}}, \binits{D.}},
\oauthor{\bsnm{{Finkbeiner}}, \binits{D.P.}},
\oauthor{\bsnm{{Holmberg}}, \binits{C.}},
\oauthor{\bsnm{{Koppenhoefer}}, \binits{J.}},
\oauthor{\bsnm{{Price}}, \binits{P.A.}},
\oauthor{\bsnm{{Rest}}, \binits{A.}},
\oauthor{\bsnm{{Saglia}}, \binits{R.P.}},
\oauthor{\bsnm{{Schlafly}}, \binits{E.F.}},
\oauthor{\bsnm{{Smartt}}, \binits{S.J.}},
\oauthor{\bsnm{{Sweeney}}, \binits{W.}},
\oauthor{\bsnm{{Wainscoat}}, \binits{R.J.}},
\oauthor{\bsnm{{Burgett}}, \binits{W.S.}},
\oauthor{\bsnm{{Chastel}}, \binits{S.}},
\oauthor{\bsnm{{Grav}}, \binits{T.}},
\oauthor{\bsnm{{Heasley}}, \binits{J.N.}},
\oauthor{\bsnm{{Hodapp}}, \binits{K.W.}},
\oauthor{\bsnm{{Jedicke}}, \binits{R.}},
\oauthor{\bsnm{{Kaiser}}, \binits{N.}},
\oauthor{\bsnm{{Kudritzki}}, \binits{R.-P.}},
\oauthor{\bsnm{{Luppino}}, \binits{G.A.}},
\oauthor{\bsnm{{Lupton}}, \binits{R.H.}},
\oauthor{\bsnm{{Monet}}, \binits{D.G.}},
\oauthor{\bsnm{{Morgan}}, \binits{J.S.}},
\oauthor{\bsnm{{Onaka}}, \binits{P.M.}},
\oauthor{\bsnm{{Shiao}}, \binits{B.}},
\oauthor{\bsnm{{Stubbs}}, \binits{C.W.}},
\oauthor{\bsnm{{Tonry}}, \binits{J.L.}},
\oauthor{\bsnm{{White}}, \binits{R.}},
\oauthor{\bsnm{{Ba{\~n}ados}}, \binits{E.}},
\oauthor{\bsnm{{Bell}}, \binits{E.F.}},
\oauthor{\bsnm{{Bender}}, \binits{R.}},
\oauthor{\bsnm{{Bernard}}, \binits{E.J.}},
\oauthor{\bsnm{{Boegner}}, \binits{M.}},
\oauthor{\bsnm{{Boffi}}, \binits{F.}},
\oauthor{\bsnm{{Botticella}}, \binits{M.T.}},
\oauthor{\bsnm{{Calamida}}, \binits{A.}},
\oauthor{\bsnm{{Casertano}}, \binits{S.}},
\oauthor{\bsnm{{Chen}}, \binits{W.-P.}},
\oauthor{\bsnm{{Chen}}, \binits{X.}},
\oauthor{\bsnm{{Cole}}, \binits{S.}},
\oauthor{\bsnm{{Deacon}}, \binits{N.}},
\oauthor{\bsnm{{Frenk}}, \binits{C.}},
\oauthor{\bsnm{{Fitzsimmons}}, \binits{A.}},
\oauthor{\bsnm{{Gezari}}, \binits{S.}},
\oauthor{\bsnm{{Gibbs}}, \binits{V.}},
\oauthor{\bsnm{{Goessl}}, \binits{C.}},
\oauthor{\bsnm{{Goggia}}, \binits{T.}},
\oauthor{\bsnm{{Gourgue}}, \binits{R.}},
\oauthor{\bsnm{{Goldman}}, \binits{B.}},
\oauthor{\bsnm{{Grant}}, \binits{P.}},
\oauthor{\bsnm{{Grebel}}, \binits{E.K.}},
\oauthor{\bsnm{{Hambly}}, \binits{N.C.}},
\oauthor{\bsnm{{Hasinger}}, \binits{G.}},
\oauthor{\bsnm{{Heavens}}, \binits{A.F.}},
\oauthor{\bsnm{{Heckman}}, \binits{T.M.}},
\oauthor{\bsnm{{Henderson}}, \binits{R.}},
\oauthor{\bsnm{{Henning}}, \binits{T.}},
\oauthor{\bsnm{{Holman}}, \binits{M.}},
\oauthor{\bsnm{{Hopp}}, \binits{U.}},
\oauthor{\bsnm{{Ip}}, \binits{W.-H.}},
\oauthor{\bsnm{{Isani}}, \binits{S.}},
\oauthor{\bsnm{{Jackson}}, \binits{M.}},
\oauthor{\bsnm{{Keyes}}, \binits{C.D.}},
\oauthor{\bsnm{{Koekemoer}}, \binits{A.M.}},
\oauthor{\bsnm{{Kotak}}, \binits{R.}},
\oauthor{\bsnm{{Le}}, \binits{D.}},
\oauthor{\bsnm{{Liska}}, \binits{D.}},
\oauthor{\bsnm{{Long}}, \binits{K.S.}},
\oauthor{\bsnm{{Lucey}}, \binits{J.R.}},
\oauthor{\bsnm{{Liu}}, \binits{M.}},
\oauthor{\bsnm{{Martin}}, \binits{N.F.}},
\oauthor{\bsnm{{Masci}}, \binits{G.}},
\oauthor{\bsnm{{McLean}}, \binits{B.}},
\oauthor{\bsnm{{Mindel}}, \binits{E.}},
\oauthor{\bsnm{{Misra}}, \binits{P.}},
\oauthor{\bsnm{{Morganson}}, \binits{E.}},
\oauthor{\bsnm{{Murphy}}, \binits{D.N.A.}},
\oauthor{\bsnm{{Obaika}}, \binits{A.}},
\oauthor{\bsnm{{Narayan}}, \binits{G.}},
\oauthor{\bsnm{{Nieto-Santisteban}}, \binits{M.A.}},
\oauthor{\bsnm{{Norberg}}, \binits{P.}},
\oauthor{\bsnm{{Peacock}}, \binits{J.A.}},
\oauthor{\bsnm{{Pier}}, \binits{E.A.}},
\oauthor{\bsnm{{Postman}}, \binits{M.}},
\oauthor{\bsnm{{Primak}}, \binits{N.}},
\oauthor{\bsnm{{Rae}}, \binits{C.}},
\oauthor{\bsnm{{Rai}}, \binits{A.}},
\oauthor{\bsnm{{Riess}}, \binits{A.}},
\oauthor{\bsnm{{Riffeser}}, \binits{A.}},
\oauthor{\bsnm{{Rix}}, \binits{H.W.}},
\oauthor{\bsnm{{R{\"o}ser}}, \binits{S.}},
\oauthor{\bsnm{{Russel}}, \binits{R.}},
\oauthor{\bsnm{{Rutz}}, \binits{L.}},
\oauthor{\bsnm{{Schilbach}}, \binits{E.}},
\oauthor{\bsnm{{Schultz}}, \binits{A.S.B.}},
\oauthor{\bsnm{{Scolnic}}, \binits{D.}},
\oauthor{\bsnm{{Strolger}}, \binits{L.}},
\oauthor{\bsnm{{Szalay}}, \binits{A.}},
\oauthor{\bsnm{{Seitz}}, \binits{S.}},
\oauthor{\bsnm{{Small}}, \binits{E.}},
\oauthor{\bsnm{{Smith}}, \binits{K.W.}},
\oauthor{\bsnm{{Soderblom}}, \binits{D.R.}},
\oauthor{\bsnm{{Taylor}}, \binits{P.}},
\oauthor{\bsnm{{Thomson}}, \binits{R.}},
\oauthor{\bsnm{{Taylor}}, \binits{A.N.}},
\oauthor{\bsnm{{Thakar}}, \binits{A.R.}},
\oauthor{\bsnm{{Thiel}}, \binits{J.}},
\oauthor{\bsnm{{Thilker}}, \binits{D.}},
\oauthor{\bsnm{{Unger}}, \binits{D.}},
\oauthor{\bsnm{{Urata}}, \binits{Y.}},
\oauthor{\bsnm{{Valenti}}, \binits{J.}},
\oauthor{\bsnm{{Wagner}}, \binits{J.}},
\oauthor{\bsnm{{Walder}}, \binits{T.}},
\oauthor{\bsnm{{Walter}}, \binits{F.}},
\oauthor{\bsnm{{Watters}}, \binits{S.P.}},
\oauthor{\bsnm{{Werner}}, \binits{S.}},
\oauthor{\bsnm{{Wood-Vasey}}, \binits{W.M.}},
\oauthor{\bsnm{{Wyse}}, \binits{R.}}:
{The Pan-STARRS1 Surveys}.
arXiv e-prints,
1612--05560
(2016)
{\href{https://arxiv.org/abs/1612.05560}{{arXiv:1612.05560}}}
{[astro-ph.IM]}
\end{botherref}
\endbibitem

\bibitem{McMahon+2013}
\begin{barticle}
\bauthor{\bsnm{{McMahon}}, \binits{R.G.}},
\bauthor{\bsnm{{Banerji}}, \binits{M.}},
\bauthor{\bsnm{{Gonzalez}}, \binits{E.}},
\bauthor{\bsnm{{Koposov}}, \binits{S.E.}},
\bauthor{\bsnm{{Bejar}}, \binits{V.J.}},
\bauthor{\bsnm{{Lodieu}}, \binits{N.}},
\bauthor{\bsnm{{Rebolo}}, \binits{R.}},
\bauthor{\bsnm{{VHS Collaboration}}}:
\batitle{{First Scientific Results from the VISTA Hemisphere Survey (VHS)}}.
\bjtitle{The Messenger}
\bvolume{154},
\bfpage{35}--\blpage{37}
(\byear{2013})
\end{barticle}
\endbibitem

\bibitem{WISE_Wright+10}
\begin{barticle}
\bauthor{\bsnm{{Wright}}, \binits{E.L.}},
\bauthor{\bsnm{{Eisenhardt}}, \binits{P.R.M.}},
\bauthor{\bsnm{{Mainzer}}, \binits{A.K.}},
\bauthor{\bsnm{{Ressler}}, \binits{M.E.}},
\bauthor{\bsnm{{Cutri}}, \binits{R.M.}},
\bauthor{\bsnm{{Jarrett}}, \binits{T.}},
\bauthor{\bsnm{{Kirkpatrick}}, \binits{J.D.}},
\bauthor{\bsnm{{Padgett}}, \binits{D.}},
\bauthor{\bsnm{{McMillan}}, \binits{R.S.}},
\bauthor{\bsnm{{Skrutskie}}, \binits{M.}},
\bauthor{\bsnm{{Stanford}}, \binits{S.A.}},
\bauthor{\bsnm{{Cohen}}, \binits{M.}},
\bauthor{\bsnm{{Walker}}, \binits{R.G.}},
\bauthor{\bsnm{{Mather}}, \binits{J.C.}},
\bauthor{\bsnm{{Leisawitz}}, \binits{D.}},
\bauthor{\bsnm{{Gautier}}, \binits{I.} \bsuffix{Thomas~N.}},
\bauthor{\bsnm{{McLean}}, \binits{I.}},
\bauthor{\bsnm{{Benford}}, \binits{D.}},
\bauthor{\bsnm{{Lonsdale}}, \binits{C.J.}},
\bauthor{\bsnm{{Blain}}, \binits{A.}},
\bauthor{\bsnm{{Mendez}}, \binits{B.}},
\bauthor{\bsnm{{Irace}}, \binits{W.R.}},
\bauthor{\bsnm{{Duval}}, \binits{V.}},
\bauthor{\bsnm{{Liu}}, \binits{F.}},
\bauthor{\bsnm{{Royer}}, \binits{D.}},
\bauthor{\bsnm{{Heinrichsen}}, \binits{I.}},
\bauthor{\bsnm{{Howard}}, \binits{J.}},
\bauthor{\bsnm{{Shannon}}, \binits{M.}},
\bauthor{\bsnm{{Kendall}}, \binits{M.}},
\bauthor{\bsnm{{Walsh}}, \binits{A.L.}},
\bauthor{\bsnm{{Larsen}}, \binits{M.}},
\bauthor{\bsnm{{Cardon}}, \binits{J.G.}},
\bauthor{\bsnm{{Schick}}, \binits{S.}},
\bauthor{\bsnm{{Schwalm}}, \binits{M.}},
\bauthor{\bsnm{{Abid}}, \binits{M.}},
\bauthor{\bsnm{{Fabinsky}}, \binits{B.}},
\bauthor{\bsnm{{Naes}}, \binits{L.}},
\bauthor{\bsnm{{Tsai}}, \binits{C.-W.}}:
\batitle{{The Wide-field Infrared Survey Explorer (WISE): Mission Description
  and Initial On-orbit Performance}}.
\bjtitle{\aj}
\bvolume{140}(\bissue{6}),
\bfpage{1868}--\blpage{1881}
(\byear{2010})
{\href{https://arxiv.org/abs/1008.0031}{{arXiv:1008.0031}}}
{[astro-ph.IM]}.
\doiurl{10.1088/0004-6256/140/6/1868}
\end{barticle}
\endbibitem

\end{thebibliography}
\end{document}